\documentclass[6pt,oneside,final,twocolumn]{IEEEtran}

\usepackage[colorlinks,urlcolor=blue,linkcolor=blue,citecolor=blue]{hyperref}

\usepackage{color,array,amsthm}

\usepackage{amsmath}
\usepackage{amsfonts}
\usepackage{amssymb}
\usepackage{amsthm}
\usepackage{commath}
\usepackage{mathrsfs}
\usepackage{graphicx,cite,calc}
\usepackage{filecontents}
\usepackage{subfigure}
\usepackage{wasysym}
\usepackage{lineno}
\usepackage{ifsym}
\usepackage{pstricks}
\usepackage{hhline}

\begin{document}

	\title{{Integrated MIMO Passive Radar Target Detection}} 
	\author {{ Amir Zaimbashi,~\IEEEmembership{Senior Member,~IEEE}, Maria Sabrina Greco,~\IEEEmembership{Fellow,~IEEE}, Fulvio Gini,~\IEEEmembership{Fellow,~IEEE}}
		\thanks{A. Zaimbashi is with the Optical and RF Communication Systems (ORCS) Laboratory, Department of Electrical Engineering, Shahid Bahonar University of Kerman, Kerman, Iran, 76169-14111.  (e-mail: a.zaimbashi@uk.ac.ir). He is currently a visiting fellow working in the Department of Information Engineering at the University of Pisa.\\ M. S. Greco is with Università di Pisa, Dipartimento di Ingegneria dell'Informazione, Via Caruso, 56122, Pisa, Italy. (email: maria.greco@unipi.it). \\ 	F. Gini is with Università di Pisa, Dipartimento di Ingegneria dell'Informazione, Via Caruso, 56122, Pisa, Italy. (email: fulvio.gini@unipi.it).\\The work of  M.S. Greco and F. Gini has been partially supported by the 
			Italian Ministry of Education and Research (MUR) in the framework of the 
			FoReLab project (Departments of Excellence).}
	}
	\markboth{}%
	\IEEEtitleabstractindextext{%

		\maketitle

		\maketitle

		\begin{abstract} Integrated passive radar (IPR) can be regarded as  next generation passive radar technology, which aims to integrate communication and radar systems. Unlike conventional passive radar, which does not prioritize communication-centric radar technology, IPR technology places a higher priority on incorporating specific radar constraints to develop waveforms that are better suited for radar applications. This paper deals with the problem of distributed MIMO IPR target detection under uncalibrated surveillance/reference receivers. We focus on a communication-centric radar system consisting of several opportunity transmitters non-overlapping in frequency with the same bandwidth and several spatially separated receivers. Five new detectors are devised according to the likelihood ratio test (LRT),  Rao, Wald, Gradient and Durbin criteria. Although, it is shown that these detectors are asymptotically equivalent, they provide different performance in the presence of  noisy reference channels.  { The invariance principle is applied in this paper to show that all uncertainties affecting threshold setting can be unified in the direct-path signal power-to-noise power ratio (DNR) of the reference channels.} To ensure effective detection threshold setting regardless of the DNR values in the reference channels, we introduce a novel strategy to adjust the level of the proposed detectors rather than their sizes. Then, we examine false alarm regulations and detection performance of the fixed-level proposed detectors to demonstrate their effectiveness compared to several existing detectors. Thus, we create a unified framework for uncalibrated MIMO IPR target detection in the presence of noisy reference channels.
		\end{abstract}
		

		\section{INTRODUCTION}
{Passive radar is a type of Joint Communication and Radar (JCR) system that utilizes the transmission from an existing communication system to detect targets, eliminating the need for a dedicated radar transmitter. In the design, the communication system is given high priority, while passive radar operation is considered an add-on with no priority. This approach can be described as a communication-centric radar system with low priority on the radar system \cite{blu,has,p5zhang}. In this special case, the transmitted signals are not specifically designed for radar applications; instead, they are intercepted and processed by a passive radar receiver employing optimized signal processing algorithms. Fortunately, modern digital transmitters for broadcast communications, such as digital video broadcast-terrestrial (DVB-T), digital audio broadcast (DAB), and 5G signals, offer waveforms better suited for radar applications than those available in the past \cite{xxx}. To further enhance radar capabilities, future communication-centric radar systems may aim to carefully manipulate communication waveforms to align with radar requirements \cite{amin, p5kang, p5fan, Liux}. This type of radar system, termed Nexus radar, exemplifies the convergence of communication and radar technologies and underscores the advanced integration of these systems in next-generation passive radar. Consequently, two types of passive radar systems employing opportunity signals will emerge: legacy and next-generation passive radars.
		\begin{itemize}
			\item  Conventional Passive Radar (CPR) technology utilizes non-radar communication or broadcast signals to detect targets (see \cite{mybook, bookGB} and references therein). Extracting useful information from these signals necessitates sophisticated signal processing due to various factors that can hinder the detection process, such as interference from direct-path and multi-path signals, and the presence of strong targets that may mask weaker ones. Designing effective passive radar systems demands careful consideration of these factors when modeling the received signal and developing target detection methods. Detecting multiple targets poses another significant challenge that requires meticulous attention in this system. These types of passive radars are categorized as legacy systems
			\item  Integrated Passive Radar (IPR), also known as next-generation passive radar, is a type of radar technology that prioritizes communication but incorporates specific radar constraints to design waveforms better suited for radar applications. This technology emphasizes the integration of communication and radar systems, resulting in communication-centric radar technology. Unlike traditional passive radars, IPR places greater emphasis on the radar system to generate suitable waveforms, enhancing the efficiency and straightforwardness of target detection algorithms due to the improved waveform quality.
		\end{itemize}}
		In contrast to the active radar, the CPR systems use illuminators of opportunity (IO) signals, which are not specifically designed for radar applications and are generally unknown to PR receivers \cite{mybook}. It implies that it is required to focus on the PR receiver to design optimal signal processing algorithms. Some systematic detection-theory frameworks have been presented in \cite{p1c2zim6}-\cite{GE}, covering signal modeling, detection algorithms, and analytical analysis for the CPR target detection problems. In \cite{p1c2zim6}, for example, the target detection problem in a single-input single-output (SISO) (i.e., bistatic CPR)  CPR system is formulated as an M-ary hypothesis test for multi-target scenarios. The proposed approach is a forward and sequential GLR-based detector, which detects targets one at a time and treats previously detected targets as interferences, enabling detection of even weak targets in multi-target scenarios. In all these works, it is assumed that a spatial and/or temporal filter is used to separate the direct-path signal and the target-path signal into two channels; the reference channel (RC) and the surveillance channel (SC). The RC comprises a delayed copy of the transmitted signal and is utilized by the passive radar to detect targets and eliminate interference in the SC.  \par 
{While the aforementioned works addressed the multi-target detection problem within the context of CPR, they presupposed the availability of a high-quality RC. In real-world scenarios, however, the direct-path signal on the RC can be contaminated by thermal noise and multipath components, potentially leading to a significant degradation in detection performance. To alleviate the detrimental effects of multipath on the RC, an effective equalization algorithm, known as the signal conditioning algorithm, must be employed \cite{mybook, bibid}. Following equalization, thermal noise may be remained the primary factor limiting system performance. As a result, we are compelled to assume that the transmitted signal is unknown. This assumption, along with the presence of strong interfering target signals on the SC that can induce masking effects, can considerably complicate the PR target detection problem. Therefore, a transition from CPR to IPR is essential, as it enhances waveform quality for radar applications, effectively simplifying signal modeling of the SC by disregarding the effects of interfering target signals. Several studies have addressed single-target detection problems in the context of IPR \cite{p1c2him1}-\cite{GE}.}	The authors of \cite{p1c2him1} have developed a generalized likelihood ratio test (GLRT) for calibrated passive multiple-input multiple-output (MIMO) radar (PMR) networks, which is called PMR-GLRT. The PMR-GLRT detector uses reference and surveillance signals to exploit all correlations within the measurement data, and it outperforms other PMR detectors that  use only some of these correlations.  In \cite{p1c2him1}, it was assumed that the noise variances in the RC and SC were known. Regardless of the simple structure of the PMR-GLRT, its performance severly degrades with the noise variance uncertainty (NVU)\footnote{The presence of NVU indicates that the actual (true) thermal noise variance deviates from the nominal (assumed) value.}.  The authors of \cite{p1c2gog} consider a calibrated single-input single-output (SISO) two-channel IPR configuration and propose a new detector based on the cross-correlation of the principal left singular vectors of the received data matrix in the RC and SC to exploit the rank-1 structure of the transmitted signal. In \cite{ref3}, GLRT-based detectors have been developed for various assumptions concerning signal and noise models in a multiple-input single-output (MISO) two-channel IPR scenario. In this setup, both the RC and the SC utilized multiantenna arrays to exploit the inherent subspace structure of the received signals. The authors of \cite{Ram} also exploit the cyclostationary property of the illuminator of opportunity signals to devise new detectors for a calibrated SISO two-channel IPR.  The detection thresholds of the above detectors are strongly influenced by the values of the direct-path signal power-to-noise power ratio (DNR) of the RC, particularly in a low DNR regime, which may result in a high false alarm rate and/or reduced target detection performance.  In practical situations, it is crucial to consider the impact of the DNR of the RC on the threshold setting, which has been overlooked in these studies. Moreover, the methods used for detection in \cite{p1c2gog,ref3,Ram} cannot be directly applied to the distributed MIMO configuration, which is the focus of our paper.
		In \cite{p1c2amir2}, a calibrated SISO two-channel PR structure is considered and new detectors are devised based on detection theory and kernel theory. In \cite{p1c2amir1}, this approach is respectively extended to calibrated single-input multiple-output (SIMO) and MIMO configurations, where new detectors have been proposed. The authors of \cite{p1c2him3}-\cite{DD} have focused on the single-channel IPR target detection problem, where they utilize only a single-channel surveillance channel without a reference channel. Therefore, these methods are referred to as single-channel target detection methods. Instead, they employed multiple surveillance channels to gather more data samples within a given observation time to address the target detection problem. Some of these works have specifically tackled the issue of uncalibrated IPR target detection. In \cite{p1c2amir3} and \cite{GE}, a framework was established within the context of single-channel  multistatic IPR under uncalibrated surveillance receivers.
		
		{ A comprehensive review of single-channel and two-channel detection methods, covering signal modeling, parameter assumptions, and algorithm development, is provided in \cite{mybook}. However, this paper deals with the problem of two-channel distributed MIMO IPR target detection under uncalibrated receivers. The main contributions of this work are summarized as follows:
		\begin{itemize}
			\item We propose five new detectors based on different criteria, including LRT, Rao, Wald, Durbin, and Gradient, designed for IPR systems. The detector design is within the framework of two channels, namely the reference and surveillance channels, establishing a framework for two-channel distributed MIMO IPR target detection under uncalibrated distributed receivers. A low-complexity implementation of the LRT-based detector is also derived for long-range PR utilizing a long integration time.
			\item In contrast to the signal-channel IPR target detection algorithms,  our approach leverages reference channels to double the number of observations, enabling us to treat the transmitter signals as unknown. Nevertheless, the integration of noisy reference channels introduces complications in the threshold selection procedure. For the first time in the literature, the invariance principle is applied in this paper to show that all uncertainties affecting threshold setting can be unified in the DNR of the reference channels. To ensure effective detection thresholds regardless of DNR values, we introduce a novel strategy to achieve a fixed level-of-test rather than a fixed size-of-test (CFAR test), as obtaining CFAR detectors is not feasible under conditions of DNR uncertainties.
			\item The proposed detectors are evaluated and compared with existing detectors in the literature. The study results provide evidence that the proposed IPR target detectors offer significant advantages over their counterparts. Among the proposed detectors, the LRT-based detector exhibits the lowest computational complexity compared to the others. However, the Rao test outperforms them in terms of both detection performance and false alarm regulation.
			\item The influence of multipath signals on the reference channels is examined. The results indicate that the proposed detectors maintain fixed levels in the presence of multipath signals on the reference channels. Importantly, we demonstrate for the first time that the concept of a fixed-level-test extends the applicability of detectors beyond conditions of DNR uncertainty to the presence of imperfect signal conditioning in the reference channels.
		\end{itemize}}
		The paper is structured as follows. In Section \ref{sec2}, the signal model of a two-channel MIMO IPR is presented, where the detection of IPR targets is formulated as a problem of composite binary hypothesis testing. The issue of invariance in the considered problem is explored in Section \ref{sec3}. New detectors based on LRT, Rao, Wald, Gradient, and Durbin criteria are developed in Section \ref{sec4}. The computational complexity of the proposed detectors is discussed in Section \ref{sec5}. Section \ref{sec6} presents the Monte-Carlo simulation results for the proposed detectors and existing methods. These results clearly demonstrate the advantages of the proposed methods over other approaches. Finally,  in Section \ref{sec7}, we summarize the findings and suggest directions for future research.
		\section{Signal Modeling}\label{sec2} 
		In this section, the focus is on detecting a target in a multistatic IPR system. The communication-centric radar system consists of $\mathrm{{\mathrm{N_{t}}}}$ transmitters and $\mathrm{N_{\rm r}}$ two-channel distributed and uncalibrated receivers. The transmitters, positioned at $\{\mathbf{t}_{{\mathrm{j}}}\}_{{\mathrm{j}}=1}^{\mathrm{{\mathrm{N_{t}}}}}$\footnote{{\it{Notation-}}The transpose, complex conjugate, and complex conjugate transpose of a matrix are represented by superscripts $(.)^{{\rm{T}}}$, $(.)^{{\rm{*}}}$, and $(.)^{{\rm{H}}}$, respectively. Scalars are denoted by non-boldface characters, vectors by boldface lowercase characters, and matrices by boldface uppercase characters. The Euclidean norm of a vector is denoted by $\|{\mathbf{a}}\|$, and the modulus of a complex number $\boldsymbol{\alpha}$ is represented by $|\boldsymbol{\alpha}|$. The Frobenius norm and $\rm{L_2}$ induced norm of a matrix $\mathbf{A}$ are represented by $\|{\mathbf{A}}\|_{\rm{F}}$ and $\|{\mathbf{A}}\|_2$, respectively. The notation $\det (\mathbf{D})$ represents the determinant of a square matrix $\mathbf{D}$. The $\rm (m,n)$ entry of matrix $\mathbf{A}$ is represented by $[\mathbf{A}]_{\rm mn}$, and the $n$-th column (row) of matrix $\mathbf{A}$ is represented by $[\mathbf{A}]_{n}$ ($[\mathbf{A}]_{n}$). The $n$-th element of vector $\mathbf{a}$ is represented by $[\mathbf{a}]_{n}$. The square diagonal matrix that operates on $\mathbf{a}$ vector and forms a diagonal matrix with the elements of the $\mathbf{a}$ vector along the diagonal is denoted by $\rm{diag}(\mathbf{a})$, and the block diagonal matrix with submatrices $\mathbf{A}^{\rm{(1)}},...,\mathbf{A}^{\rm{(P)}}$ along the diagonal and zeros elsewhere is represented by $\rm{Diag}(\mathbf{A}^{\rm{(1)}},...,\mathbf{A}^{\rm{(P)}})$. The identity matrix of size ${\rm{L}}$ is represented by $\mathbf{I}_{\rm{L}}$, and the all-zeros matrix of size ${\rm{M\times N}}$ is represented by $\mathbf{0}_{\rm{M\times N}}$. The expectation operator is denoted by $\rm E\{.\}$, and the real and imaginary parts of a complex number $\zeta$ are represented by $\Re(\zeta)$ and $\Im (\zeta)$. $\otimes$ is the Kronecher product operator. A complex (proper) Gaussian-distributed vector with mean $\boldsymbol{\mu}$ and covariance matrix $\mathbf{R}$ is denoted by $\mathbf{x}\sim\mathcal{CN}(\boldsymbol{\mu},\mathbf{R})$. $\mathrm{\mathbf{X} \sim\mathcal{C N}_{N \times M}(\mathbf{A}, \mathbf{B}, \mathbf{C})}$ denotes a complex (proper) Gaussian-distributed matrix $\mathbf{X}$ with mean $\mathbf{A} \in \mathbb{C}^{N \times M}$ and $\operatorname{Cov}[\operatorname{vec}(\mathbf{X})]=\mathrm{\mathbf{B}} \otimes \mathrm{\mathbf{C}}$. The eigenvector associated with the largest eigenvalue of matrix $\mathbf{A}$ (i.e., $\lambda_{\mathrm{max}}(\mathbf{A})$) is represented by $\mathbf{e}_{1}(\mathbf{A})$. The supremum of $f(x)$ over $\mathcal{A}$, denoted as $\sup_{\mathcal{A}}(f(x))$, is the smallest upper bound of the set of all values that $f(x)$ takes on in $\mathcal{A}$. Similarly, the infimum of $g(x)$ over $\mathcal{B}$, denoted as $\inf_{\mathcal{B}}(g(x))$, is the greatest lower bound of the set of all values that $g(x)$ takes on in $\mathcal{B}$. Finally, $\mathsf{j}=\sqrt{-1}$ is defined.}}, emit signals $\{\mathbf{s}_{{\mathrm{j}}}\}_{{\mathrm{j}}=1}^{\mathrm{{\mathrm{N_{t}}}}}$ to illuminate a scene that contains a target.  The signals transmitted by all transmitters have the same bandwidth  $\rm B$ and do not overlap in frequency, where $f_{\rm j}^{(c)}$ represents the carrier frequency of the ${\mathrm{j}}$-th transmitter. The target is located at position $\mathbf{r}_{\text{T}}$ and is moving with a velocity of $\mathbf{v}_{\text{T}}$. The $\mathrm{k}$-th receiver is located at position $\mathbf{r}_{{\mathrm{k}}}$. All transmitters and receivers are stationary and their positions are known.   At each receiver, the received signal is demodulated to baseband, and then sampled at a rate of ${f_s}$ to generate sampled signals of length $\mathrm{L}$, where $\mathrm{L=T}f_{\rm s}$ and $\mathrm{T}$ represents the integration time. The sampled complex baseband signals obtained from the ${\mathrm{k}}$-th reference  and surveillance receivers in the ${\mathrm{j}}$-th frequency channel are represented by $\mathbf{x}_{\mathrm{ jk}}$ and $\mathbf{y}_{\mathrm{jk}}$, expressed respectively as
	\begin{align}\label{eq2x}
		\left\{
		\begin{array}{ll}
			{\mathbf{y}}_{{\rm jk}}={\beta}_{{\rm jk}}{{{\rm \ }{\mathbf{D}}}_{{\rm d}_{{\rm jk}}}(0){\mathbf{s}}}_{{\rm j}}{\rm +}{\mathbf{e}}_{{\rm jk}},  \\
			{\mathbf{x}}_{{\rm jk}}={{\alpha}}_{{\rm jk}}{{{\rm \ }{\mathbf{D}}}_{{\rm p}_{{\rm jk}}}({\rm f_{jk}){\mathbf{s}}}}_{{\rm j}}{\rm +}{\mathbf{n}}_{{\rm jk}}
		\end{array}
		\right.
	\end{align}
	where  ${\mathbf{s}}_{{\rm j}}\in\mathbb{C}^{\mathrm{L}\times1}$ is in the time-domain vector signal transmitted by the $\rm j$-th transmitter, and consists of $\mathrm{L}$ unknown complex values. The coefficient ${\beta}_{{\rm jk}}$ is used to account for the complex scaling involved in transmitting signals from the $j$th transmitter to the $k$th receiver. It takes into consideration various aspects of signal propagation such as the distance between  transmitter and receiver, the nature of transmitting and receiving antennas, and the characteristics of the medium through which the signal travels.  ${\mathbf{D}}_{{\rm \tau}}({ f)\in\mathbb{C}^{\mathrm{L}\times \mathrm{L}}}$ is a matrix-based operator that performs time delay and Doppler shift on a signal. The time delay operation shifts the signal by a time interval of $\tau$, while the Doppler shift operation changes the frequency of the signal by ${f}$. This unitary operator is defined as ${\mathbf{D}}_{{\rm \tau}}({\rm f)=\mathbf{B}(f\hskip 0.05cm f_{\rm s}^{-1})\mathbf{F}^{H}\mathbf{B}(-\tau\Delta{f})\mathbf{F}}$ with the matrix $\mathbf{B}(a)$ being a diagonal matrix of size $\mathrm{L}\times \mathrm{L}$ with diagonal entries given by the expression $[\mathbf{B}(a)]_{ii}=e^{-\mathsf{j}2\pi(i-1)a}$, where $a$ is a scalar and $i$ ranges from $1$ to $\mathrm{L}$. The matrix $\mathbf{F}$ is an $\mathrm{L}$-point Discrete Fourier Transform (DFT) matrix of size $\mathrm{L}\times \mathrm{L}$. Its elements are given by $[\mathbf{F}]_{i,l}={\mathrm{L}}^{-0.5}e^{-\mathsf{j}2\pi(i-1)\frac{\Delta{f}}{f_{\rm s}}(l-1)}$ for $i,l=1,...,\mathrm{L}$, where $\Delta{f}=\frac{f_{\rm s}}{\mathrm{L}}$ is the frequency bin of $\mathrm{L}$-point DFT. For example, the term ${\mathbf{D}}_{{\rm p}_{{\rm jk}}}({\rm f_{jk}})$ represents a unitary linear operator that applies a time delay of ${{\rm p}_{{\rm jk}}}$ and a Doppler shift of ${\rm f_{jk}}$ to the input signal. Similarly, the term ${{{\rm \ }{\mathbf{D}}}_{{\rm d}_{{\rm jk}}}(0)}$ represents a unitary linear operator that applies a time delay of ${{\rm d}_{{\rm jk}}}$ and a zero Doppler shift to the input signal. The propagation delays of both the direct-path and the target-path can be respectively obtained from ${\rm d_{{\rm jk}}}  ={c}^{-1}{||\mathbf{r}_{\text{k}}-\mathbf{t}_{{\mathrm{j}}}||}$ and  ${\rm p_{{\rm jk}}}  ={c}^{-1}({||\mathbf{r}_{\text{T}}-\mathbf{t}_{{\mathrm{j}}}||+||\mathbf{r}_{\text{T}}-\mathbf{r}_{{\mathrm{k}}}||})$, 
	where $c$ is the speed of light. Similarly, target-path Doppler
	shifts can be obtained as ${\rm f_{jk}}={\lambda_{\rm  j}}^{-1}{\mathbf{v}_{\text{T}}^{\rm T}\big(\frac{\mathbf{r}_{\text{T}}-\mathbf{t}_{\rm  j}}{||\mathbf{r}_{\text{T}}-\mathbf{t}_{\rm j}||}+\frac{\mathbf{r}_{\text{T}}-\mathbf{r}_{\rm  k}}{||\mathbf{r}_{\text{T}}-\mathbf{r}_{\rm  k}||}\big)}$ 
	where the wavelength of the ${\mathrm{j}}$-th transmitter is calculated as $\lambda_{\rm j}=\frac{c}{f_{\rm j}^{(c)}}$ \cite{mybook}. The coefficient ${{\alpha}}_{{\rm jk}}$ is a complex value that accounts for various factors affecting the propagation of the signal from the $\rm j$-th transmitter to the $\rm  k$-th receiver, including the gains of the antennas, the effects of channel propagation in the path from the transmitter to the receiver, and the reflectivity of the target.
	${\mathbf{n}}_{{\rm jk}}\in\mathbb{C}^{\rm L\times1}$ (${\mathbf{e}}_{{\rm jk}}\in\mathbb{C}^{\rm L\times1}$) is formed by the
	thermal noise samples of the $\rm j$-th frequency channel at the $\rm k$-th
	surveillance (reference) receiver, modeled as zero-mean uncorrelated Gaussian noise with unknown variance
	$\rm \sigma_{jk}^{2}$. The unknown noise variances $\rm \sigma_{jk}^{2}$
	can in practice differ in each frequency/transmitter channel for $\rm k=1,...,N_{\rm r}$
	and $\rm j=1,...,{\mathrm{N_{t}}}$, then $\rm \sigma_{jk}^{2}\ne\sigma_{mn}^{2}$
	for $\rm k\ne n$ and $\rm j\ne m$. { Therefore, we are focusing on receivers that are fully uncalibrated. In each surveillance receiver (or reference receiver), we employ multi-channel tuning with carrier frequencies corresponding to the transmitted signals, assuming that these frequency channels exhibit distinct noise variances. Furthermore, we make the assumption that $\rm N_{\rm r}$ distributed receivers are also uncalibrated.} 
	
	Let us stack the observation from $\rm N_{\rm r}$ receivers due to the $\rm j$-th transmitter to form ${\mathbf{X}}_{{\rm j}}{\mathrm{=}}{[{\mathbf{x}}_{{\rm j1}}{\mathbf{,}\dots,}{\mathbf{x}}_{{\rm j}{\rm N}_{{\rm r}}}]}$ and ${\mathbf{Y}}_{{\rm j}}{\mathrm{=}}{[{\mathbf{y}}_{{\rm j1}}{\mathbf{,}\dots,}{\mathbf{y}}_{{\rm j}{\rm N}_{{\rm r}}}]}$. By defining
	${\boldsymbol{\alpha}}_{{\rm j}}{\mathrm{=}}{[{\alpha}_{{\rm j1}}{\rm ,\dots,}{\alpha}_{{\rm j}{\rm N}_{{\rm r}}}]}^{{\rm T}}$,
	${\boldsymbol{\beta}}_{{\rm j}}{\mathrm{=}}{[{\beta}_{{\rm j1}}{\rm ,\dots,}{\beta}_{{\rm j}{\rm N}_{{\rm r}}}]}^{{\rm T}}$,
	${\mathbf{N}}_{{\rm{j}}}=[{\mathbf{n}}_{{\rm j1}}{\mathbf{,}\dots,}{\mathbf{n}}_{{\rm j}{\rm N}_{{\rm r}}}]$, ${\mathbf{E}}_{{\rm{j}}}=[{\mathbf{e}}_{{\rm j1}}{\mathbf{,}\dots,}{\mathbf{e}}_{{\rm j}{\rm N}_{{\rm r}}}]$,
	${\boldsymbol{\mathcal{D}}}_{{\rm d}_{{\rm j}}}=[ {\mathbf{D}}_{{\rm d}_{{\rm j1}}}, {\rm \dots}, {\mathbf{D}}_{{\rm d}_{{\rm j}{\rm N}_{{\rm r}}}}] $
	and ${\boldsymbol{\mathcal{D}}}_{{\rm p}_{{\rm j}}}=[ {\mathbf{D}}_{{\rm p}_{{\rm j1}}}, {\rm \dots}, {\mathbf{D}}_{{\rm p}_{{\rm j}{\rm N}_{{\rm r}}}}] $, we can rewrite (\ref{eq2x}) compactly as
	\begin{equation}\label{eq2y}
		\left\{
		\begin{array}{ll}
			\begin{split}
				{\mathbf{Y}}_{{\rm j}}&={\boldsymbol{\mathcal{D}}}_{{\rm d}_{{\rm j}}}(\rm{diag}({\boldsymbol{\beta}}_{j})\otimes{\mathbf{s}}_{\rm{j}})+{\mathbf{E}}_{{\rm{j}}},  \\
				{\mathbf{X}}_{{\rm j}}&={\boldsymbol{\mathcal{D}}}_{{\rm p}_{{\rm j}}}(\rm{diag}({\boldsymbol{\alpha}}_{j})\otimes{\mathbf{s}}_{\rm{j}})+{\mathbf{N}}_{{\rm{j}}}.
			\end{split}
		\end{array}
		\right.
	\end{equation}
	where ${\boldsymbol{\mathcal{D}}}_{{\rm d}_{{\rm j}}}\in\mathbb{C}^{\rm L\times LN_r}$ and ${\boldsymbol{\mathcal{D}}}_{{\rm p}_{{\rm j}}}\in\mathbb{C}^{\rm L\times LN_r}$.
	Let us partition the unknown parameters vector $\boldsymbol{\mathrm{\theta}}$ as $\boldsymbol{\mathrm{\theta}}=[\boldsymbol{\mathrm{\theta}}_{\rm r},\boldsymbol{\mathrm{\theta}}_{\rm s}]^{\rm T}$, where the first subvector $\boldsymbol{\mathrm{\theta}}_{\rm r}$ includes the parameters under test and the second subvector contains the nuisance parameters. In our case, $\boldsymbol{\mathrm{\theta}}_{\rm r}=\boldsymbol{\boldsymbol{\alpha}}\in\mathbb{C}^{\rm \mathrm{N_{\rm r}}\mathrm{{\mathrm{N_{t}}}}\times1}$, where  $\boldsymbol{\boldsymbol{\alpha}}=[\boldsymbol{\boldsymbol{\alpha}}_{1}^{\rm T},...,\boldsymbol{\boldsymbol{\alpha}}_{\mathrm{{\mathrm{N_{t}}}}}^{\rm T}]^{\rm T}$ with ${\boldsymbol{\boldsymbol{\alpha}}}_{{\rm j}}{\mathrm{=}}{[{\boldsymbol{\alpha}}_{{\rm j1}}{\rm ,\dots,}{\boldsymbol{\alpha}}_{{\rm j}{\rm N}_{{\rm r}}}]}^{{\rm T}}$, and meanwhile $\boldsymbol{\mathrm{\theta}}_{\rm s}\in\mathbb{C}^{\rm((\mathrm{N_{\rm r}}+L)\mathrm{{\mathrm{N_{t}}}}+1)\times1}$ can be partitioned as $\boldsymbol{\mathrm{\theta}}_{\rm s}=[\boldsymbol{\beta}^{\rm T},\mathbf{s}^{\rm T},\boldsymbol{\sigma}^{\rm T}]^{\rm T}$, where $\mathrm{{\boldsymbol{\beta}}=[{\boldsymbol{\beta}}_{1}^{\rm T},...,{\boldsymbol{\beta}}_{\mathrm{{\mathrm{N_{t}}}}}^{\rm T}]^{\rm T}}$ with ${\boldsymbol{\beta}}_{{\rm j}}{\mathrm{=}}{[{\beta}_{{\rm j1}}{\rm ,\dots,}{\beta}_{{\rm j}{\rm N}_{{\rm r}}}]}^{{\rm T}}$,   $\mathrm{{\mathbf{s}}=[{\mathbf{s}}_{1}^{\rm T},...,{\mathbf{s}}_{\mathrm{{\mathrm{N_{t}}}}}^{\rm T}]^{\mathrm{T}}}$, and $\mathrm{{\boldsymbol{\sigma}}=[{\boldsymbol{\sigma}^T_{1}},...,{\boldsymbol{\sigma}^T_{N_t}}]^{\rm T}}$ with ${\boldsymbol{\sigma}_{\rm j}}=[{{\sigma}}^2_{\rm j1},...,{{\sigma}}^2_{\rm jN_r}]^{\rm T}$.

{In our proposed approach, we need to employ a centralized data fusion strategy, which requires accurate synchronization between receivers. However, this aspect is beyond the scope of the current paper. The primary goal of this paper is to determine the presence or absence of a target.} We denote the hypothesis $\mathcal{H}_0$ as the absence of the target and the alternative hypothesis $\mathcal{H}_1$ as the presence of the target in the received surveillance channels. The IPR target detection problem can be formulated as a binary hypothesis-testing problem, represented by
	\begin{subequations}\label{eq31bxt}
		\begin{align}\label{eq31bb}
			\mathcal{H}_0:\left\{
			\begin{array}{ll}
				{\mathbf{Y}}_{{\rm j}}={\boldsymbol{\mathcal{D}}}_{{\rm d}_{{\rm j}}}(\rm{diag}({\boldsymbol{\beta}}_{j})\otimes{\mathbf{s}}_{\rm{j}})+{\mathbf{E}}_{{\rm{j}}},  \\
				{\mathbf{X}}_{{\rm j}}={\mathbf{N}}_{{\rm{j}}},\\
			\end{array}
			\right.
			\qquad  \\
			\mathcal{H}_1:\left\{
			\begin{array}{ll}
				{\mathbf{Y}}_{{\rm j}}={\boldsymbol{\mathcal{D}}}_{{\rm d}_{{\rm j}}}(\rm{diag}({\boldsymbol{\beta}}_{j})\otimes{\mathbf{s}}_{\rm{j}})+{\mathbf{E}}_{{\rm{j}}},  \\
				{\mathbf{X}}_{{\rm j}}={\boldsymbol{\mathcal{D}}}_{{\rm p}_{{\rm j}}}(\rm{diag}({\boldsymbol{\alpha}}_{j})\otimes{\mathbf{s}}_{\rm{j}})+{\mathbf{N}}_{{\rm{j}}},
			\end{array}
			\right.
			\qquad  
		\end{align}
	\end{subequations}
\section{Invariance Issues}\label{sec3} 
	Let $\mathcal{G}$ be a group of transformations that map the sample space onto itself. In general, a hypothesis-testing problem  remains invariant under $\mathcal{G}$ provided that the distributions belong to the same family and the parameter spaces are preserved \cite{leh}-\cite{ciu}. In our case, we can demonstrate  that the detection problem in (\ref{eq31bxt}) is also invariant under the transformation group $\mathcal{G}$, defined by
	\begin{equation}\label{eqt8sx}
		\mathcal{G}=\left\{\it{g}: \it{g}({\mathbf{Z}}_{{\rm j}})={\mathbf{Z}}_{{\rm j}}
		\begin{split}
			\left[\begin{array}{cc}
				\rm{diag}(\boldsymbol{\gamma}_{\rm{j}})&\mathbf{0}_{\rm{N_r \times N_r}}\\
				\mathbf{0}_{\rm{N_r \times N_r}} & \rm{diag}(\boldsymbol{\gamma}_{\rm{j}})
			\end{array}\right]
		\end{split}\right\}
	\end{equation}
	where ${\mathbf{Z}}_{{\rm j}}=[{\mathbf{X}}_{{\rm j}}, {\mathbf{Y}}_{{\rm j}}]\in\mathbb{C}^{\rm L\times 2N_r}$ and $\boldsymbol{\gamma}_{\rm{j}}=[{\gamma}_{\rm j1},...,{\gamma}_{\rm jN_r}]^{\rm T}$ with $\{\gamma_{\rm jk}\}_{{\mathrm{k}}=1}^{\mathrm{N_{\rm r}}}$ being complex and arbitrary values. The  transformation (\ref{eqt8sx}) is a group since it is a closed set and is associative, and contains both the identity and inverse elements. The transformed data under $\mathcal{G}$ is distributed as  
	\begin{subequations}\label{eq31bx}
		\begin{align}\label{eq31bb}
			\mathcal{H}_0:\left\{
			\begin{array}{ll}
				\it{g}({\mathbf{Y}}_{{\rm j}})\sim\mathcal{C N}_{\rm{N \times N_r}}({\boldsymbol{\mathcal{D}}}_{{\rm d}_{{\rm j}}}(\rm{diag}({\boldsymbol{\bar{\beta}}_{j}})\otimes{\mathbf{\bar s}}_{\rm{j}}), \mathbf{I}_L, diag(\boldsymbol{\mathrm{\bar \sigma}_{j}}))\\
				\it{g}({\mathbf{X}}_{{\rm j}})\sim\mathcal{C N}_{\rm{N \times N_r}}({\boldsymbol{0}}_{\rm{N \times N_r}}, \mathbf{I}_{\rm L}, \mathrm{diag(\boldsymbol{\mathrm{\bar\sigma}_{j}})})\\
			\end{array}
			\right.
			\qquad  \\
			\mathcal{H}_1:\left\{
			\begin{array}{ll}
				\it{g}({\mathbf{Y}}_{{\rm j}})\sim\mathcal{C N}_{\rm{N \times N_r}}({\boldsymbol{\mathcal{D}}}_{{\rm d}_{{\rm j}}}(\rm{diag}({\boldsymbol{\bar\beta}}_{j})\otimes{\mathbf{\bar s}}_{\rm{j}}), \mathbf{I}_L, diag(\boldsymbol{\mathrm{\bar \sigma}_{j}}))   \\
				\it{g}({\mathbf{X}}_{{\rm j}})\sim\mathcal{C N}_{\rm{N \times N_r}}({\boldsymbol{\mathcal{D}}}_{{\rm p}_{{\rm j}}}(\rm{diag}({\boldsymbol{\bar \alpha}}_{j})\otimes{\mathbf{\bar s}}_{\rm{j}}), \mathbf{I}_L, diag(\boldsymbol{\mathrm{\bar \sigma}_{j}})) 
			\end{array}
			\right.
			\qquad  
		\end{align}
	\end{subequations}
	where ${\boldsymbol{\bar\alpha}}_{j}$, ${\boldsymbol{\bar\beta}}_{j}$,  $\boldsymbol{\mathrm{\bar \sigma}_{j}}$ and ${\mathbf{\bar s}}_{\rm{j}}$ can be termed as induced vector parameters under transformation $\mathcal{G}$,  given by
	\begin{subequations}\label{eq31bxY}
		\begin{align}
			{\boldsymbol{\bar\alpha}}_{j}&=\rm diag(\boldsymbol{\mathrm{\gamma}_{j}}){\boldsymbol{\alpha}}_{j},\\
			{\boldsymbol{\bar\beta}}_{j}&=\rm diag(\boldsymbol{\mathrm{\gamma}_{j}}){\boldsymbol{\beta}}_{j},\\
			{\boldsymbol{\bar\sigma}}_{j}&=\rm diag(\boldsymbol{\gamma}_{\rm{j}}^*\odot\boldsymbol{\mathrm{\gamma}_{j}}){\boldsymbol{\sigma}}_{j},\\
			{\mathbf{\bar s}}_{\rm{j}}&={\mathbf{s}}_{\rm{j}}.
		\end{align}
	\end{subequations}
	We can utilize this to identify the induced transformation groups on the parameter space, given by
	\begin{equation}\label{eqt8s}
		\mathcal{\bar{G}}=\left\{\it{\bar{g}}: \it{\bar{g}}(\boldsymbol{\mathrm{\theta}})=\rm{diag}(\boldsymbol{\mathrm t})\boldsymbol{\mathrm{\theta}}\right\}
	\end{equation}
	where $\boldsymbol{\mathrm t}=[\boldsymbol{\gamma}^{\rm T},\boldsymbol{\gamma}^{\rm T},\mathbf{1}_{\rm{LN_t}}^{\rm T},{\boldsymbol{\gamma}^{\rm H}}\odot{\boldsymbol{\gamma}^{\rm T}]^{\rm T}}$ where $\mathrm{{\boldsymbol{\gamma}}=[{\boldsymbol{\gamma}^T_{1}},\dots,{\boldsymbol{\gamma}^T_{N_t}}]^{\mathrm{T}}}$. 
	Since ${\boldsymbol{\beta}}_{j}$, ${\boldsymbol{\alpha}}_{j}$, and $\boldsymbol{\mathrm{ \sigma}_{j}}$ are all unknown vectors, we can consider ${\boldsymbol{\bar\beta}}_{j}$, ${\boldsymbol{\bar\alpha}}_{j}$, and $\boldsymbol{\mathrm{\bar \sigma}_{j}}$ as new unknowns. This means that the distribution family of the transformed data does not change. Thus, {we can state that} the hypothesis-testing problem (\ref{eq31bxt}) is invariant under the scaling group $\mathcal{G}$.\par
	In statistics, the equivariant property refers to the invariance of an estimator under certain transformations of the data. Specifically, a statistical estimator is said to be equivariant if it retains its value under certain group transformations of the data. In the context of MLE, the ML estimator is equivariant under group transformations of the underlying probability distribution \cite{leh}. This means that if we transform the data using a group transformation $\mathcal{G}$, then the MLE of the parameters will be transformed by the corresponding
	induced transformation in the parameter space, denoted by $\bar{\mathcal{G}}$. In our case, this implies
	\begin{equation}\label{eq31bxYx}
		\hat{\boldsymbol{\mathrm{\theta}}}(\it{g}(\{{\mathbf{Z}}_{{\rm j}}\}_{{\mathrm{j}}=\rm 1}^{\mathrm{{\mathrm{N_{t}}}}}))=\it{\bar{g}}(\hat{\boldsymbol{\mathrm{\theta}}}(\{{\mathbf{Z}}_{{\rm j}}\}_{{\mathrm{j}}=\rm 1}^{\mathrm{{\mathrm{N_{t}}}}}))
	\end{equation}
	where $\hat{\boldsymbol{\mathrm{\theta}}}$ is the MLE of the unknown parameter vector $\boldsymbol{\mathrm{\theta}}$. This property is called the equivariant property of the MLE in the sequel. Specifically, this property means that the MLEs of ${\mathbf{s}}_{\mathrm{j}}$ are the same whether we use the original data or transformed data. The MLEs obtained from the original and transformed data are referred to as $\hat{\mathbf{s}}_{\mathrm{j}}$ and $\hat{\mathbf{\bar s}}_{\mathrm{j}}$, respectively, i.e., $\hat{\mathbf{s}}_{\mathrm{j}}=\hat{\mathbf{\bar s}}_{\mathrm{j}}$.  \\
	\section{Detectors Design}\label{sec4}
	The goal of this paper is not to design waveforms for an integrated passive radar system. However, the signal modeling of the problem being studied indicates the need of incorporating specific radar constraints in order to avoid the masking effect of strong interference, including multipath echo signals and direct-path signals as well as interfering targets. This study develops some new detectors to investigate the problem of distributed MIMO IPR target detection when dealing with noisy reference channels. These detectors will be based on the principles of LRT, Rao, Wald, Gradient, and Durbin tests \cite{GRA}-\cite{tre}.
	Therefore, we tentatively derive the detectors to provide a framework for the two-channel distributed MIMO IPR target detection problem, i.e.,
	\begin{equation}\begin{split}
			\label{eq14}
			\Lambda_{\mathrm{L}}({\mathbf{z}})=\sup_{\boldsymbol{\sigma}, \mathbf{\beta}, \mathbf{\alpha}, 
				\mathbf{s}} \mathcal{L}\left(\boldsymbol{\mathrm{\theta}}_{\mathrm{r}}, \boldsymbol{\mathrm{\theta}}_{\mathrm{s}} ; {{\mathbf{z}}}\right)\
			-  \sup_{\boldsymbol{\sigma }, \mathbf{\beta }, 
				\mathbf{s} } \mathcal{L}\left(\boldsymbol{\mathrm{\theta}}_{\mathrm{r}}=\mathbf{0}, \boldsymbol{\mathrm{\theta}}_{\mathrm{s}} ; {{\mathbf{z}}}\right) 
	\end{split}\end{equation}
	\begin{equation}
		\resizebox{.87\hsize}{!}{$	\begin{split}\label{eq15}
				\left\{\begin{array}{l}
					\Lambda_{\mathrm{UW}}(\mathbf{z})=(\hat{\boldsymbol{\mathrm{\theta}}}_{\mathrm{r}, 1}-\boldsymbol{\mathrm{\theta}}_{\mathrm{r}, 0})^{\mathrm{H}}\big([\boldsymbol{\mathcal{J}}^{-\mathbf{1}}(\hat{\boldsymbol{\mathrm{\theta}}}_{\mathrm{1}})]_{\mathrm{rr}}\big)^{-\mathbf{1}}(\hat{\boldsymbol{\mathrm{\theta}}}_{\mathrm{r}, 1}-\boldsymbol{\mathrm{\theta}}_{\mathrm{r}, 0})\\
					\Lambda_{\mathrm{AW}}(\mathbf{z})=(\hat{\boldsymbol{\mathrm{\theta}}}_{\mathrm{r}, 1}-\boldsymbol{\mathrm{\theta}}_{\mathrm{r}, 0})^{\mathrm{H}}\big([\boldsymbol{\mathcal{J}}^{-\mathbf{1}}(\hat{\boldsymbol{\mathrm{\theta}}}_0)]_{\mathrm{rr}}\big)^{-\mathbf{1}}(\hat{\boldsymbol{\mathrm{\theta}}}_{\mathrm{r}, 1}-\boldsymbol{\mathrm{\theta}}_{\mathrm{r}, 0})
				\end{array}\right. 
			\end{split}$}
	\end{equation}
	\begin{equation}\begin{split}
			\label{eq16}
			\resizebox{.87\hsize}{!}{$	\Lambda_{\mathrm{G}}(\mathbf{z})=2\mathcal{Q}{\bigg(\big[\frac{\partial{\mathcal {L }}(\boldsymbol{\mathrm{\theta}}_{\mathrm{r}}, \boldsymbol{\mathrm{\theta}}_{\mathrm{s}} ; {{\mathbf{z}}})}{\partial \boldsymbol{\mathrm{\theta}}_{\mathrm{r}}^*}\big]^{\mathrm{H}}\Bigg|_{(\boldsymbol{\mathrm{\theta}}_{\mathrm{\rm r}}, \boldsymbol{\mathrm{\theta}}_{\mathrm{s}})=(\hat{\boldsymbol{\mathrm{\theta}}}_{\mathrm{r}, 0}, \hat{\boldsymbol{\mathrm{\theta}}}_{\mathrm{s}, 0})}(\hat{\boldsymbol{\mathrm{\theta}}}_{\mathrm{r} ,{1}}-\theta_{\mathrm{r}, 0})\bigg)} $}
	\end{split}\end{equation}
	\begin{equation}\begin{split}
			\label{eq17X}
			\resizebox{.87\hsize}{!}{$\Lambda_{\mathrm{R}}(\mathbf{z})=2\bigg[\frac{\partial \mathcal{L}\left(\boldsymbol{\mathrm{\theta}}_{\mathrm{r}}, \boldsymbol{\mathrm{\theta}}_{\mathrm{s}} ; {{\mathbf{z}}}\right)}{\partial \boldsymbol{\mathrm{\theta}}_{\mathrm{r}}^*}\bigg]^{\mathrm{H}}\big[\boldsymbol{\mathcal{J}}^{-\mathbf{1}}(\boldsymbol{\mathrm{\theta}})\big]_{\mathrm{rr}}\bigg[\frac{\partial \mathcal{L}\left(\boldsymbol{\mathrm{\theta}}_{\mathrm{r}}, \boldsymbol{\mathrm{\theta}}_{\mathrm{s}} ; {{\mathbf{z}}}\right)}{\partial \boldsymbol{\mathrm{\theta}}_{\mathrm{r}}^*}\bigg]\Bigg|_{\left(\boldsymbol{\mathrm{\theta}}_{\mathrm{\rm r}}, \boldsymbol{\mathrm{\theta}}_{\mathrm{s}}\right)=(\hat{\boldsymbol{\mathrm{\theta}}}_{\mathrm{r}, 0}, \hat{\boldsymbol{\mathrm{\theta}}}_{\mathrm{s}, 0})}  $}
	\end{split}\end{equation}
	\begin{equation}\begin{split}
			\label{eq18x}
			\resizebox{.87\hsize}{!}{$	\Lambda_{\mathrm{D}}(\mathbf{z})=(\hat{\boldsymbol{\mathrm{\theta}}}_{\mathrm{r}, 10}-\boldsymbol{\mathrm{\theta}}_{\mathrm{r}, 0})^{\mathrm{H}}\big([\boldsymbol{\mathcal{J}}(\hat{\boldsymbol{\mathrm{\theta}}}_0)]_{\mathrm{rr}}[\boldsymbol{\mathcal{J}}^{-1}(\hat{\boldsymbol{\mathrm{\theta}}}_0)]_{\mathrm{rr}}[\boldsymbol{\mathcal{J}}(\hat{\boldsymbol{\mathrm{\theta}}}_0)]_{\mathrm{rr}}\big)(\hat{\boldsymbol{\mathrm{\theta}}}_{\mathrm{r}, 10}-\boldsymbol{\mathrm{\theta}}_{\mathrm{r}, 0})$}
	\end{split}\end{equation}
	Here, $\Lambda_{\mathrm{L}}$, $\Lambda_{\mathrm{UW}}$, $\Lambda_{\mathrm{AW}}$, $\Lambda_{\mathrm{G}}$, $\Lambda_{\mathrm{R}}$, and $\Lambda_{\mathrm{D}}$ stand for LRT (L), usual Wald (UW), alternative Wald (AW), Gradient (G), Rao (R) and Durbin (D) test statistics, respectively.  In the usual Gradient test statistic, $\mathcal{Q}{\left( .\right)}$ is defined as a real part operator, denoted as $\Re{\left( .\right)}$. However, it can be replaced with an absolute operator $| . |$ to achieve higher detection performance, especially for low values of DNR in the reference channels. Furthermore,
	$\hat{\boldsymbol{\mathrm{\theta}}}_{\mathrm{0}}$ and $\hat{\boldsymbol{\mathrm{\theta}}}_{\mathrm{1}}$ are respectively the MLE of $\boldsymbol{\mathrm{\theta}}$ under $\mathcal{H}_{0}$ and $\mathcal{H}_{1}$,  $\hat{\boldsymbol{\mathrm{\theta}}}_{s,0}$ ($\hat{\boldsymbol{\mathrm{\theta}}}_{r,0}$) and $\hat{\boldsymbol{\mathrm{\theta}}}_{s,1}$ ($\hat{\boldsymbol{\mathrm{\theta}}}_{r,1}$) are respectively MLE of $\boldsymbol{\mathrm{\theta}}_{\rm s}$ ($\boldsymbol{\mathrm{\theta}}_{\rm r}$) under $\mathcal{H}_{0}$ and $\mathcal{H}_{1}$, and $\hat{\boldsymbol{\mathrm{\theta}}}_{\mathrm{r}, 10}$ is the MLE of $\boldsymbol{\mathrm{\theta}}_{\rm r}$ under $\mathcal{H}_{1}$ with $\boldsymbol{\mathrm{\theta}}_{\rm s}=\hat{\boldsymbol{\mathrm{\theta}}}_{s,0}$.
	In (\ref{eq16}) and (\ref{eq17X}), $\frac{\partial}{\partial\boldsymbol{\mathrm{\theta}}^*_{\rm r}}$
	denotes the gradient operator with respect to $\boldsymbol{\mathrm{\theta}}^*_{\rm r}$. Finally,   $\boldsymbol{\mathcal{J}}(\boldsymbol{\mathrm{\theta}})$ is the Fisher information matrix (FIM), and ${[\boldsymbol{\mathcal{J}}(\boldsymbol{\mathrm{\theta}})]}_{\mathrm{rr}}$ and ${[\boldsymbol{\mathcal{J}}^{-1}(\boldsymbol{\mathrm{\theta}})]}_{\mathrm{rr}}$respectively refer to the submatrix located at the top-left corner of the FIM and its inverse evaluated at $\boldsymbol{\mathrm{\theta}}$. The log-likelihood function (LLF), denoted by $\mathcal{L}\left(\boldsymbol{\mathrm{\theta}}_{\rm r}, \boldsymbol{\mathrm{\theta}}_{\rm s};\mathbf{z}\right)$, is given by
	\begin{equation}
		\begin{split}\label{eq13}
			\mathcal{L}\left(\boldsymbol{\mathrm{\theta}}_{\rm r}, \boldsymbol{\mathrm{\theta}}_{\rm s} ; \boldsymbol{z}\right)=-2 \mathrm{LN}_{\mathrm{t}} \mathrm{N}_{\mathrm{r}} \ln \pi-2\mathrm{L} \sum_{\mathrm{j}=1}^{\mathrm{N}_{\mathrm{t}}} \sum_{\mathrm{k}=1}^{\mathrm{N}_{\mathrm{r}}} \ln \sigma_{\mathrm{jk}}^2\\-\sum_{\mathrm{j}=1}^{\mathrm{N}_{\mathrm{t}}} \sum_{\mathrm{k}=1}^{\mathrm{N}_{\mathrm{r}}} \frac{\lVert \boldsymbol{x}_{\mathrm{jk}}-\alpha_{\mathrm{jk}} \mathbf{s}_{\mathrm{j}}\rVert^2}{\sigma_{\mathrm{jk}}^2}-\sum_{\mathrm{j}=1}^{\mathrm{N}_{\mathrm{t}}} \sum_{\mathrm{k}=1}^{\mathrm{N}_{\mathrm{r}}} \frac{\lVert \boldsymbol{y}_{\mathrm{jk}}-\beta_{\mathrm{jk}} \mathbf{s}_{\mathrm{j}}\rVert^2}{\sigma_{\mathrm{jk}}^2}
	\end{split}\end{equation}
	Here,  $\mathrm{\mathbf{z}}\mathrm{=}[ 
	\mathbf{x}^{\mathrm{T}}\mathrm{,}\mathbf{y}^{\mathrm{T}} 
	]^{\mathrm{T}}$, where $\mathrm{\mathbf{x=}}[ 
	\mathbf{x}_{\mathrm{\mathbf{1}}}^{\mathrm{T}}\mathrm{\mathbf{\mathellipsis.}}\mathbf{x}_{\mathrm{N}_{\mathrm{t}}}^{\mathrm{T}} 
	]^{\mathrm{T}}$, $\mathbf{x}_{\mathrm{j}}\mathrm{\mathbf{=}}[ 
	\mathbf{x}_{\mathrm{j1}}^{\mathrm{T}}\mathrm{\mathbf{,\mathellipsis,}}\mathbf{x}_{\mathrm{j}\mathrm{N}_{\mathrm{r}}}^{\mathrm{T}} 
	]^{\mathrm{T}}$, $\mathrm{\mathbf{y=}}[ 
	\mathbf{y}_{\mathrm{\mathbf{1}}}^{\mathrm{T}}\mathrm{\mathbf{\mathellipsis.}}\mathbf{y}_{\mathrm{N}_{\mathrm{t}}}^{\mathrm{T}} 
	]^{\mathrm{T}}$, $\mathbf{y}_{\mathrm{j}}\mathrm{=}[ 
	\mathbf{y}_{\mathrm{j1}}^{\mathrm{T}}\mathrm{\mathbf{,\mathellipsis,}}\mathbf{y}_{\mathrm{j}\mathrm{N}_{\mathrm{r}}}^{\mathrm{T}} 
	]^{\mathrm{T}}$.  In (\ref{eq13}),  ${\boldsymbol{x}}_{{\rm j}}={[{\boldsymbol{x}}_{{\rm j1}}^{{\rm T}}{,\dots,}{\boldsymbol{x}}_{{\rm j}{\rm N}_{{\rm r}}}^{{\rm T}}]}^{{\rm T}}$ with
	$\boldsymbol{x}_{\mathrm{jk}}={\mathrm{\, 
			\mathbf{D}}}_{\mathrm{p}_{\mathrm{jk}}}^{\mathrm{H}}\mathbf{x}_{\mathrm{jk}}$, ${\boldsymbol{y}}_{{\rm j}}={[{\boldsymbol{y}}_{{\rm j1}}^{{\rm T}}{,\dots,}{\boldsymbol{y}}_{{\rm j}{\rm N}_{{\rm r}}}^{{\rm T}}]}^{{\rm T}}$ with $\boldsymbol{y}_{\mathrm{jk}}={\mathrm{\, 
			\mathbf{D}}}_{\mathrm{d}_{\mathrm{jk}}}^{\mathrm{H}}\mathbf{y}_{\mathrm{jk}}$, $\boldsymbol{n}_{\mathrm{jk}}={\mathrm{\, 
			\mathbf{D}}}_{\mathrm{p}_{\mathrm{jk}}}^{\mathrm{H}}\mathbf{n}_{\mathrm{jk}}\sim\mathcal{CN}(\mathbf{0},\sigma_{\mathrm{jk}}^2\mathbf{I}_{\rm L})$, and  $\boldsymbol{e}_{\mathrm{jk}}={\mathrm{\, 
			\mathbf{D}}}_{\mathrm{d}_{\mathrm{jk}}}^{\mathrm{H}}\mathbf{e}_{\mathrm{jk}}\sim\mathcal{CN}(\mathbf{0},\sigma_{\mathrm{jk}}^2\mathbf{I}_{\rm L})$. Furthermore, let us define
	\begin{equation}\label{eqx1}
		\boldsymbol{\boldsymbol{\boldsymbol{\mathcal{X}}}}_{{\rm j}}{\mathrm{=}}{[\boldsymbol{x}_{{\rm j1}}{\mathbf{,}\dots,}\boldsymbol{x}_{{\rm j}{\rm N}_{{\rm r}}}]}\in\mathbb{C}^{\rm L\times N_r}
	\end{equation}
	\begin{equation}\label{eqx2}
		\boldsymbol{\boldsymbol{\boldsymbol{\mathcal{Y}}}}_{{\rm j}}{\mathrm{=}}{[\boldsymbol{y}_{{\rm j1}}{\mathbf{,}\dots,}\boldsymbol{y}_{{\rm j}{\rm N}_{{\rm r}}}]}\in\mathbb{C}^{\rm L\times N_r}
	\end{equation}
	\begin{equation}\label{eqx3}
		\mathrm{\mathbf{\Xi }}_{\mathrm{j}}^{-1}\mathrm{=}\mathrm{diag}\bigg( 
		\frac{\mathrm{1}}{\lVert \boldsymbol{y}_{\mathrm{j1}} 
			\rVert^{\mathrm{2}}\mathrm{+}\lVert  \boldsymbol{x}_{{\rm j1}} 
			\rVert^{\mathrm{2}}}\mathrm{,\, \mathellipsis ,}\frac{\mathrm{1}}{\lVert  
			\boldsymbol{y}_{\mathrm{j}\mathrm{N}_{\mathrm{r}}} 
			\rVert^{\mathrm{2}}\mathrm{+}\lVert  
			\boldsymbol{x}_{\mathrm{j}\mathrm{N}_{\mathrm{r}}} 
			\rVert^{\mathrm{2}}} \bigg)
	\end{equation}
	\newtheorem{prop}{Proposition}
	\begin{prop}
		The LRT (or GLRT) detector statistic $\Lambda_{\mathrm{L}}({\boldsymbol{\boldsymbol{\mathcal{X}}}}_{\mathrm{j}}, {\boldsymbol{\boldsymbol{\mathcal{Y}}}}_{\mathrm{j}})$ can be obtained as
		\begin{equation}\label{eq17x}
			\begin{split}
				\Lambda_{\mathrm{L}}({\boldsymbol{\boldsymbol{\mathcal{X}}}}_{\mathrm{j}}, {\boldsymbol{\boldsymbol{\mathcal{Y}}}}_{\mathrm{j}})&=\sum_{\mathrm{j}=1}^{\mathrm{N}_{\mathrm{t}}} \lambda_{\max }(\boldsymbol{\boldsymbol{\boldsymbol{\mathcal{Y}}}}_{\mathrm{j}} \boldsymbol{\Xi}_{\mathrm{j}}^{-1} \boldsymbol{\boldsymbol{\boldsymbol{\mathcal{Y}}}}_{\mathrm{j}}^{\mathrm{H}}+\boldsymbol{\boldsymbol{\mathcal{X}}}_{\mathrm{j}} \boldsymbol{\Xi}_{\mathrm{j}}^{-1} \boldsymbol{\boldsymbol{\mathcal{X}}}_{\mathrm{j}}^{\mathrm{H}})\\&-\mathrm{\sum_{j=1}^{N_t} \lambda_{\max }(\boldsymbol{\boldsymbol{\boldsymbol{\mathcal{Y}}}}_{\mathrm{j}} \boldsymbol{\Xi}_{\mathrm{j}}^{-1} \boldsymbol{\boldsymbol{\boldsymbol{\mathcal{Y}}}}_{\mathrm{j}}^{\mathrm{H}})}
			\end{split}
		\end{equation}
		\begin{proof} 
			The derivation of the test statistic (\ref{eq17x}) is given in Appendix \ref{A}.
		\end{proof}
	\end{prop}
	There are some practical difficulties in order to obtain the $\mathrm{\boldsymbol{\boldsymbol{\boldsymbol{\mathcal{Y}}}}}_{\mathrm{j}}\mathrm{\mathbf{\Xi}}_{\mathrm{j}}^{-1}\mathrm{\boldsymbol{\boldsymbol{\boldsymbol{\mathcal{Y}}}}}_{\mathrm{j}}^\mathrm{H}$ and $\mathrm{\boldsymbol{\boldsymbol{\mathcal{X}}}}_{\mathrm{j}}\mathrm{\mathbf{\Xi}}_{\mathrm{j}}^{-1}\mathrm{\boldsymbol{\boldsymbol{\mathcal{X}}}}_{\mathrm{j}}^\mathrm{H}$ matrices, which are $\rm L\times L$ matrices. In practical situations, $\rm L$ takes on a large value to obtain a long-range PR system. The proposed LRT-based detector can be efficiently implemented as
	\begin{equation}
		\begin{split}
			\label{eq23}
			\resizebox{.89\hsize}{!}{$		\Lambda_{\mathrm{L}}({\boldsymbol{\boldsymbol{\mathcal{X}}}}_{\mathrm{j}}, {\boldsymbol{\boldsymbol{\mathcal{Y}}}}_{\mathrm{j}})=\sum_{\mathrm{j}=1}^{\mathrm{N}_{\mathrm{t}}}\Big( \lambda_{\max }( 
				\mathrm{\boldsymbol{\mathcal{Z}}}_{\mathrm{j}}^\mathrm{H}\mathrm{\boldsymbol{\mathcal{Z}}}_{\mathrm{j}})- \lambda_{\max }( 
				\mathrm{\boldsymbol{\mathcal{R}}}_{\mathrm{j}}^\mathrm{H}\mathrm{\boldsymbol{\mathcal{R}}}_{\mathrm{j}})\Big)\begin{array}{c}{\mathcal{H}}_1 \\ \gtrless  \\ {\mathcal{H}}_0 \end{array}\eta_L$}
		\end{split}
	\end{equation}
	where $	\mathrm{\boldsymbol{\boldsymbol{\mathcal{R}}}}_{\mathrm{j}}\mathrm{\mathrm{=}}\mathrm{\boldsymbol{\boldsymbol{\boldsymbol{\mathcal{Y}}}}}_{\mathrm{j}}\mathrm{\mathbf{\Xi 
	}}_{\mathrm{j}}^{-\frac{1}{2}}$, $\mathrm{\boldsymbol{\mathcal{T}}}_{\mathrm{j}}\mathrm{\mathrm{=}}\mathrm{\boldsymbol{\boldsymbol{\mathcal{X}}}}_{\mathrm{j}}\mathrm{\mathbf{\Xi }}_{\mathrm{j}}^{-\frac{1}{2}}$ and $\mathrm{\boldsymbol{\mathcal{Z}}}_{\mathrm{j}}\mathrm{=}[ 
	\mathrm{\boldsymbol{\mathcal{R}}}_{\mathrm{j}}\mathrm{\mathbf{,}}\mathrm{\boldsymbol{\mathcal{T}}}_{\mathrm{j}} 
	]$.   
	Here, $\eta_L$ represents the detection threshold of the proposed LRT-based detector to be set according to a desired false alarm probability, denoted as $p_{_{\rm fa}}$.\par
	Let us now demonstrate that the LRT test exhibits the CFAR property against the NVU. To do so, we resort to invariance property. We have shown that the distribution family would not be changed by applying the group transformation (\ref{eqt8sx}). Additionally, by using (\ref{eq31bx}), we have
	\begin{subequations}\label{eq31bxYzs}
		\begin{align}
			\it{g}({\boldsymbol{\boldsymbol{\mathcal{Y}}}}_{\mathrm{j}})&={\boldsymbol{\boldsymbol{\mathcal{Y}}}}_{\mathrm{j}}\rm{diag}(\boldsymbol{\gamma}_{\rm{j}}),\\
			\it{g}({\boldsymbol{\boldsymbol{\mathcal{X}}}}_{\mathrm{j}})&={\boldsymbol{\boldsymbol{\mathcal{X}}}}_{\mathrm{j}}\rm{diag}(\boldsymbol{\gamma}_{\rm{j}}),\\
			\mathrm{\mathbf{\Xi }}_{\mathrm{j}}^{-1}(\it{g}({\boldsymbol{\boldsymbol{\mathcal{X}}}}_{\mathrm{j}}), \it{g}({\boldsymbol{\boldsymbol{\mathcal{Y}}}}_{\mathrm{j}}))&=\rm{diag}(\boldsymbol{\gamma}_{\rm{j}})^{-1}\mathrm{\mathbf{\Xi }}_{\mathrm{j}}^{-1}({\boldsymbol{\boldsymbol{\mathcal{X}}}}_{\mathrm{j}}, {\boldsymbol{\boldsymbol{\mathcal{Y}}}}_{\mathrm{j}})\rm{diag}(\boldsymbol{\gamma}_{\rm{j}})^{-\rm H}
		\end{align}
	\end{subequations}
	Substituting (\ref{eq31bxYzs}) into (\ref{eq23}) implies that $	\mathrm{\Lambda_{\mathrm{L}}(\it{g}({\boldsymbol{\boldsymbol{\mathcal{X}}}}_{\mathrm{j}}), \it{g}({\boldsymbol{\boldsymbol{\mathcal{Y}}}}_{\mathrm{j}}))}=\mathrm{\Lambda_{\mathrm{L}}({\boldsymbol{\boldsymbol{\mathcal{X}}}}_{\mathrm{j}}, {\boldsymbol{\boldsymbol{\mathcal{Y}}}}_{\mathrm{j}})}$.
	{It follows that LRT test statistic is invariant to the transformation group $\mathcal{G}$.}
	\begin{prop}
		The usual Wald test statistic $\Lambda_{\mathrm{SW}}(\mathbf{z})$ is zero, indicating that the usual Wald test is null. However, the test statistic of $\Lambda_{\mathrm{AW}}(\mathbf{z})$ can be developed as
		\begin{equation}\label{eq25}
			\resizebox{.87\hsize}{!}{$\Lambda_{\mathrm{AW}}({\boldsymbol{\boldsymbol{\mathcal{X}}}}_{\mathrm{j}}, {\boldsymbol{\boldsymbol{\mathcal{Y}}}}_{\mathrm{j}})=	\sum\limits_{\mathrm{j=1}}^\mathrm{N_t}{ {			\frac{\lVert  \hat{\mathbf{s}}_{\mathrm{j,0}} \rVert^{\mathrm{4}}}{\lVert  
							\hat{\mathbf{s}}_{\mathrm{j,1}} \rVert^{\mathrm{4}}}
						\sum\limits_{\mathrm{k=1}}^\mathrm{N_r}{
							{\frac{\hat{\mathbf{s}}_{\mathrm{j,1}}^{\mathrm{H}}\frac{\boldsymbol{x}_{\mathrm{jk}}\boldsymbol{x}_{\mathrm{jk}}^{\mathrm{H}}}{\lVert  
										\boldsymbol{y}_{\mathrm{jk}} \rVert^{\mathrm{2}}\mathrm{+}\lVert  
										\boldsymbol{x}_{\mathrm{jk}} 
										\rVert^{\mathrm{2}}}\hat{\mathbf{s}}_{\mathrm{j,1}}}{\hat{\mathbf{s}}_{\mathrm{j}, 0}^{\mathrm{H}}\big(\mathbf{I}-\frac{\boldsymbol{y}_{\mathrm{jk}} \boldsymbol{y}_{\mathrm{jk}}^{\mathrm{H}}}{\lVert \boldsymbol{x}_{\mathrm{jk}}\rVert^2+\lVert \boldsymbol{y}_{\mathrm{jk}}\rVert^2}\big) \hat{\mathbf{s}}_{\mathrm{j}, 0}}}}}}$}
		\end{equation}
		where
		\begin{equation}\label{eq17a}
			\hat{\mathbf{s}}_{\mathrm{j,0}}\mathrm{=}\mathrm{\mathbf{e}}_{\mathrm{1}}(\boldsymbol{\boldsymbol{\boldsymbol{\mathcal{Y}}}}_{\mathrm{j}} \boldsymbol{\Xi}_{\mathrm{j}}^{-1} \boldsymbol{\boldsymbol{\boldsymbol{\mathcal{Y}}}}_{\mathrm{j}}^{\mathrm{H}}),
		\end{equation}
		\begin{equation}\label{eq17b}
			\hat{\mathbf{s}}_{\mathrm{j,1}}\mathrm{=}\mathrm{\mathbf{e}}_{\mathrm{1}}(\mathrm{\boldsymbol{\boldsymbol{\mathcal{Y}}}}_{\mathrm{j}}\mathrm{\mathbf{\Xi 
			}}_{\mathrm{j}}^{-1}\mathrm{\boldsymbol{\boldsymbol{\mathcal{Y}}}}_{\mathrm{j}}^\mathrm{H}\mathrm{+}\mathrm{\boldsymbol{\boldsymbol{\mathcal{X}}}}_{\mathrm{j}}\mathrm{\mathbf{\Xi }}_{\mathrm{j}}^{-1}\mathrm{\boldsymbol{\boldsymbol{\mathcal{X}}}}_{\mathrm{j}}^\mathrm{H}).
		\end{equation}
		\begin{proof} The derivation of the test statistic (\ref{eq25}) is given in Appendix \ref{D}, where the FIM and MLE of the unknowns are presented in Appendices \ref{B} and \ref{C}, respectively.
		\end{proof}
	\end{prop}
	\begin{prop}
		The usual and alternative Gradient test statistics are respectively given by
		\begin{equation}
			\label{eq26x}
			\resizebox{.87\hsize}{!}{$	\Lambda_{\mathrm{UG}}({\boldsymbol{\boldsymbol{\mathcal{X}}}}_{\mathrm{j}}, {\boldsymbol{\boldsymbol{\mathcal{Y}}}}_{\mathrm{j}})=2\mathcal{\Re}{\Bigg(	\sum\limits_{\mathrm{j=1}}^\mathrm{N_t}{ {			\frac{\lVert  \hat{\mathbf{s}}_{\mathrm{j,0}} \rVert^{\mathrm{2}}}{\lVert  
								\hat{\mathbf{s}}_{\mathrm{j,1}} \rVert^{\mathrm{2}}}
							\sum\limits_{\mathrm{k=1}}^\mathrm{N_r}{
								{\frac{\hat{\mathbf{s}}_{\mathrm{j,1}}^{\mathrm{H}}\frac{\boldsymbol{x}_{\mathrm{jk}}\boldsymbol{x}_{\mathrm{jk}}^{\mathrm{H}}}{\lVert  
											\boldsymbol{y}_{\mathrm{jk}} \rVert^{\mathrm{2}}\mathrm{+}\lVert  
											\boldsymbol{x}_{\mathrm{jk}} 
											\rVert^{\mathrm{2}}}\hat{\mathbf{s}}_{\mathrm{j,0}}}{\hat{\mathbf{s}}_{\mathrm{j}, 0}^{\mathrm{H}}\big(\mathbf{I}-\frac{\boldsymbol{y}_{\mathrm{jk}} \boldsymbol{y}_{\mathrm{jk}}^{\mathrm{H}}}{\lVert \boldsymbol{x}_{\mathrm{jk}}\rVert^2+\lVert \boldsymbol{y}_{\mathrm{jk}}\rVert^2}\big) \hat{\mathbf{s}}_{\mathrm{j}, 0}}}}}}\Bigg)}$},
		\end{equation}
		\begin{equation}\label{eq26y}
			\resizebox{.87\hsize}{!}{$
				\Lambda_{\mathrm{AG}}({\boldsymbol{\boldsymbol{\mathcal{X}}}}_{\mathrm{j}}, {\boldsymbol{\boldsymbol{\mathcal{Y}}}}_{\mathrm{j}})=2\Bigg|{	\sum\limits_{\mathrm{j=1}}^\mathrm{N_t}{ {			\frac{\lVert  \hat{\mathbf{s}}_{\mathrm{j,0}} \rVert^{\mathrm{2}}}{\lVert  
								\hat{\mathbf{s}}_{\mathrm{j,1}} \rVert^{\mathrm{2}}}
							\sum\limits_{\mathrm{k=1}}^\mathrm{N_r}{
								{\frac{\hat{\mathbf{s}}_{\mathrm{j,1}}^{\mathrm{H}}\frac{\boldsymbol{x}_{\mathrm{jk}}\boldsymbol{x}_{\mathrm{jk}}^{\mathrm{H}}}{\lVert  
											\boldsymbol{y}_{\mathrm{jk}} \rVert^{\mathrm{2}}\mathrm{+}\lVert  
											\boldsymbol{x}_{\mathrm{jk}} 
											\rVert^{\mathrm{2}}}\hat{\mathbf{s}}_{\mathrm{j,0}}}{\hat{\mathbf{s}}_{\mathrm{j}, 0}^{\mathrm{H}}\big(\mathbf{I}-\frac{\boldsymbol{y}_{\mathrm{jk}} \boldsymbol{y}_{\mathrm{jk}}^{\mathrm{H}}}{\lVert \boldsymbol{x}_{\mathrm{jk}}\rVert^2+\lVert \boldsymbol{y}_{\mathrm{jk}}\rVert^2}\big) \hat{\mathbf{s}}_{\mathrm{j}, 0}}}}}}}\Bigg|.
				$}
		\end{equation}
		\begin{proof} The derivation of these test statistics are given in Appendix \ref{F}.
		\end{proof}
	\end{prop}
	\begin{prop} The test statistics for both the Rao and Durbin (RD) tests are identical and can be expressed as
		\begin{equation}\label{eq24}
			\Lambda_{\mathrm{RD}}({\boldsymbol{\boldsymbol{\mathcal{X}}}}_{\mathrm{j}}, {\boldsymbol{\boldsymbol{\mathcal{Y}}}}_{\mathrm{j}})=\sum_{\mathrm{j}=1}^{\mathrm{N}_{\mathrm{t}}} \sum_{\mathrm{k}=1}^{\mathrm{N}_{\mathrm{r}}} \frac{\hat{\mathbf{s}}_{\mathrm{j}, 0}^{\mathrm{H}} \frac{\boldsymbol{x}_{\mathrm{jk}} \boldsymbol{x}_{\mathrm{jk}}^{\mathrm{H}}}{\lVert \boldsymbol{x}_{\mathrm{jk}}\rVert^2+\lVert \boldsymbol{y}_{\mathrm{jk}}\rVert^2} \hat{\mathbf{s}}_{\mathrm{j}, 0}}{\hat{\mathbf{s}}_{\mathrm{j}, 0}^{\mathrm{H}}\big(\mathbf{I}-\frac{\boldsymbol{y}_{\mathrm{jk}} \boldsymbol{y}_{\mathrm{jk}}^{\mathrm{H}}}{\lVert \boldsymbol{x}_{\mathrm{jk}}\rVert^2+\lVert \boldsymbol{y}_{\mathrm{jk}}\rVert^2}\big) \hat{\mathbf{s}}_{\mathrm{j}, 0}}
		\end{equation}
		\begin{proof} The derivation of Rao's test statistic is given in Appendix \ref{E}, while that of the Durbin test is given in Appendix \ref{G}.  
		\end{proof}
	\end{prop}
	It is worth noting that the alternative Wald, usual and alternative Gradient, and Rao (or Durbin) detectors can be unified as
	\begin{equation}\label{eq25zxs}
		\resizebox{.87\hsize}{!}{$	\Lambda({\boldsymbol{\boldsymbol{\mathcal{X}}}}_{\mathrm{j}}, {\boldsymbol{\boldsymbol{\mathcal{Y}}}}_{\mathrm{j}})=\mathcal{P}\bigg(	\sum\limits_{\mathrm{j=1}}^\mathrm{N_t}{ {			\frac{\lVert  \hat{\mathbf{s}}_{\mathrm{j,0}} \rVert^{\mathrm{p}}}{\lVert  
						\hat{\mathbf{s}}_{\mathrm{j,1}} \rVert^{\mathrm{p}}}
					\sum\limits_{\mathrm{k=1}}^\mathrm{N_r}{
						{\frac{\hat{\mathbf{s}}_{\mathrm{j,a}}^{\mathrm{H}}\frac{\boldsymbol{x}_{\mathrm{jk}}\boldsymbol{x}_{\mathrm{jk}}^{\mathrm{H}}}{\lVert  
									\boldsymbol{y}_{\mathrm{jk}} \rVert^{\mathrm{2}}\mathrm{+}\lVert  
									\boldsymbol{x}_{\mathrm{jk}} 
									\rVert^{\mathrm{2}}}\hat{\mathbf{s}}_{\mathrm{j,b}}}{\hat{\mathbf{s}}_{\mathrm{j}, 0}^{\mathrm{H}}\big(\mathbf{I}-\frac{\boldsymbol{y}_{\mathrm{jk}} \boldsymbol{y}_{\mathrm{jk}}^{\mathrm{H}}}{\lVert \boldsymbol{x}_{\mathrm{jk}}\rVert^2+\lVert \boldsymbol{y}_{\mathrm{jk}}\rVert^2}\big) \hat{\mathbf{s}}_{\mathrm{j}, 0}}}}}}\bigg)$}
	\end{equation}
	where
	\begin{equation}\label{eq2y}
		(p, a, b, \mathcal{P})=\left\{
		\begin{array}{ll}
			\big(4, 1, 1, \mathcal{P}(r)=r\big) & \mathrm{AW},\\
			\big(2, 1, 0, \mathcal{P}(r)=2\mathcal{\Re}(r)\big) & \mathrm{UG},\\
			\big(2, 1, 0, \mathcal{P}(r)=2\big|r\big|\big) & \mathrm{AG},\\
			\big(0, 0, 0, \mathcal{P}(r)=r\big) & \mathrm{RD}.
		\end{array}
		\right.
	\end{equation}
	In (\ref{eq25zxs}), we can use different thresholds  $\eta_{\rm AW}$, $\eta_{\rm UG}$, $\eta_{\rm AU}$, and $\eta_{\rm RD}$ for the alternative Wald (AW), usual Gradient (UG), alternative Gradient (AG) and Rao-Durbin (RD) detectors, respectively. These thresholds are set to ensure a desired false alarm probability  $p_{_{\rm fa}}$. \par
	Let us now examine whether these unified detectors are CFAR. By using (\ref{eq31bxYzs}) in (\ref{eq17a}) and (\ref{eq17b}), we can find that $\hat{\mathbf{s}}_{\mathrm{j,0}}(\it{g}\big({\boldsymbol{\boldsymbol{\mathcal{Y}}}}_{\mathrm{j}})\big)\mathrm{=} \hat{\mathbf{s}}_{\mathrm{j,0}}({\boldsymbol{\boldsymbol{\mathcal{Y}}}}_{\mathrm{j}})$ and $\hat{\mathbf{s}}_{\mathrm{j,1}}\big(\it{g}({\boldsymbol{\boldsymbol{\mathcal{X}}}}_{\mathrm{j}}),\it{g}({\boldsymbol{\boldsymbol{\mathcal{Y}}}}_{\mathrm{j}})\big)\mathrm{=} \hat{\mathbf{s}}_{\mathrm{j,1}}({\boldsymbol{\boldsymbol{\mathcal{X}}}}_{\mathrm{j}}, {\boldsymbol{\boldsymbol{\mathcal{Y}}}}_{\mathrm{j}})$. 
	 In addition, both terms $\frac{\boldsymbol{x}_{\mathrm{jk}}\boldsymbol{x}_{\mathrm{jk}}^{\mathrm{H}}}{\lVert  
		\boldsymbol{y}_{\mathrm{jk}} \rVert^{\mathrm{2}}\mathrm{+}\lVert  
		\boldsymbol{x}_{\mathrm{jk}} 
		\rVert^{\mathrm{2}}}$ and $\frac{\boldsymbol{y}_{\mathrm{jk}}\boldsymbol{y}_{\mathrm{jk}}^{\mathrm{H}}}{\lVert  
		\boldsymbol{y}_{\mathrm{jk}} \rVert^{\mathrm{2}}\mathrm{+}\lVert  
		\boldsymbol{x}_{\mathrm{jk}} 
		\rVert^{\mathrm{2}}}$ remain unchanged when applying the group transformation (\ref{eqt8sx}), implying that $	\mathrm{\Lambda_{\mathrm{AW}}(\it{g}({\boldsymbol{\boldsymbol{\mathcal{X}}}}_{\mathrm{j}}), \it{g}({\boldsymbol{\boldsymbol{\mathcal{Y}}}}_{\mathrm{j}}))}=\mathrm{\Lambda_{\mathrm{AW}}({\boldsymbol{\boldsymbol{\mathcal{X}}}}_{\mathrm{j}}, {\boldsymbol{\boldsymbol{\mathcal{Y}}}}_{\mathrm{j}})}$, $	\mathrm{\Lambda_{\mathrm{UG}}(\it{g}({\boldsymbol{\boldsymbol{\mathcal{X}}}}_{\mathrm{j}}), \it{g}({\boldsymbol{\boldsymbol{\mathcal{Y}}}}_{\mathrm{j}}))}=\mathrm{\Lambda_{\mathrm{UG}}({\boldsymbol{\boldsymbol{\mathcal{X}}}}_{\mathrm{j}}, {\boldsymbol{\boldsymbol{\mathcal{Y}}}}_{\mathrm{j}})}$, $	\mathrm{\Lambda_{\mathrm{AG}}(\it{g}({\boldsymbol{\boldsymbol{\mathcal{X}}}}_{\mathrm{j}}), \it{g}({\boldsymbol{\boldsymbol{\mathcal{Y}}}}_{\mathrm{j}}))}=\mathrm{\Lambda_{\mathrm{AG}}({\boldsymbol{\boldsymbol{\mathcal{X}}}}_{\mathrm{j}}, {\boldsymbol{\boldsymbol{\mathcal{Y}}}}_{\mathrm{j}})}$ and $	\mathrm{\Lambda_{\mathrm{RD}}(\it{g}({\boldsymbol{\boldsymbol{\mathcal{X}}}}_{\mathrm{j}}), \it{g}({\boldsymbol{\boldsymbol{\mathcal{Y}}}}_{\mathrm{j}}))}=\mathrm{\Lambda_{\mathrm{RD}}({\boldsymbol{\boldsymbol{\mathcal{X}}}}_{\mathrm{j}}, {\boldsymbol{\boldsymbol{\mathcal{Y}}}}_{\mathrm{j}})}$. 
{Therefore, the test statistics of the unified detectors remain invariant under the transformation group $\mathcal{G}$.}
{\section{Two-channel IPR  Threshold Setting}\label{sec4b}
In contrast to single-channel PR, the threshold setting for target detection in two-channel PR is inherently influenced by the quality of the direct-path signal in the reference channels. Consequently, the DNR in the reference channels can impact the threshold setting. This section delves into a discussion of this novel aspect. The implications of this new aspect have already been incorporated into the transformation group $\mathcal{G}$, as discussed in Section \ref{sec3}. To gain further insight, let us define the average DNR of the reference channels as
	\begin{align}\label{eq5a2}
		\mathrm{DNR_{\text{avg}}=\frac{1}{\mathrm{N_{\rm r}}\mathrm{{\mathrm{N_{t}}}}}\sum_{j=1}^{\mathrm{{\mathrm{N_{t}}}}}\sum_{k=1}^{\mathrm{N_{\rm r}}}\frac{\abs{\beta_{jk}}^{2}}{\sigma_{jk}^{2}}}.
	\end{align}
	
	It can be demonstrated that the transformation $\mathcal{G}$ preserves a constant $\mathrm{DNR_{\text{avg}}}$. This conclusion, raised from the invariance characteristic of our hypothesis testing problem, means that the proposed detectors possess the CFAR property against NVU as long as the DNR\textsubscript{avg} remains unchanged. This indicates that the proposed detectors have a conditional CFAR property under NVU conditions. From (\ref{eq5a2}), it can be inferred that, in addition to $\mathrm{\sigma_{jk}^{2}}$, the values of $\mathrm{\abs{\beta_{jk}}^{2}}$ can also influence  $\mathrm{DNR_{\text{avg}}}$, and changing the value of $\mathrm{DNR_{\text{avg}}}$ could potentially impact the detection thresholds of the proposed detectors. This suggests that, instead of exclusively focusing on regulating the false alarm probability (FAP) of the proposed detectors against NVU, it is crucial to shift our attention to the more encompassing parameter $\mathrm{DNR_{\text{avg}}}$, which incorporates the effects of NVU. This behavior can also be observed in the works of \cite{p1c2him1}-\cite{Ram}, which address the two-channel IPR target detection problem with the noisy reference channels. However, the authors of these studies did not recognize or discuss this aspect.
	
	  In practical situations where the PR designer lacks control over the specifications of the illuminators or the MIMO channels between transmitters and receivers, it becomes crucial to handle or deal with this issue. Additionally, the DNR\textsubscript{avg} of the RC is influenced by the placements of the distributed receivers, which, in turn, affects the thresholds. Hence, it becomes essential to address this concern.  To do so, we propose a strategy to control the levels of the tests rather than their sizes when there are uncertainties regarding $\mathrm{DNR_{\text{avg}}}$.
	  For the LRT-based detector it can be readily shown that the Gram matrices ${{\boldsymbol{\mathcal{R}}}^{{\mathrm{H}}}_{\mathrm{j}}}{\boldsymbol{\mathcal{R}}}_{\mathrm{j}}$ and $\mathrm{\boldsymbol{\mathcal{Z}}}_{\mathrm{j}}^{\mathrm{H}}\mathrm{\boldsymbol{\mathcal{Z}}}_{\mathrm{j}}$
	  follow non-central uncorrelated complex Wishart distributions under the null hypothesis. Thus, the random variables ${\mathrm{\lambda }}_{\mathrm{max }}( \mathrm{\boldsymbol{\mathcal{Z}}}_{\mathrm{j}}^{\mathrm{H}}\mathrm{\boldsymbol{\mathcal{Z}}}_{\mathrm{j}})$ and ${\mathrm{\lambda }}_{\mathrm{max }}({{\boldsymbol{\mathcal{R}}}^{{\mathrm{H}}}_{\mathrm{j}}}{\boldsymbol{\mathcal{R}}}_{\mathrm{j}})$ exhibit non-independence due to the fact that ${{\boldsymbol{\mathcal{R}}}^{{\mathrm{H}}}_{\mathrm{j}}}{\boldsymbol{\mathcal{R}}}_{\mathrm{j}}$
	  forms one of the diagonal blocks of $\mathrm{\boldsymbol{\mathcal{Z}}}_{\mathrm{j}}^{\mathrm{H}}\mathrm{\boldsymbol{\mathcal{Z}}}_{\mathrm{j}}$
	  \cite{ALLO}. This non-independence poses a significant challenge in deriving a closed-form expression for the FAP of (\ref{eq23}). The same difficulty arises with other proposed tests.
	   Even though the closed-form solutions for the false alarm probabilities of the proposed detectors are not easy to obtain, we thereby use the idea of the well-known energy detector (ED)\footnote{An analytical solution for designing a fixed-level ED under conditions of NVU has been provided in \cite{zimc1} for spectrum sensing in calibrated cognitive radio.} to control the level of the proposed detectors even under conditions of uncertainty regarding the parameter $\mathrm{DNR_{\text{avg}}}$. In this approach, the empirical FAP ($P_{fa}$) is constrained to the desired FAP ($p_{_{\rm fa}}$), i.e., $P_{fa}\le p_{_{\rm fa}}$ \cite{zimc1, cas}. As a result, we refer to the proposed detectors as $p_{fa}$-fixed-level detectors instead of CFAR detectors (i.e., $P_{fa}=p_{_{\rm fa}}$), as obtaining CFAR detectors is not feasible under conditions of $\mathrm{DNR_{\text{avg}}}$ uncertainties. It can be shown that regulating the FAP of the proposed detectors only requires knowledge of the minimum achievable value of $\mathrm{DNR_{\text{avg}}}$, as illustrated in Section \ref{sec6}. 
}
	\section{Computational Complexity Comparison of Proposed Detection Algorithms}\label{sec5} 
	In this section, we  evaluate the computational complexity of proposed target detection techniques in order to compare their performance in real-world implementations. For the LRT-based detector, the main computational complexity arises from the calculation of $\lambda_{\max }(\mathrm{\boldsymbol{\mathcal{Z}}}_{\mathrm{j}}^\mathrm{H}\mathrm{\boldsymbol{\mathcal{Z}}}_{\mathrm{j}})$, which involves $\mathcal{O}(4\rm{LN_r^2 N_t+4N_r^3N_t})$ computations. The complexity of this detector is directly proportional to the number of integration time samples (i.e., $\rm{L}$). On the other hand, for the other detectors, the main computational complexity is associated with the calculation of the eigenvector corresponding to the largest eigenvalues, requiring $\mathcal{O}(\rm{LN_r^2 N_t+L^2N_r N_t+L^3N_t})$ computations.
	The results indicate that the LRT-based detector has the lowest complexity, followed by the Rao, Gradient, and Wald tests.
	\section{Performance Assessment}\label{sec6} 
	In this section, we present some Monte Carlo (MC) simulation results to assess the effectiveness of the proposed detectors as well as to compare them to some existing detection approaches. This can help us to select a proper detector based on factors such as performance, computational complexity, robustness, and the trade off between performance and computational complexity.
	\subsection{Simulation Setup}
	As mentioned before, the goal of this paper is not to design waveform for an integrated passive radar system. Thus, for simplicity, we generate $\mathbf{s}_{j}$ according to $\mathbf{s}_{j}=\exp\{-\mathsf{j}\boldsymbol{\boldsymbol{\phi}}_{j}\}$, where the vectors $\boldsymbol{\boldsymbol{\phi}}_{j}\in\mathcal{R}^{\mathrm{L\times1}}$ have independent and identically distributed (IID) components, uniformly distributed on the interval $[0,2\pi)$. Moreover, unless
	otherwise stated, we assume that the number of integration time samples is equal to 1024 (i.e., $\mathrm{L=1024}$), and there are three distributed receivers (represented by $\mathrm{N_r=3}$) and two transmitters (represented by $\mathrm{N_t=2}$). Thus, we consider $\mathrm{{\boldsymbol{\sigma}}=[{\boldsymbol{\sigma}^T_{1}}, {\boldsymbol{\sigma}^T_{2}}]^{\rm T}}$ with $\boldsymbol{\sigma}_{1}=\sigma_{0}^{2}[1, 0.75, 1.3]^{\rm T}$
	and $\boldsymbol{\sigma}_{2}=1.15\times\sigma_{0}^{2}[1, 0.75, 1.3]^{\rm T}$, where, we use $\sigma_{0}^{2}=8.28\times10^{-13}$ {Watt}, resulting in a noise variance of about -121 dB. Besides, we assume that three receivers are located at $\mathbf{r}_{\rm 1}=[1,30]^{\rm T}$
	km, $\mathbf{r}_{\rm 2}=[5,50]^{\rm T}$ km and $\mathbf{r}_{\rm 3}=[12,80]^{\rm T}$
	km, two transmitters are at $\mathbf{t}_{\rm 1}=[30,10]^{\rm T}$ km and $\mathbf{t}_{\rm 2}=[40,50]^{\rm T}$
	km, and a target is at $\mathbf{r}_{\rm T}=[35,45]^{\rm T}$ km {with the velocity of $\mathbf{v}_{\rm T}=[100,100]^{\rm T}$ m/s}. Since the closed-form expressions of the probability density function (pdf) for the test statistics under each hypothesis are unavailable, the performance assessment is carried out by resorting to MC simulations. To obtain a good estimation accuracy {for the nominal $p_{_{\rm fa}}$ of $10^{-4}$}, the probability of detection ($\rm P_d$) and empirical false alarm probability $\rm P_{fa}$ are estimated based on $10^4$ and $5\times10^6$ independent MC trials, respectively. Similar to (\ref{eq5a2}), the average signal power-to-noise power ratio of the SC can be defined as
	\begin{align}\label{eq5a1}
		\mathrm{SNR_{\text{avg}}=\frac{1}{\mathrm{\mathrm{N_{\rm r}}\mathrm{{\mathrm{N_{t}}}}}}\sum_{j=1}^{\mathrm{{\mathrm{N_{t}}}}}\sum_{k=1}^{\mathrm{N_{\rm r}}}\frac{\abs{{\alpha}_{jk}}^{2}}{\sigma_{jk}^{2}}},
	\end{align}
	\subsection{FAP Sensitivity on DNR\textsubscript{avg}}
	In Table \ref{table2}, the detection threshold values of the proposed detectors are reported to achieve a desired FAP of $p_{\rm fa}=10^{-4}$ for two values of DNR\textsubscript{avg}, namely -10 dB and 0 dB. It can be observed that the detection thresholds of the proposed detector, except for the Rao detector, strongly depend on the values of DNR\textsubscript{avg}. Here, the threshold of the Rao detector slightly decreases with increasing DNR\textsubscript{avg}. This implies that the LRT, Wald and Gradient detectors are non-robust detectors against the DNR\textsubscript{avg} values. Therefore, a PR designer must be aware of the DNR\textsubscript{avg} value to set the detection thresholds of non-robust detectors. In real-world scenarios, this may not always be feasible and may result in errors.\par
\begin{figure}[!]
		\centerline{\includegraphics[width=2.8in, height=2.2in]{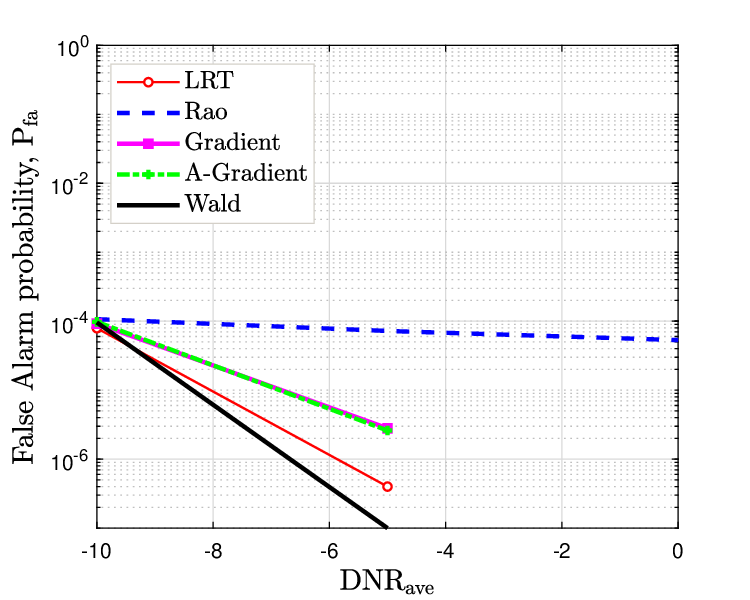}}
		\caption{Probability of false alarm as a function of $\rm{{DNR}_{avg}}$ when $\mathrm{N_t=2}$, $\mathrm{N_r=3}$, $\rm L=1024$, $\rm{DNR_{avg}^{\eta}}= -10$ dB and  $p_{_{\rm fa}}=10^{-4}$.}
		\label{fig6}
	\end{figure}
\begin{figure}[!]
		\centerline{\includegraphics[width=2.8in, height=2.2in]{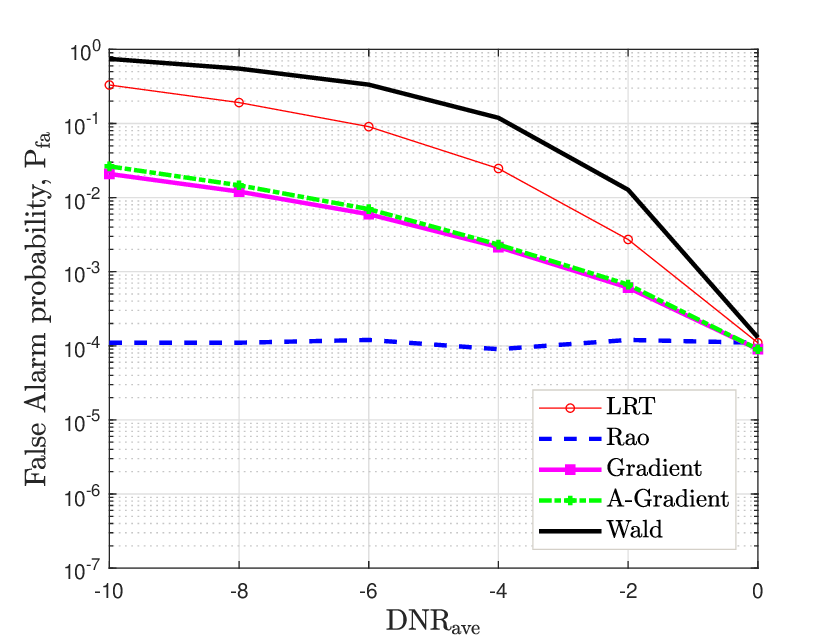}}
		\caption{Probability of false alarm as a function of $\rm{{DNR}_{avg}}$ when $\mathrm{N_t=2}$, $\mathrm{N_r=3}$, $\rm L=1024$, $\rm{DNR_{avg}^{\eta}}= 0$ dB and  $p_{_{\rm fa}}=10^{-4}$}
		\label{fig7}
	\end{figure}

	\begin{table}[t!]
		\centering
		\caption{Detection Threshold Values for a Desired False Alarm Probability of $p_{_{\rm fa}}=10^{-4}$}
		\label{table2}
		\small 		
		\begin{tabular}{c||c|c} 
			\hline Detector & $\rm{{DNR}_{avg}}=-10$ dB& $\rm{{DNR}_{avg}}=0$ dB \\ 
			\hline \hline LRT & 0.0411 & 0.0096 \\
			\hline Wald & 0.4991 & 0.0431 \\
			\hline Rao (Durbin) & 0.0231 & 0.0227 \\
			\hline Gradient & 0.0896 & 0.0311 \\
			\hline A-Gradient & 0.0916 & 0.0311 \\
			\hline
		\end{tabular}
	\end{table}
	However, we can resort to the new concept of the level-of-test to set the level of the non-robust detectors rather than their sizes. To illustrate, let us represent the true averaged value of reference channel DNR as DNR\textsubscript{avg}, while $\rm{DNR}_{avg}^{\eta}$ represents the value of DNR\textsubscript{avg} by which the detection threshold of a detector is adjusted. Assume that the received DNR\textsubscript{avg} for the considered radar-setup scenarios is in the range of [-10, 0] dB. To find a solution for setting the detection thresholds of the non-robust detectors, the FAP as a function of DNR\textsubscript{avg} is plotted in Figs. \ref{fig6} and \ref{fig7} when thresholds are set according to $\rm{{DNR}_{avg}^{\eta}}$= -10 dB and $\rm{{DNR}_{avg}^{\eta}}$= 0 dB, respectively. It is important to highlight that the FAP of the Rao-based detector remains relatively consistent, irrespective of the values of DNR\textsubscript{avg}. Specifically,  the threshold of the Rao detector slightly decreases as DNR\textsubscript{avg} increases. This highlights another robustness of the Rao score test when dealing with uncertainties in the PR systems. For the non-robust detectors, {our results demonstrate that setting the detection thresholds based on the minimum value of DNR\textsubscript{avg} (i.e., $\rm{{DNR}_{avg}^{\eta}}$= -10 dB) enables us to achieve fixed detection levels, ensuring that false alarm probabilities remain below the desired FAP of $p_{_{\rm fa}}=10^{-4}$. Conversely, setting the detection thresholds based on the maximum value of DNR\textsubscript{avg} (i.e., $\rm{{DNR}_{avg}^{\eta}}$= 0 dB) leads to an excessive number of false alarms, which are not tolerable in real-world applications.}
	\subsection{Detection Performance Comparison of Proposed Detection Algorithms}
{This subsection delves into the detection performance of the proposed detectors under two scenarios to elucidate the effects of $\rm{DNR_{avg}}$ and detection threshold adjustments on detector performance. In the first scenario, we set $\rm{DNR_{avg}^\eta}$
equal to the true value of $\rm{DNR_{avg}}$, enabling us to construct detectors of fixed size. This approach allows us to assess the impact of $\rm{DNR_{avg}}$
on detection performance under conditions of complete certainty. In the second scenario, we assume $\rm{DNR_{avg}^\eta}$ is set at -10 dB while the true value of $\rm{DNR_{avg}}$ is 0 dB. Our aim in this case is to examine the degradation in detection performance that arises from employing larger detection thresholds to establish the level of tests under the considered worst-case condition.}

\begin{figure}[!]
		\centerline{\includegraphics[width=2.8in, height=2.2in]{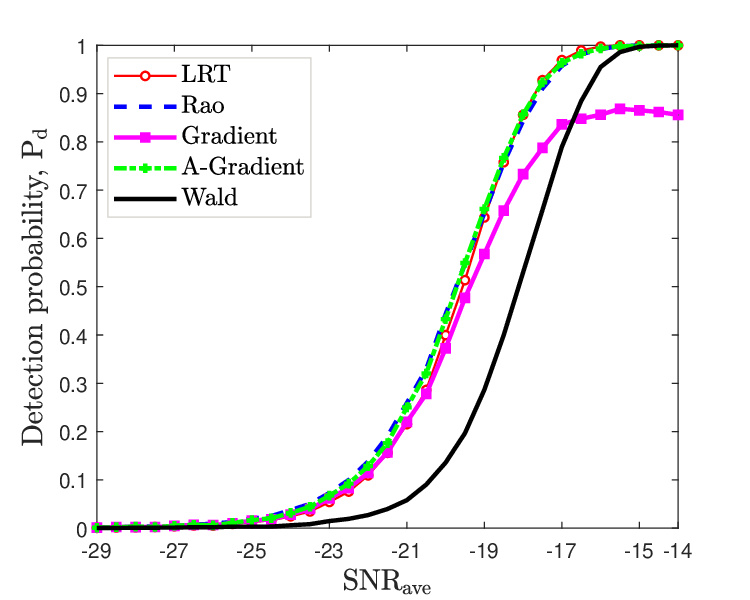}}
		\caption{Probability of detection as a function of $\rm{{SNR}_{avg}}$ when $\mathrm{N_t=2}$, $\mathrm{N_r=3}$, $\rm L=1024$, $\rm{{DNR}_{avg}}=\rm{DNR_{avg}^{\eta}}= -10$ dB and  $p_{_{\rm fa}}=10^{-4}$.}
		\label{fig8}
	\end{figure}

\begin{figure}[!]
		\centerline{\includegraphics[width=2.8in, height=2.2in]{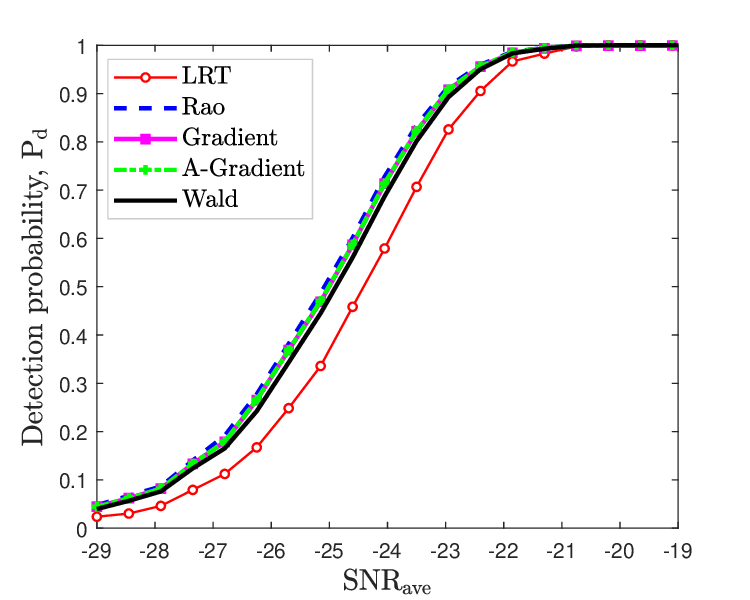}}
		\caption{Probability of detection as a function of $\rm{{SNR}_{avg}}$ when $\mathrm{N_t=2}$, $\mathrm{N_r=3}$, $\rm L=1024$,  $\rm{{DNR}_{avg}}=\rm{DNR_{avg}^{\eta}}= 0$ dB and  $p_{_{\rm fa}}=10^{-4}$.}
		\label{fig9}
	\end{figure}

\begin{figure}[!]
		\centerline{\includegraphics[width=2.8in, height=2.2in]{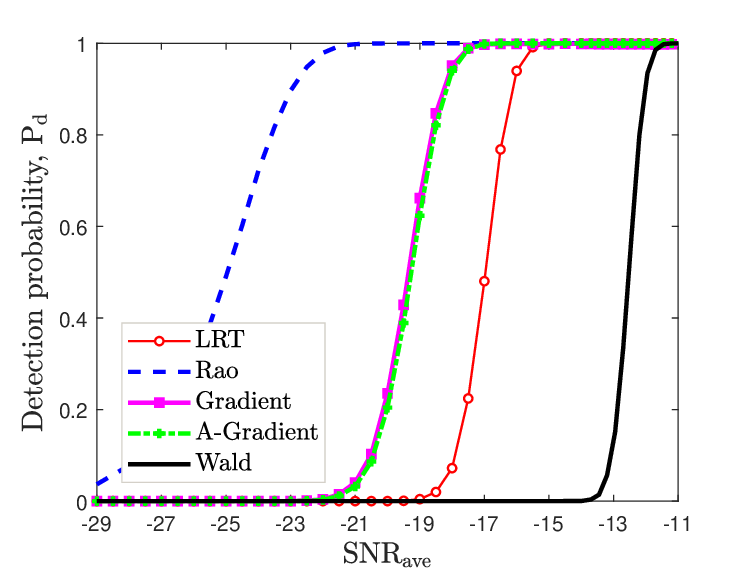}}
		\caption{Probability of detection as a function of $\rm{{SNR}_{avg}}$ when $\mathrm{N_t=2}$, $\mathrm{N_r=3}$, $\rm L=1024$,  $\rm{{DNR}_{avg}}= 0$ dB, $\rm{DNR_{avg}^{\eta}}= -10$ dB and  $p_{_{\rm fa}}=10^{-4}$.}
		\label{fig11}
	\end{figure}

The results of the first scenario are illustrated in Figs. \ref{fig8}-\ref{fig9} for two values of $\rm{DNR_{avg}^\eta}=\rm{DNR_{avg}}$, namely -10 dB and 0 dB, respectively. Several key observations can be made. Firstly, as expected, the detection performance improves with an increase in $\rm{DNR_{avg}}$. Secondly, the Wald and Gradient detectors exhibit enhanced target detection capability with higher $\rm{DNR_{avg}}$. Notably, when $\rm{DNR_{avg}}$ is high, the performances of the Wald and Gradient detectors are comparable to that of the Rao test. However, in scenarios where the reference channels are noisy and $\rm{DNR_{avg}}$ is low, these two detectors demonstrate varying levels of performance despite their asymptotic equivalence. Lastly, regardless of the values of $\rm{DNR_{avg}}$, the proposed Rao detector performs better than other detectors. For the second scenario, the results are shown in Fig. \ref{fig11}. The Rao-based detector consistently outperforms the other detectors, demonstrating superior performance across all simulation scenarios. The Gradient, A-Gradient, LRT, and Wald tests follow in terms of detection performance. A comparison between the results shown in Figs. \ref{fig9} and \ref{fig11} highlights a substantial decline in performance for the non-robust detectors, whereas the performance of the Rao detector remains unchanged. {This clearly emphasizes the importance of finding robust detectors capable of handling practical scenarios under conditions of uncertainty. The proposed Rao (or Durbin) detector offers such a solution.} 
	\subsection{Influence of Multipath Signal on Reference Channels}
\begin{figure}[!]
		\centerline{\includegraphics[width=2.8in, height=2.2in]{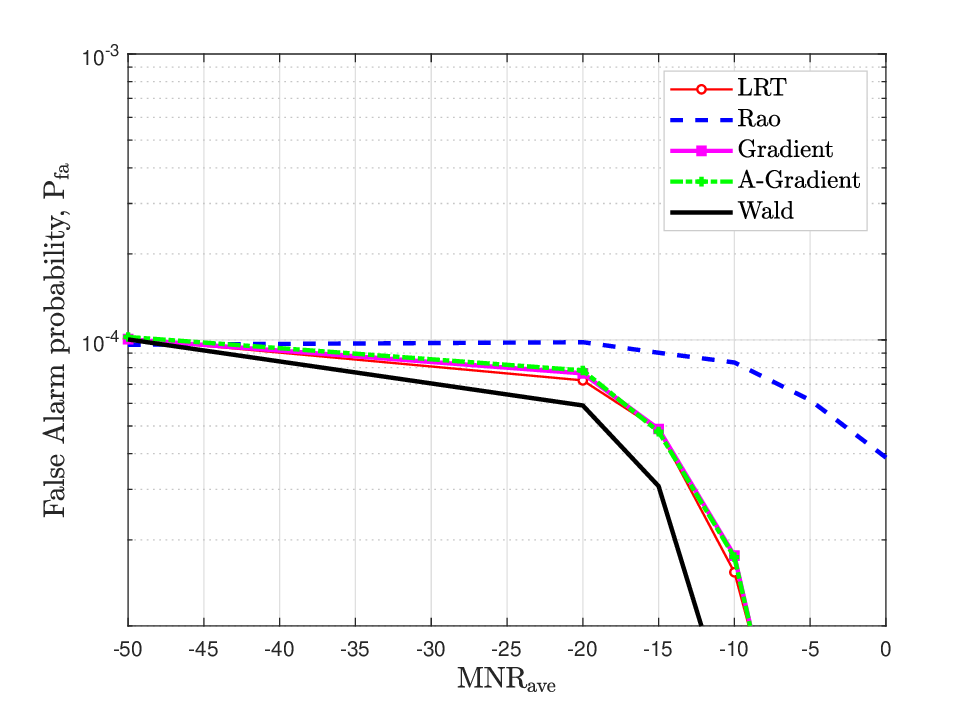}}
		\caption{Probability of false alarm as a function of $\rm{{MNR}_{avg}}$ when $\mathrm{N_t=2}$, $\mathrm{N_r=3}$, $\rm L=1024$, $\rm{{DNR}_{avg}}=\rm{DNR_{avg}^{\eta}}= -10$ dB, and  $p_{_{\rm fa}}=10^{-4}$.}
		\label{fig12}
	\end{figure}

\begin{figure}[!]
		\centerline{\includegraphics[width=2.8in, height=2.2in]{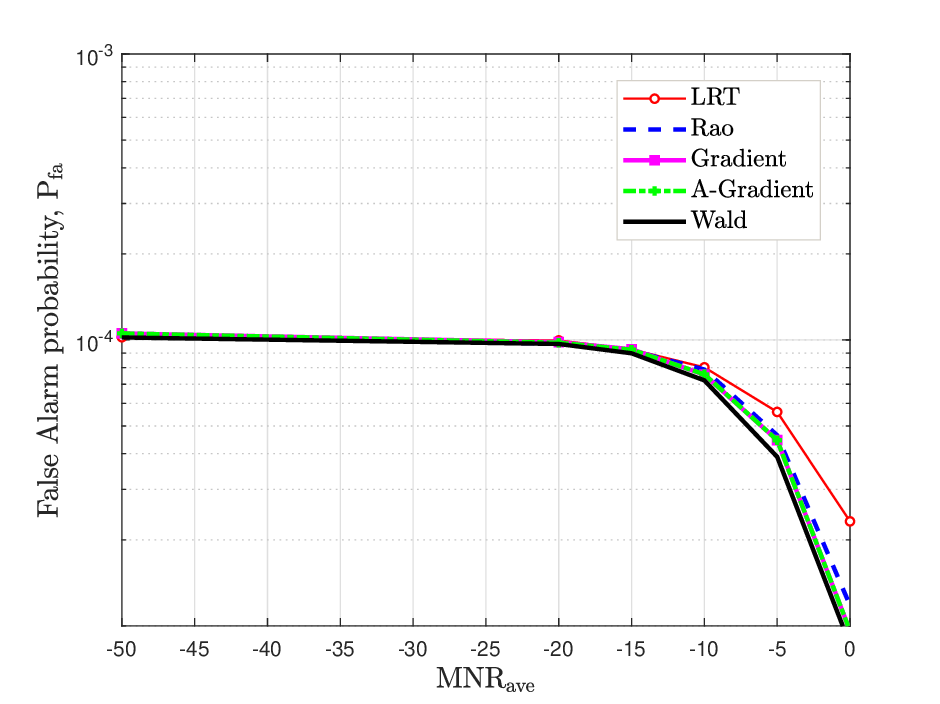}}
		\caption{Probability of false alarm as a function of $\rm{{MNR}_{avg}}$ when $\mathrm{N_t=2}$, $\mathrm{N_r=3}$, $\rm L=1024$, $\rm{{DNR}_{avg}}=\rm{DNR_{avg}^{\eta}}= 0$ dB and  $p_{_{\rm fa}}=10^{-4}$.}
		\label{fig13}
	\end{figure}

\begin{figure}[!]
		\centerline{\includegraphics[width=2.8in, height=2.2in]{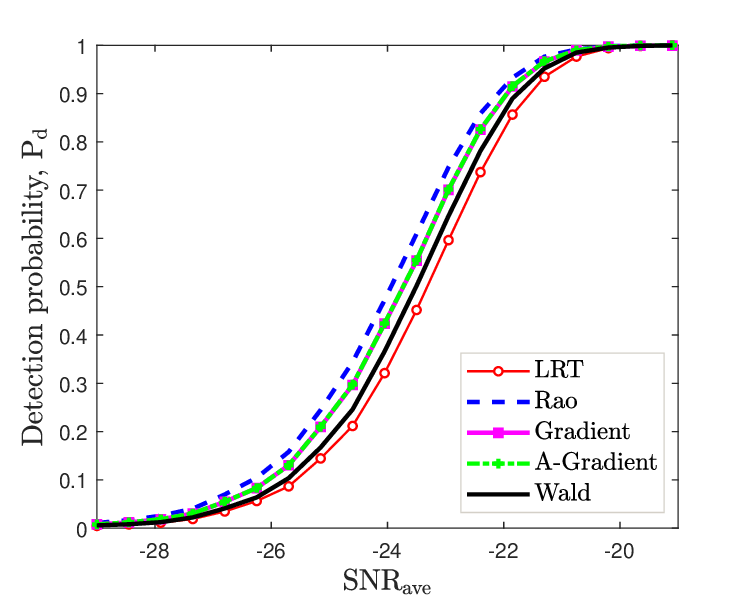}}
		\caption{Probability of detection as a function of $\rm{{SNR}_{avg}}$ when $\mathrm{N_t=2}$, $\mathrm{N_r=3}$, $\rm L=1024$,  $\rm{{DNR}_{avg}}=\rm{DNR_{avg}^{\eta}}= 0$ dB, $\rm{{MNR}_{avg}} = 0$ dB, and  $p_{_{\rm fa}}=10^{-4}$.}
		\label{fig9multi}
	\end{figure}

	{Eliminating multipath echoes in the reference channels remains a significant challenge in signal processing for CPR and IPR systems. Our signal model assumes the absence of multipath echo signals in the reference channels. This assumption implies that signal conditioning in the reference channels has been optimized to ensure noise levels comparable to those of the direct-path signals. To assess the robustness of the proposed detectors to multipath echo on reference channels}, we consider a MIMO IPR scenario with three receivers located at $\mathbf{r}_{1}=[1,30]^{\mathrm{T}}$
	km, $\mathbf{r}_{2}=[5,50]^{\mathrm{T}}$ km, and $\mathbf{r}_{3}=[12,80]^{\mathrm{T}}$
	km. There are two transmitters located at $\mathbf{t}_{1}=[30,10]^{\mathrm{T}}$ km and $\mathbf{t}_{2}=[40,50]^{\mathrm{T}}$
	km, as well as a clutter scatterer (multipath) at $\mathbf{r}_{\mathrm{M}}=[10,15]^{\mathrm{T}}$ km. 
	Similar to (\ref{eq5a2}), the average multipath signal power-to-noise power ratio of the RC can be defined as
	\begin{align}\label{eq5a1}
		\mathrm{MNR_{\text{avg}}=\frac{1}{\mathrm{\mathrm{N_{\rm r}}\mathrm{{\mathrm{N_{t}}}}}}\sum_{j=1}^{\mathrm{{\mathrm{N_{t}}}}}\sum_{k=1}^{\mathrm{N_{\rm r}}}\frac{\abs{{\zeta}_{jk}}^{2}}{\sigma_{jk}^{2}}},
	\end{align}
	The complex amplitudes ${\zeta}_{{\rm jk}}$ account for various factors affecting the propagation of the signal from the $\rm j$-th transmitter to the $\rm k$-th receiver. These factors include the gains of the antennas, the effects of channel propagation in the path from the transmitter to the receiver, and the reflectivity of the clutter scatterer. In Figs. \ref{fig12} and \ref{fig13}, the FAP is plotted as a function of $\rm{MNR}_{\text{avg}}$, when thresholds are set to $\rm{DNR}_{\text{avg}}^{\eta} = -10$ dB and $\rm{DNR}_{\text{avg}}^{\eta} = 0$ dB, respectively. Since the antenna pattern of the reference channels is steered towards the transmitters, it makes sense to assume that $\rm{MNR}_{\text{avg}}\ll\rm{DNR}_{\text{avg}}$. Under these practical situations, it can be observed that the proposed detectors achieve fixed sizes in the presence of multipath echo signals in the reference channels. In Figs. \ref{fig12} and \ref{fig13}, we also consider the cases of large $\rm{MNR}_{\text{avg}}$ and show that the proposed detectors are able to control their levels based on the desired FAP of $10^{-4}$. This clearly indicates the superior performance of the proposed detectors in the presence of multipath echoes in the reference channels. Among the proposed detectors, the Rao detector attempts to adjust its size rather than its level, particularly in the low $\rm{DNR}_{\text{avg}}$ regime. In terms of detection performance, when $\rm{DNR}_{\text{avg}}^{\eta} = 0$ dB and $\rm{MNR}_{\text{avg}}=0$ dB, the detection performance of the proposed detectors is examined in Fig. \ref{fig9multi}. The results of Fig. \ref{fig9multi} indicate that all the curves in Fig. \ref{fig9} are just shifted to the right by about 1 dB. Meanwhile, the gap between the LRT and other detectors is reduced. This and the decline in detection performance can be explained by referring to Fig. \ref{fig13}, which illustrates how the detection thresholds of the proposed detectors have been increased in order to regulate the test levels.
	\subsection{Detection Performance Comparison between Proposed and Existing Detectors}
\begin{figure}[!]
		\centerline{\includegraphics[width=2.8in, height=2.2in]{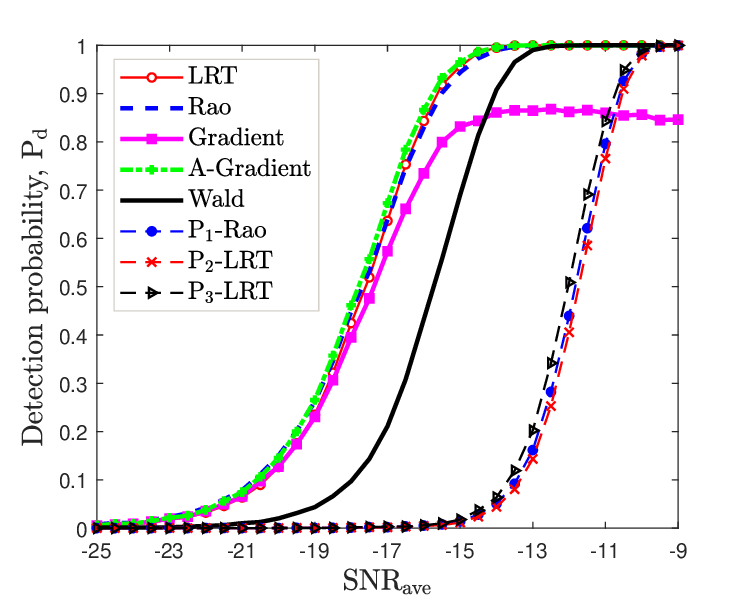}}
		\caption{Probability of detection as a function of $\rm{{SNR}_{avg}}$ when $\mathrm{N_t=1}$, $\mathrm{N_r=3}$ (i.e., SIMO configuration),  $\rm L=1024$, $\rm{{DNR}_{avg}^{\eta}}= -10$ dB, $\rm{{DNR}_{avg}}= -10 $ dB, and $p_{_{\rm fa}}=10^{-4}$.}
		\label{fig14}
	\end{figure}

\begin{figure}[!]
		\centerline{\includegraphics[width=2.8in, height=2.2in]{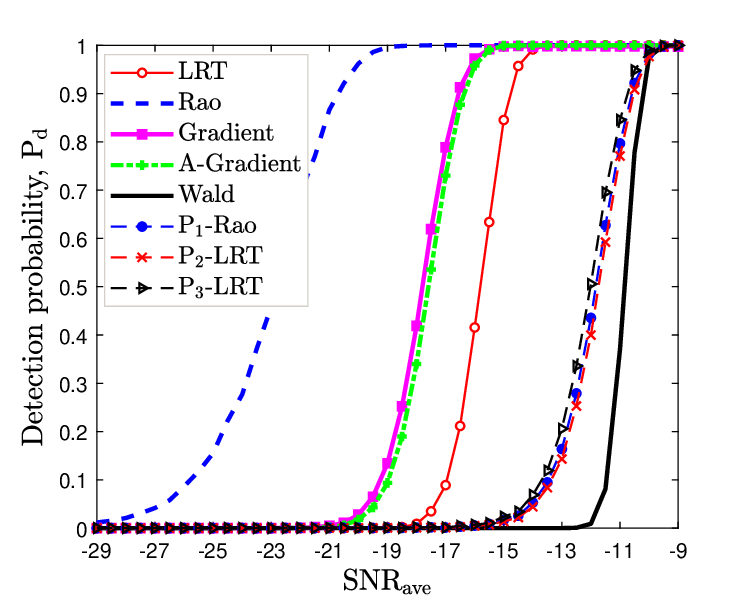}}
		\caption{Probability of detection as a function of $\rm{{SNR}_{avg}}$ when $\mathrm{N_t=1}$, $\mathrm{N_r=3}$ (i.e., SIMO configuration),  $\rm L=1024$, $\rm{{DNR}_{avg}^{\eta}}=-10$ dB, $\rm{{DNR}_{avg}}= 0$ dB and $p_{_{\rm fa}}=10^{-4}$.}
		\label{fig16}
	\end{figure}

	In \cite{p1c2amir3}, a framework for the single-channel SIMO IPR target detection problem is developed, utilizing the same signal model (\ref{eq2x}) on the surveillance channel. However, it examines three different target detection problems (i.e., $\rm P_1$, $\rm P_2$, and $\rm P_3$) based on different assumptions concerning the unknown parameter space. The Rao and LRT principles have been exploited to propose three new detectors, referred to as $\rm P_1$-Rao, $\rm P_2$-LRT, and $\rm P_3$-LRT, given by
	\begin{align}\label{eqex1}
		\Lambda_{\mathrm{\rm P_1-Rao}}=2\mathrm{tr}\big\{(\widehat{\boldsymbol{\mathrm R}}_{\boldsymbol{\boldsymbol{\boldsymbol{\mathcal{X}}}}_{{\rm j}}}{{\widehat{\boldsymbol{\mathrm{\Sigma}}}}_0}^{-1}-\mathbf{I}_{\rm N_r})^2\big\}
	\end{align}
	\begin{align}\label{eqex2}
		\Lambda_{\mathrm{\rm P_2-LRT}}=\rm det^{-1}\big({{\widehat{\boldsymbol{\mathrm{\Sigma}}}}_0}^{-\frac{1}{2}}\widehat{\boldsymbol{\mathrm R}}_{\boldsymbol{\boldsymbol{\boldsymbol{\mathcal{X}}}}_{{\rm j}}}{{\widehat{\boldsymbol{\mathrm{\Sigma}}}}_0}^{-\frac{1}{2}}\big)
	\end{align}
	\begin{align}\label{eqex4}
		\Lambda_{\mathrm{\rm P_3-LRT}}={\mathrm{\lambda }}_{\rm max}\big({{\widehat{\boldsymbol{\mathrm{\Sigma}}}}_0}^{-\frac{1}{2}}\widehat{\boldsymbol{\mathrm R}}_{\boldsymbol{\boldsymbol{\boldsymbol{\mathcal{X}}}}_{{\rm j}}}{{\widehat{\boldsymbol{\mathrm{\Sigma}}}}_0}^{-\frac{1}{2}}\big)
	\end{align}
	where ${{\widehat{\boldsymbol{\mathrm{\Sigma}}}}_0}=\mathrm{Diag}(\widehat{\boldsymbol{\mathrm R}}_{\boldsymbol{\boldsymbol{\boldsymbol{\mathcal{X}}}}_{{\rm j}}})$ and $\widehat{\boldsymbol{\mathrm R}}_{\boldsymbol{\boldsymbol{\boldsymbol{\mathcal{X}}}}_{{\rm j}}}={\boldsymbol{\boldsymbol{\boldsymbol{\mathcal{X}}}}^{\rm H}_{{\rm j}}}{\boldsymbol{\boldsymbol{\boldsymbol{\mathcal{X}}}}_{{\rm j}}}$. {We use these detectors to ensure a fair comparison, as they utilize the same signal model as our proposed detectors on the surveillance channels.  Furthermore, to the best of the authors' knowledge, there is no two-channel detection method addressing the sensitivity of threshold setting on DNR\textsubscript{avg}. Additionally, this enables us to compare two-channel IPR detection methods with single-channel ones. It is shown in \cite{mybook} that the $\rm P_1$-Rao test is also a locally most powerful invariant test (LMPIT) in the single-channel case. } For comparative purposes, we consider a SIMO configuration with $\mathrm{N_t=1}$ and $\mathrm{N_r=3}$, an integration sample size of $\mathrm{L=1024}$, and a desired FAP of $\mathrm{p_{_{\rm fa}}=10^{-4}}$. It is worth noting that detectors $\mathrm{P_1}$-Rao, $\mathrm{P_2}$-LRT, and $\mathrm{P_3}$-LRT are not affected by the quality of the reference channels, {as they are categorized as single-channel detectors.} {In Fig. \ref{fig14}, we investigate a scenario with complete certainty (i.e., $\rm{{DNR}_{avg}^{\eta}}=\rm{{DNR}_{avg}}= -10 $ dB) and compare the performance of the aforementioned detectors.} The results clearly demonstrate that the proposed detectors exhibit significantly better performance compared to the $\mathrm{P_1}$-Rao, $\mathrm{P_2}$-LRT, and $\mathrm{P_3}$-LRT detectors. In the uncertainty scenario illustrated in Fig. \ref{fig16}, however, the detection performance of the proposed Wald test exhibits a significant degradation. In contrast, the other proposed detectors, specifically the Rao test, demonstrate superior performance compared to $\mathrm{P_1}$-LRT, $\mathrm{P_2}$-LRT, and $\mathrm{P_3}$-LRT. {The superior performance of the proposed detectors can be attributed to their utilization of reference channels, which facilitates two-channel target detection.}
	\section{Conclusion and Future Works}\label{sec7}
	In this study, we developed a framework for a two-channel uncalibrated MIMO passive radar network operating at multiple frequencies, where the reference channels were noisy.
	We proposed five new detectors based on different criteria, including LRT, Rao, Wald, Durbin, and Gradient. Our results demonstrate that the detection thresholds of the proposed detectors, except for the Rao test, are influenced by the DNR\textsubscript{avg} of the reference channels, rendering them non-robust detectors. {To effectively establish detection thresholds for proposed non-robust detectors, we introduced a novel threshold-setting strategy that aims to achieve a fixed level-of-test while accounting for uncertainty in DNR\textsubscript{avg} values of the reference channels.  The LRT-based detector exhibits the lowest computational complexity among the proposed detectors, making it particularly well-suited for very long-range integrated MIMO passive radar systems. However, the Rao test, despite its higher computational demands, demonstrates superior performance in terms of both detection performance and false alarm regulation curves. Furthermore, our simulations demonstrated the effectiveness of the proposed detectors in maintaining a fixed  level of false alarm probability even in the presence of multipath signal in the reference channels, rendering them robust to imperfect signal conditioning.
	Finally, we evaluated the performance of the proposed two-channel detectors against state-of-the-art single-channel detectors, including a locally most powerful invariant test among them. The results demonstrate that the proposed detectors achieve superior detection performance compared to the existing single-channel detection methods, even in scenarios with significant uncertainty (i.e., 10 dB) between the true value of DNR\textsubscript{avg} and that used for setting the detection threshold. This exceptional resilience to reference channel uncertainty represents a significant advancement in the field of two-channel IPR detection, as no existing work in the literature has previously demonstrated such remarkable robustness under these conditions.}

	The direct treatment of multipath signals as interference when solving the detection problem is often impractical. This explains why most passive radar researchers have not considered multipath signals when designing target detection algorithms \cite{p1c2him1}-\cite{GE}. Admittedly, the residual multipath signals are non-negligible in the two-channel MIMO PR when considering practical received signals. In addition to including the reference channel to formulate the target detection problem, a promising solution to eliminate residuals from multipath signals is to separately use the reference channel to estimate the transmitted signals. This enables us to employ the interference cancellation method, as introduced in \cite{p1c2zim3}, to remove direct-path and multipath signals. The success of this approach depends on the quality of the reference channels and the chosen detection methods. Our study in this paper indicates that the proposed detectors perform robustly in the presence of multipath signals in the reference channels. Further investigation in this area is imperative to advance our research in this direction, and such studies are already in progress. Furthermore, this expansion would facilitate the examination of how waveform design impacts interference elimination and the detection of multiple targets in scenarios involving a communication-centric radar system.
	\appendices{ }
	\section{LRT Detector Derivation}\label{A}
	According to (\ref{eq14}), the LRT statistic can be represented as
	\begin{equation}\label{eq3a1}
		\mathrm{\Lambda }\left(\boldsymbol{\mathrm{z}}\right)={\mathrm{\Lambda }}_{\mathrm{FT}}\left(\boldsymbol{\mathrm{z}}\right)-{\mathrm{\Lambda }}_{\mathrm{ST}}\left(\boldsymbol{\mathrm{z}}\right)
	\end{equation}
	where the first term (FT) and the second term (ST) of the LRT statistic can be computed as
	\begin{equation}\label{eq3a2}
		\resizebox{.87\hsize}{!}{${\mathrm{\Lambda }}_{\mathrm{FT}}\left(\boldsymbol{\mathrm{z}}\right)=\mathop{\mathrm{sup}}_{{\left\{{\mathrm{\sigma }}^{\mathrm{2}}_{\mathrm{jk}}\right\}}^{\left({\mathrm{N}}_{\mathrm{r}},{\mathrm{N}}_{\mathrm{t}}\right)\ }_{\left(\mathrm{j,k}\right)\mathrm{=(1,1)}},{\left\{{\mathrm{\beta }}_{\mathrm{jk}}\right\}}^{\left({\mathrm{N}}_{\mathrm{r}},{\mathrm{N}}_{\mathrm{t}}\right)\ }_{\left(\mathrm{j,k}\right)\mathrm{=(1,1)}},{\left\{{\mathbf{s}}_{\mathrm{j}}\right\}}^{{\mathrm{N}}_{\mathrm{t}}}_{\mathrm{j=1}}\mathrm{\ }\mathrm{\ }}\mathcal{L}\left({\boldsymbol{\mathrm{\theta }}}_{\boldsymbol{\mathrm{r}}}\boldsymbol{=}\boldsymbol{0},{\boldsymbol{\mathrm{\theta }}}_{\mathbf{s}}\mathrm{;}\boldsymbol{\ }\boldsymbol{\mathrm{z}}\right)$},
	\end{equation} 
	\begin{equation}\label{eq3a3}
		\resizebox{.87\hsize}{!}{$	{\mathrm{\Lambda }}_{\mathrm{ST}}\left(\boldsymbol{\mathrm{z}}\right)=\mathop{\mathrm{sup}}_{{\mathrm{\sigma }}^{\mathrm{2}},{\left\{{\mathrm{\beta }}_{\mathrm{jk}}\right\}}^{\left({\mathrm{N}}_{\mathrm{r}},{\mathrm{N}}_{\mathrm{t}}\right)\ }_{\left(\mathrm{j,k}\right)\mathrm{=(1,1)}},\ \ {\left\{{\mathrm{\alpha }}_{\mathrm{jk}}\right\}}^{\left({\mathrm{N}}_{\mathrm{r}},{\mathrm{N}}_{\mathrm{t}}\right)\ }_{\left(\mathrm{j,k}\right)\mathrm{=(1,1)}},{\left\{{\mathbf{s}}_{\mathrm{j}}\right\}}^{{\mathrm{N}}_{\mathrm{t}}}_{j=1}\mathrm{\ }\mathrm{\ }}\mathcal{L}\left({\boldsymbol{\mathrm{\theta }}}_{\boldsymbol{\mathrm{r}}},{\boldsymbol{\mathrm{\theta }}}_{\mathbf{s}}\mathrm{;}\boldsymbol{\ }\boldsymbol{\mathrm{z}}\right)$}.
	\end{equation}
	The FT can be obtained as shown at the top of the next page in (\ref{eq3a4}).
	\newcounter{tempequationcounter}
	\begin{figure*}[!]
		\normalsize
		\setcounter{tempequationcounter}{\value{equation}}
		\begin{IEEEeqnarray}{rCl}
			\setcounter{equation}{38}
			\label{eq3a4}
			\resizebox{.93\hsize}{!}{${\mathrm{\Lambda }}_{\mathrm{FT}}\left(\boldsymbol{\mathrm{z}}\right)\mathrm{=}-\mathrm{2}\mathrm{L}{\mathrm{N}}_{\mathrm{t}}{\mathrm{N}}_{\mathrm{r}}{\mathrm{ln} \mathrm{\pi }}+\mathop{\mathrm{sup}}_{{\left\{{\mathbf{s}}_{\mathrm{j}}\right\}}^{{\mathrm{N}}_{\mathrm{t}}}_{\mathrm{j=1}}\mathrm{\ }\mathrm{\ }}\left\{\mathop{\mathrm{sup}}_{{\left\{{\mathrm{\sigma }}^{\mathrm{2}}_{\mathrm{jk}}\right\}}^{\left({\mathrm{N}}_{\mathrm{r}},{\mathrm{N}}_{\mathrm{t}}\right)\ }_{\left(\mathrm{j,k}\right)\mathrm{=(1,1)}}}\left\{\mathop{\mathrm{sup}}_{{\left\{{\mathrm{\beta }}_{\mathrm{jk}}\right\}}^{\left({\mathrm{N}}_{\mathrm{r}},{\mathrm{N}}_{\mathrm{t}}\right)\ }_{\left(\mathrm{j,k}\right)\mathrm{=(1,1)}}} {\mathrm{-2L}\ }\sum^{{\mathrm{N}}_{\mathrm{t}}}_{\mathrm{j=1}}{\sum^{{\mathrm{N}}_{\mathrm{r}}}_{\mathrm{k=1}}}   \left\{{{\mathrm{ln} {\mathrm{\sigma }}^{\mathrm{2}}_{\mathrm{jk}}\ }}\mathrm{-}{{\frac{{\lVert \boldsymbol{x}_{\mathrm{jk}}\rVert}^{\mathrm{2}}}{{\mathrm{\sigma }}^{\mathrm{2}}_{\mathrm{jk}}}}}\mathrm{-}{{\frac{{\lVert \boldsymbol{y}_{\mathrm{jk}}\mathrm{-}{\mathrm{\beta }}_{\mathrm{jk}}{\mathbf{s}}_{\mathrm{j}}\rVert}^{\mathrm{2}}}{{\mathrm{\sigma }}^{\mathrm{2}}_{\mathrm{jk}}}}}\right\}\right\}\right\}$}
		\end{IEEEeqnarray}
		\hrulefill
	\end{figure*}
	The supremum of the inner optimization problem of (\ref{eq3a4}) with respect to ${\mathrm{\beta }}_{\mathrm{jk}}$ can be obtained for  
	\begin{equation}\label{eq3a5}
		{{\mathrm{\beta }}}_{\mathrm{jk}}\mathrm{=}\frac{{{\mathbf{s}}_{\mathrm{j}}}^{\mathrm{H}}\boldsymbol{y}_{\mathrm{jk}}}{{\lVert {\mathbf{s}}_{\mathrm{j}}\rVert}^{\mathrm{2}}}
	\end{equation} 
	Substituting (\ref{eq3a5}) into (\ref{eq3a4}) results in (\ref{eq3a6c}), as shown at the top of the next page.
	\begin{figure*}[!]
		\normalsize
		\setcounter{tempequationcounter}{\value{equation}}
		\begin{IEEEeqnarray}{rCl}
			\setcounter{equation}{39}\label{eq3a6c}
			{\mathrm{\Lambda }}_{\mathrm{FT}}\left(\boldsymbol{\mathrm{z}}\right)=-\mathrm{2}\mathrm{L}{\mathrm{N}}_{\mathrm{t}}{\mathrm{N}}_{\mathrm{r}}{\mathrm{ln} \mathrm{\pi}}+\mathop{\mathrm{sup}}_{{\left\{{\mathbf{s}}_{\mathrm{j}}\right\}}^{{\mathrm{N}}_{\mathrm{t}}}_{\mathrm{j=1}}}\Bigg\{\mathop{\mathrm{sup}}_{{\left\{{\mathrm{\sigma }}^{\mathrm{2}}_{\mathrm{jk}}\right\}}^{\left({\mathrm{N}}_{\mathrm{r}},{\mathrm{N}}_{\mathrm{t}}\right)\ }_{\left(\mathrm{j,k}\right)\mathrm{=(1,1)}}\mathrm{\ }\mathrm{\ }}\bigg\{\mathrm{-2L}\ \sum^{{\mathrm{N}}_{\mathrm{t}}}_{\mathrm{j=1}}{\sum^{{\mathrm{N}}_{\mathrm{r}}}_{\mathrm{k=1}}\bigg({{\mathrm{ln} {\mathrm{\sigma }}^{\mathrm{2}}_{\mathrm{jk}}\ }}}-{\frac{\left({\lVert \boldsymbol{y}_{\mathrm{jk}}\rVert}^{\mathrm{2}}\mathrm{-}\frac{{\left|{{\mathbf{s}}_{\mathrm{j}}}^{\mathrm{H}}\boldsymbol{y}_{\mathrm{jk}}\right|}^{\mathrm{2}}}{{\lVert {\mathbf{s}}_{\mathrm{j}}\rVert}^{\mathrm{2}}}\right)\mathrm{+}{\lVert \boldsymbol{x}_{\mathrm{jk}}\rVert}^{\mathrm{2}}}{{\mathrm{\sigma }}^{\mathrm{2}}_{\mathrm{jk}}}}\bigg)\bigg\}\Bigg\}
		\end{IEEEeqnarray}
		\hrulefill
	\end{figure*}
	Again, the supremum of the inner optimization problem in (\ref{eq3a6c}) with respect to ${\mathrm{\sigma }}^{\mathrm{2}}_{\mathrm{jk}}$ can be obtained for
	\begin{equation}\label{eq3a6z}
		{\mathrm{\sigma }}^{\mathrm{2}}_{\mathrm{jk}}\mathrm{=}\frac{{\lVert \boldsymbol{y}_{\mathrm{jk}}\rVert}^{\mathrm{2}}\mathrm{-}\frac{{\left|{{\mathbf{s}}_{\mathrm{j}}}^{\mathrm{H}}\boldsymbol{y}_{\mathrm{jk}}\right|}^{\mathrm{2}}}{{\lVert {\mathbf{s}}_{\mathrm{j}}\rVert}^{\mathrm{2}}}\mathrm{+}{\lVert \boldsymbol{x}_{\mathrm{jk}}\rVert}^{\mathrm{2}}}{\mathrm{2L}}\end{equation}
	Substituting (\ref{eq3a6z}) into (\ref{eq3a6c}) produces 
	\begin{equation}\label{eq3a6df}
		\resizebox{.87\hsize}{!}{$	\begin{split}
				{\mathrm{\Lambda }}_{\mathrm{FT}}\left(\boldsymbol{\mathrm{z}}\right)&=\boldsymbol{\mathrm{-}}\mathrm{2}\mathrm{L}{\mathrm{N}}_{\mathrm{t}}{\mathrm{N}}_{\mathrm{r}}{\mathrm{ln} \mathrm{\pi }\ }-2\mathrm{L}\sum^{{\mathrm{N}}_{\mathrm{t}}}_{\mathrm{j=1}}{\sum^{{\mathrm{N}}_{\mathrm{r}}}_{\mathrm{k=1}}{{\mathrm{ln} \left(\frac{{\lVert \boldsymbol{y}_{\mathrm{jk}}\rVert}^{\mathrm{2}}+{\lVert \boldsymbol{x}_{\mathrm{jk}}\rVert}^{\mathrm{2}}}{\mathrm{2L}}\right)\ }}}\\&-\mathrm{2L}{\mathop{\mathrm{inf}}_{{\left\{{\mathbf{s}}_{\mathrm{j}}\right\}}^{{\mathrm{N}}_{\mathrm{t}}}_{\mathrm{j=1}}\mathrm{\ }} \bigg\{\sum^{{\mathrm{N}}_{\mathrm{t}}}_{\mathrm{j=1}}{\sum^{{\mathrm{N}}_{\mathrm{r}}}_{\mathrm{k=1}}{{\mathrm{ln} \bigg(\mathrm{1-}\frac{{{\mathbf{s}}_{\mathrm{j}}}^{\mathrm{H}}\frac{\boldsymbol{y}_{\mathrm{jk}}\boldsymbol{y}^{\mathrm{H}}_{\mathrm{jk}}}{{\lVert \boldsymbol{y}_{\mathrm{jk}}\rVert}^{\mathrm{2}}+{\lVert \boldsymbol{x}_{\mathrm{jk}}\rVert}^{\mathrm{2}}}{\mathbf{s}}_{\mathrm{j}}}{{{\mathbf{s}}_{\mathrm{j}}}^{\mathrm{H}}{\mathbf{s}}_{\mathrm{j}}}\bigg)\ }}}\bigg\}}\\&\mathrm{-}\mathrm{2L}{\mathrm{N}}_{\mathrm{t}}\mathrm{N}_{\mathrm{r}}
			\end{split}$}
	\end{equation}
	By using (\ref{eqx2}) and  (\ref{eqx3}), the FT of the LRT statistic can be simplified as
	\begin{equation}\label{eq3a6h}
		\resizebox{.87\hsize}{!}{$	\begin{split}
				{\mathrm{\Lambda }}_{\mathrm{FT}}\left(\boldsymbol{\mathrm{z}}\right)&=-\mathrm{2}\mathrm{L}{\mathrm{N}}_{\mathrm{t}}{\mathrm{N}}_{\mathrm{r}}{\mathrm{ln} \mathrm{\pi }\ }-2\mathrm{L}\sum^{{\mathrm{N}}_{\mathrm{t}}}_{\mathrm{j=1}}{\sum^{{\mathrm{N}}_{\mathrm{r}}}_{\mathrm{k=1}}{{\mathrm{ln} \bigg(\frac{{\lVert \boldsymbol{y}_{\mathrm{jk}}\rVert}^{\mathrm{2}}+{\lVert \boldsymbol{x}_{\mathrm{jk}}\rVert}^{\mathrm{2}}}{\mathrm{2L}}\bigg)\ }}}\\&+\mathrm{2L}\mathrm{\varrho }\sum^{{\mathrm{N}}_{\mathrm{t}}}_{\mathrm{j=1}}{{\mathrm{\lambda }}_{\mathrm{max}}({\boldsymbol{\mathcal{Y}}}_{\mathrm{j}}{\boldsymbol{\mathrm{\Xi }}}^{-1}_{\mathrm{j}}{{\boldsymbol{\mathcal{Y}}}_{\mathrm{j}}}^\mathrm{H})}\mathrm{-}\mathrm{2L}{\mathrm{N}}_{\mathrm{t}}\mathrm{N}_{\mathrm{r}}
			\end{split}$}
	\end{equation}
	where we have used the approximation of
	\begin{equation}\label{eq3a6dd}
		{\mathrm{ln} \bigg(\mathrm{1-}\frac{{{\mathbf{s}}_{\mathrm{j}}}^{\mathrm{H}}\frac{\boldsymbol{y}_{\mathrm{jk}}\boldsymbol{y}^{\mathrm{H}}_{\mathrm{jk}}}{{\lVert \boldsymbol{y}_{\mathrm{jk}}\rVert}^{\mathrm{2}}+{\lVert \boldsymbol{x}_{\mathrm{jk}}\rVert}^{\mathrm{2}}}{\mathbf{s}}_{\mathrm{j}}}{{{\mathbf{s}}_{\mathrm{j}}}^{\mathrm{H}}{\mathbf{s}}_{\mathrm{j}}}\bigg)\ }\mathrm{\approx }\mathrm{-}\mathrm{\varrho }\frac{{{\mathbf{s}}_{\mathrm{j}}}^{\mathrm{H}}\frac{\boldsymbol{y}_{\mathrm{jk}}\boldsymbol{y}^{\mathrm{H}}_{\mathrm{jk}}}{{\lVert \boldsymbol{y}_{\mathrm{jk}}\rVert}^{\mathrm{2}}+{\lVert \boldsymbol{x}_{\mathrm{jk}}\rVert}^{\mathrm{2}}}{\mathbf{s}}_{\mathrm{j}}}{{{\mathbf{s}}_{\mathrm{j}}}^{\mathrm{H}}{\mathbf{s}}_{\mathrm{j}}}\end{equation}
	where $\mathrm{\varrho }\mathrm{>0}$.
	The second term of the LRT statistic can be represented by (\ref{eq3a6xv}), given at the top of the next page.
	\begin{figure*}[!]
		\normalsize
		\setcounter{tempequationcounter}{\value{equation}}
		\begin{IEEEeqnarray}{rCl}
			\setcounter{equation}{44}
			\label{eq3a6xv}
			\resizebox{.93\hsize}{!}{${\mathrm{\Lambda }}_{\mathrm{ST}}\left(\boldsymbol{\mathrm{z}}\right)\mathrm{=}-\mathrm{2}\mathrm{L}{\mathrm{N}}_{\mathrm{t}}{\mathrm{N}}_{\mathrm{r}}{\mathrm{ln} \mathrm{\pi }}+\mathop{\mathrm{sup}}_{{\left\{{\mathbf{s}}_{\mathrm{j}}\right\}}^{{\mathrm{N}}_{\mathrm{t}}}_{\mathrm{j=1}}\mathrm{\ }\mathrm{\ }}\Bigg\{\mathop{\mathrm{sup}}_{{\left\{{\mathrm{\sigma }}^{\mathrm{2}}_{\mathrm{jk}}\right\}}^{\left({\mathrm{N}}_{\mathrm{r}},{\mathrm{N}}_{\mathrm{t}}\right)\ }_{\left(\mathrm{j,k}\right)\mathrm{=(1,1)}}}\bigg\{\mathop{\mathrm{sup}}_{{\left\{{\mathrm{\beta }}_{\mathrm{jk}}\right\}}^{\left({\mathrm{N}}_{\mathrm{r}},{\mathrm{N}}_{\mathrm{t}}\right)\ }_{\left(\mathrm{j,k}\right)\mathrm{=(1,1)}},\ {\left\{{\mathrm{\alpha }}_{\mathrm{jk}}\right\}}^{\left({\mathrm{N}}_{\mathrm{r}},{\mathrm{N}}_{\mathrm{t}}\right)\ }_{\left(\mathrm{j,k}\right)\mathrm{=(1,1)}}}\mathrm{-2L}\ \sum^{{\mathrm{N}}_{\mathrm{t}}}_{\mathrm{j=1}}{\sum^{{\mathrm{N}}_{\mathrm{r}}}_{\mathrm{k=1}}\Big\{{{\mathrm{ln} {\mathrm{\sigma }}^{\mathrm{2}}_{\mathrm{jk}}\ }}\mathrm{-}{{\frac{{\lVert \boldsymbol{x}_{\mathrm{jk}}\mathrm{-}{\mathrm{\alpha }}_{\mathrm{jk}}{\mathbf{s}}_{\mathrm{j}}\rVert}^{\mathrm{2}}}{{\mathrm{\sigma }}^{\mathrm{2}}_{\mathrm{jk}}}}}\mathrm{-}{{\frac{{\lVert \boldsymbol{y}_{\mathrm{jk}}\mathrm{-}{\mathrm{\beta }}_{\mathrm{jk}}{\mathbf{s}}_{\mathrm{j}}\rVert}^{\mathrm{2}}}{{\mathrm{\sigma }}^{\mathrm{2}}_{\mathrm{jk}}}}}\Big\}}\bigg\}\Bigg\}$}
		\end{IEEEeqnarray}
		\hrulefill
	\end{figure*}
	The supremum of the inner optimization problem of (\ref{eq3a6xv}) with respect to ${{\mathrm{\beta }}_{\mathrm{jk}}}$ and ${{\mathrm{\alpha }}_{\mathrm{jk}}}$ can be obtained for (\ref{eq3a5}) and  
	\begin{equation}\label{eq3a6y}
		{{\mathrm{\alpha }}_{\mathrm{jk}}}\mathrm{=}\frac{{{\mathbf{s}}_{\mathrm{j}}}^{\mathrm{H}}\boldsymbol{x}_{\mathrm{jk}}}{{\lVert {\mathbf{s}}_{\mathrm{j}}\rVert}^{\mathrm{2}}}
	\end{equation}
Substituting (\ref{eq3a5}) and (\ref{eq3a6y}) into (\ref{eq3a6xv}) produces (\ref{eq3a6zq}), as shown at the top of the next page.
\begin{figure*}[!]
		\normalsize
		\setcounter{tempequationcounter}{\value{equation}}
		\begin{IEEEeqnarray}{rCl}
			\setcounter{equation}{46}
			\label{eq3a6zq}
			\resizebox{.93\hsize}{!}{$	{\mathrm{\Lambda }}_{\mathrm{ST}}\left(\boldsymbol{\mathrm{z}}\right)=\mathop{\mathrm{sup}}_{{\left\{{\mathbf{s}}_{\mathrm{j}}\right\}}^{{\mathrm{N}}_{\mathrm{t}}}_{\mathrm{j=1}}}\Bigg\{\mathop{\mathrm{sup}}_{{\left\{{\mathrm{\sigma }}^{\mathrm{2}}_{\mathrm{jk}}\right\}}^{\left({\mathrm{N}}_{\mathrm{r}},{\mathrm{N}}_{\mathrm{t}}\right)\ }_{\left(\mathrm{j,k}\right)\mathrm{=(1,1)}}\mathrm{\ }\mathrm{\ }}\bigg\{-\mathrm{2}\mathrm{L}{\mathrm{N}}_{\mathrm{t}}{\mathrm{N}}_{\mathrm{r}}{\mathrm{ln} \mathrm{\pi }\mathrm{-2L}\ }\sum^{{\mathrm{N}}_{\mathrm{t}}}_{\mathrm{j=1}}{\sum^{{\mathrm{N}}_{\mathrm{r}}}_{\mathrm{k=1}}{{\mathrm{ln} {\mathrm{\sigma }}^{\mathrm{2}}_{\mathrm{jk}}\ }}}-\sum^{{\mathrm{N}}_{\mathrm{t}}}_{\mathrm{j=1}}{\sum^{{\mathrm{N}}_{\mathrm{r}}}_{\mathrm{k=1}}{\frac{{\lVert \boldsymbol{y}_{\mathrm{jk}}\rVert}^{\mathrm{2}}\mathrm{-}\frac{{\left|{{\mathbf{s}}_{\mathrm{j}}}^{\mathrm{H}}\boldsymbol{y}_{\mathrm{jk}}\right|}^{\mathrm{2}}}{{\lVert {\mathbf{s}}_{\mathrm{j}}\rVert}^{\mathrm{2}}}\mathrm{+}{\lVert \boldsymbol{x}_{\mathrm{jk}}\rVert}^{\mathrm{2}}\mathrm{-}\frac{{\left|{{\mathbf{s}}_{\mathrm{j}}}^{\mathrm{H}}\boldsymbol{x}_{\mathrm{jk}}\right|}^{\mathrm{2}}}{{\lVert {\mathbf{s}}_{\mathrm{j}}\rVert}^{\mathrm{2}}}}{{\mathrm{\sigma }}^{\mathrm{2}}_{\mathrm{jk}}}}}\bigg\}\Bigg\}$} 
		\end{IEEEeqnarray}
		\hrulefill
	\end{figure*}
Again, the supremum of the inner optimization problem of (\ref{eq3a6zq}) with respect to ${\mathrm{\sigma }}^{\mathrm{2}}_{\mathrm{jk}}$ can be obtained for  
	\begin{equation}\label{eq3a6vv}
		{\mathrm{\sigma }}^{\mathrm{2}}_{\mathrm{jk}}\mathrm{=}\frac{{\lVert \boldsymbol{y}_{\mathrm{jk}}\rVert}^{\mathrm{2}}\mathrm{-}\frac{{\left|{{\mathbf{s}}_{\mathrm{j}}}^{\mathrm{H}}\boldsymbol{y}_{\mathrm{jk}}\right|}^{\mathrm{2}}}{{\lVert {\mathbf{s}}_{\mathrm{j}}\rVert}^{\mathrm{2}}}\mathrm{+}{\lVert \boldsymbol{x}_{\mathrm{jk}}\rVert}^{\mathrm{2}}\mathrm{-}\frac{{\left|{{\mathbf{s}}_{\mathrm{j}}}^{\mathrm{H}}\boldsymbol{x}_{\mathrm{jk}}\right|}^{\mathrm{2}}}{{\lVert {\mathbf{s}}_{\mathrm{j}}\rVert}^{\mathrm{2}}}}{\mathrm{2L}} 
	\end{equation}
	By using (\ref{eqx1})- (\ref{eqx3}), and a similar approximation of (\ref{eq3a6dd}), we can simplify the ST of the LRT statistic to obtain \footnote{It can be shown that there is no closed-form solution for the ST when the noise variances of the RC and SC are assumed to be different. This is discussed in Chapter 11 of \cite{mybook} for the calibrated SIMO setup, which can be easily extended to the uncalibrated MIMO configuration. }
	\begin{equation}\label{eq3a6s}
		\resizebox{.87\hsize}{!}{$	\begin{split}
				{\mathrm{\Lambda }}_{\mathrm{ST}}\left(\boldsymbol{\mathrm{z}}\right)&=-\mathrm{2}\mathrm{L}{\mathrm{N}}_{\mathrm{t}}{\mathrm{N}}_{\mathrm{r}}{\mathrm{ln} \mathrm{\pi }\ }-2\mathrm{L}\sum^{{\mathrm{N}}_{\mathrm{t}}}_{\mathrm{j=1}}{\sum^{{\mathrm{N}}_{\mathrm{r}}}_{\mathrm{k=1}}{{\mathrm{ln} \Big(\frac{{\lVert \boldsymbol{y}_{\mathrm{jk}}\rVert}^{\mathrm{2}}+{\lVert \boldsymbol{x}_{\mathrm{jk}}\rVert}^{\mathrm{2}}}{\mathrm{2L}}\Big)\ }}}\\&+\mathrm{2L}\mathrm{\varrho }\sum^{{\mathrm{N}}_{\mathrm{t}}}_{\mathrm{j=1}}{{\mathrm{\lambda }}_{\mathrm{max}}({\boldsymbol{\mathcal{Y}}}_{\mathrm{j}}{\boldsymbol{\mathrm{\Xi }}}^{-1}_{\mathrm{j}}{{\boldsymbol{\mathcal{Y}}}_{\mathrm{j}}}^\mathrm{H}\mathrm{+}{\boldsymbol{\mathrm{\chi }}}_{\mathrm{j}}{\boldsymbol{\mathrm{\Xi }}}^{-1}_{\mathrm{j}}{{\boldsymbol{\mathrm{\chi }}}_{\mathrm{j}}}^\mathrm{H})}\mathrm{-}\mathrm{2L}{\mathrm{N}}_{\mathrm{t}}\mathrm{N}_{\mathrm{r}}
			\end{split}$}
	\end{equation}
				The test statistic of the LRT can finally be obtained as (\ref{eq17x}).
					\section{Fisher Information Matrix Calculation}\label{B}
					In this appendix, we provide the technical derivations needed to obtain the Fisher information matrix (FIM), which is required for the Wald, Rao, and Durbin tests. The matrix $\boldsymbol{\mathcal{J}}(\boldsymbol{\mathrm{\theta}})$ in (\ref{eq15}), (\ref{eq17X}), and (\ref{eq18x}) represents the FIM, defined as
					\begin{equation}\label{eqb4}
						\boldsymbol{\mathcal{J}}\left( \boldsymbol{\mathrm{\theta}} \right)\mathrm{=E}\Bigg\{ \bigg[ 
						\frac{\mathrm{\partial \mathcal{L}}\left(\boldsymbol{\mathrm{\theta}}_{\rm r}, \boldsymbol{\mathrm{\theta}}_{\rm s} ; \mathbf{z}\right)\mathrm{\, 
						}}{\mathrm{\partial }\boldsymbol{\mathrm{\theta}}^{\mathrm{\mathbf{\ast }}}} 
						\bigg]\bigg[ \frac{\mathrm{\partial \mathcal{L}}\left(\boldsymbol{\mathrm{\theta}}_{\rm r}, \boldsymbol{\mathrm{\theta}}_{\rm s} ; \mathbf{z}\right)}{\mathrm{\partial }\boldsymbol{\mathrm{\theta}}^{\mathrm{\mathbf{\ast }}}} 
						\bigg]^{\mathrm{H}} \Bigg\}
					\end{equation}
					By representing the unknown parameters vector $\boldsymbol{\mathrm{\theta}}$ as $\boldsymbol{\mathrm{\theta}}=[\boldsymbol{\mathrm{\theta}}_{\rm r},\boldsymbol{\mathrm{\theta}}_{\rm s}]^{\rm T}$, we can express the FIM for our case more explicitly as
					\begin{equation}\label{eqb5}
						\begin{split}\boldsymbol{\mathcal{J}}(\boldsymbol{\mathrm{\theta}})&=\bigg[\begin{array}{cc}
								\boldsymbol{\mathcal{J}}_{\mathrm{rr}}(\boldsymbol{\mathrm{\theta}}) & \boldsymbol{\mathcal{J}}_{\mathrm{rs}}(\boldsymbol{\mathrm{\theta}})\\
								\boldsymbol{\mathcal{J}}_{\mathrm{sr}}(\boldsymbol{\mathrm{\theta}}) & \boldsymbol{\mathcal{J}}_{\mathrm{ss}}(\boldsymbol{\mathrm{\theta}})
							\end{array}\bigg]\end{split}
					\end{equation}
					
					Due to the statistical independence of the received data due to different transmitters, we can deduce that $\boldsymbol{\mathcal{J}}_{\mathrm{rr}}(\boldsymbol{\mathrm{\theta}})$, $\boldsymbol{\mathcal{J}}_{\mathrm{rs}}(\boldsymbol{\mathrm{\theta}})$, $\boldsymbol{\mathcal{J}}_{\mathrm{sr}}(\boldsymbol{\mathrm{\theta}})$, and $\boldsymbol{\mathcal{J}}_{\mathrm{ss}}(\boldsymbol{\mathrm{\theta}})$ possess block-diagonal structures. One therefore only needs to calculate their diagonal matrices, represented by $\boldsymbol{\mathcal{J}}_{\mathrm{rr}}^{(\rm j)}(\boldsymbol{\mathrm{\theta}})$, $\boldsymbol{\mathcal{J}}_{\mathrm{rs}}^{(\rm j)}(\boldsymbol{\mathrm{\theta}})$, $\boldsymbol{\mathcal{J}}_{\mathrm{sr}}^{(\rm j)}(\boldsymbol{\mathrm{\theta}})$ and $\boldsymbol{\mathcal{J}}_{\mathrm{ss}}^{(\rm j)}(\boldsymbol{\mathrm{\theta}})$ for $\rm j=1, ..., N_t$, respectively.   
					Besides, we have $\boldsymbol{\mathcal{J}}_{\mathrm{sr}}(\boldsymbol{\mathrm{\theta}})=\boldsymbol{\mathcal{J}}^{\mathrm{H}}_{\mathrm{rs}}(\boldsymbol{\mathrm{\theta}})$. In order to calculate  $\boldsymbol{\mathcal{J}}_{{\mathrm{rr}}}^{\rm{(j)}}\left({{\boldsymbol{\mathrm{\theta}}}}\right)$ in which $\boldsymbol{\mathrm{\theta}}_{\rm r}=\boldsymbol{\alpha}$, we have
					\begin{equation}\label{eqb8}
						\begin{split}
							\resizebox{.87\hsize}{!}{$		{\big[\boldsymbol{\mathcal{J}}_{\mathrm{rr}}^{\boldsymbol{\mathrm{(}}{\mathrm{j}}\boldsymbol{\mathrm{)}}}(\boldsymbol{\mathrm{\theta}})\big]}_{\mathrm{mn}}=\mathrm{E}\left\{ {\bigg[\frac{{\mathrm{\partial}}\mathcal{L}\left({\boldsymbol{\mathrm{\theta}}}_{\mathrm{r}},{\boldsymbol{\mathrm{\theta}}}_{\mathrm{s}}\mathrm{;}\boldsymbol{\mathrm{\ }}{\boldsymbol{\mathrm{z}}}_{\mathrm{j}}\right)}{\mathrm{\partial}{\boldsymbol{\mathrm{\boldsymbol{\alpha}}}}_{\mathrm{j}}^{\mathrm{*}}}\bigg]}_{\mathrm{m}}{{\bigg[\frac{{\mathrm{\partial}}\mathcal{L}\left({\boldsymbol{\mathrm{\theta}}}_{\mathrm{r}},{\boldsymbol{\mathrm{\theta}}}_{\mathrm{s}}\mathrm{;}\boldsymbol{\mathrm{\ }}{\boldsymbol{\mathrm{z}}}_{\mathrm{j}}\right)}{\mathrm{\partial}{\boldsymbol{\mathrm{\boldsymbol{\alpha}}}}_{\mathrm{j}}^{\mathrm{*}}}\bigg]}_{\mathrm{n}}^{\mathrm{H}}}_{\ }\right\}$ }
						\end{split}
					\end{equation}
					where 
					\begin{equation}\label{eqb9}
						\bigg[ \frac{\mathrm{\partial \mathcal{L}}\left( \mathrm{\mathbf{\theta 
							}}_{\mathrm{r}}\mathrm{\mathbf{,}}\mathrm{\mathbf{\theta 
							}}_{\mathrm{s}}\mathrm{\mathbf{;\, }}\mathrm{\mathbf{z}}_{\mathrm{j}} 
							\right)}{\mathrm{\partial }\mathrm{\mathbf{\boldsymbol{\alpha} 
							}}_{\mathrm{j}}^{\mathrm{\ast }}} \bigg]_{\mathrm{m}}\mathrm{=\, 
						}\frac{\mathrm{1}}{\mathrm{\sigma }^{\mathrm{2}}_{\mathrm{jm}}}( 
						\mathrm{\mathbf{s}}_{\mathrm{j}}^{\mathrm{H}}\mathrm{\boldsymbol{x}}_{\mathrm{jm}}\mathrm{-}\mathrm{\boldsymbol{\alpha} 
						}_{\mathrm{jm}}\mathrm{\mathbf{s}}_{\mathrm{j}}^{\mathrm{H}}\mathrm{\mathbf{s}}_{\mathrm{j}} 
						)
					\end{equation}
					Using $\mathrm{\boldsymbol{x}}_{\mathrm{jk}}\sim\mathcal{CN}({\mathrm{\boldsymbol{\alpha}}}_{\mathrm{jk}}\mathbf{s}_{\mathrm{j}},{\sigma}_{\mathrm{jk}}^{2}\mathbf{I}_{\rm L})$, after some algebra, we find that
					\begin{equation}\label{eqb10}
						\boldsymbol{\mathcal{J}}_{{\mathrm{rr}}}^{\mathrm{(j)}}\left({{\boldsymbol{\mathrm{\theta}}}}\right)={\left\Vert {{\mathbf{s}}}_{\boldsymbol{\mathrm{j}}}\right\Vert }^{\mathrm{2}}{\boldsymbol{\mathfrak{O}}}^2_{\rm j}
					\end{equation}
					where
					\begin{equation}\label{eqb10x}
						{\boldsymbol{\mathfrak{O}}}^2_{\rm j}\boldsymbol{=}\mathrm{diag}\bigg(\frac{\mathrm{1}}{{{\mathrm{\sigma }}^{\mathrm{2}}_{\mathrm{j1}}}_{\mathrm{\ }}}\mathrm{,\dots ,\ }\frac{\mathrm{1}}{{\mathrm{\sigma }}^{\mathrm{2}}_{{\mathrm{jN}}^{\mathrm{\ }}_{\mathrm{r}}}}\bigg)
					\end{equation}
					To compute $\boldsymbol{\mathcal{J}}_{\mathrm{ss}}(\boldsymbol{\mathrm{\theta}})$,
					we have
					\begin{equation}\label{eqb11}
						\begin{split}
							\resizebox{.87\hsize}{!}{${\big[\mathbf{\boldsymbol{\mathcal{J}}}_{\rm ss}(\boldsymbol{\mathrm{\theta}})\big]}_{\mathrm{mn}}=\mathrm{E}\left\{ {\bigg[\frac{{\mathrm{\partial}}\mathcal{L}\left({\boldsymbol{\mathrm{\theta}}}_{\mathrm{r}},{\boldsymbol{\mathrm{\theta}}}_{\mathrm{s}}\mathrm{;}\boldsymbol{\mathrm{\ }}{\boldsymbol{\mathrm{z}}}_{\mathrm{j}}\right)}{\mathrm{\partial}{\boldsymbol{\mathrm{\theta}}}_{\mathbf{s}}^{\boldsymbol{*}}}\bigg]}_{\mathrm{m}}{{\bigg[\frac{{\mathrm{\partial}}\mathcal{L}\left({\boldsymbol{\mathrm{\theta}}}_{\mathrm{r}},{\boldsymbol{\mathrm{\theta}}}_{\mathrm{s}}\mathrm{;}\boldsymbol{\mathrm{\ }}{\boldsymbol{\mathrm{z}}}_{\mathrm{j}}\right)}{\mathrm{\partial}{\boldsymbol{\mathrm{\theta}}}_{\mathbf{s}}^{\boldsymbol{*}}}\bigg]}_{\mathrm{n}}^{\mathrm{H}}}_{\ }\right\}$ }
						\end{split}
					\end{equation}
					In our case, the nuisance vector parameter can be expressed as $\boldsymbol{\mathrm{\theta}}_{\rm s}=[\boldsymbol{\beta}^{\rm T},\mathbf{s}^{\rm T},\boldsymbol{\sigma}^{\rm T}]^{\rm T}$. Thus, the expression for $\boldsymbol{\mathcal{J}}_{\mathrm{ss}}(\boldsymbol{\mathrm{\theta}})$ can be further elaborated as follows
					\begin{equation}\label{eqb12}
						\begin{array}{lll}
							\boldsymbol{\mathcal{J}}_{\mathrm{ss}}(\boldsymbol{\mathrm{\theta}})& =\left[\begin{array}{ccc}
								\mathbf{\boldsymbol{\mathcal{I}}}_{{\boldsymbol{\mathrm{\beta}}}{\boldsymbol{\mathrm{\beta}}}}(\boldsymbol{\mathrm{\theta}}) & \mathbf{\boldsymbol{\mathcal{I}}}_{{\boldsymbol{\mathrm{\beta}}}{\mathbf{s}}}(\boldsymbol{\mathrm{\theta}}) & \mathbf{\boldsymbol{\mathcal{I}}}_{{\boldsymbol{\mathrm{\beta}}}{\sigma}}(\boldsymbol{\mathrm{\theta}})\\
								\mathbf{\boldsymbol{\mathcal{I}}}_{{\mathbf{s}}{\boldsymbol{\mathrm{\beta}}}}(\boldsymbol{\mathrm{\theta}}) & \mathbf{\boldsymbol{\mathcal{I}}}_{{\mathbf{s}}{\mathbf{s}}}(\boldsymbol{\mathrm{\theta}}) & \mathbf{\boldsymbol{\mathcal{I}}}_{{\mathbf{s}}{\sigma}}(\boldsymbol{\mathrm{\theta}})\\
								\mathbf{\boldsymbol{\mathcal{I}}}_{\sigma{\boldsymbol{\mathrm{\beta}}}}(\boldsymbol{\mathrm{\theta}}) & \mathbf{\boldsymbol{\mathcal{I}}}_{\sigma{\mathbf{s}}}(\boldsymbol{\mathrm{\theta}}) & \mathbf{\boldsymbol{\mathcal{I}}}_{{\sigma}{\sigma}}(\boldsymbol{\mathrm{\theta}})
							\end{array}\right]
						\end{array}
					\end{equation}
					Further developments require calculating the blocks of $\boldsymbol{\mathcal{J}}_{\mathrm{ss}}(\boldsymbol{\mathrm{\theta}})$. Similarly, submatrices of $\boldsymbol{\mathcal{J}}_{\mathrm{ss}}(\boldsymbol{\mathrm{\theta}})$ are block-diagonal matrices with diagonal matrices of $\mathbf{\boldsymbol{\mathcal{I}}}_{{\boldsymbol{\mathrm{\beta}}}{\boldsymbol{\mathrm{\beta}}}_{\ }^{\mathrm{\ }}}^{(\rm j)}(\boldsymbol{\mathrm{\theta}})$, $\mathbf{\boldsymbol{\mathcal{I}}}_{{\boldsymbol{\mathrm{\beta}}}{\mathbf{s}}}^{(\rm j)}(\boldsymbol{\mathrm{\theta}})$, $\mathbf{\boldsymbol{\mathcal{I}}}_{{\boldsymbol{\mathrm{\beta}}}{\boldsymbol{\sigma}}}^{(\rm j)}(\boldsymbol{\mathrm{\theta}})$, $\mathbf{\boldsymbol{\mathcal{I}}}_{{\mathbf{s}}{\mathbf{s}}_{\ }}^{(\rm j)}(\boldsymbol{\mathrm{\theta}})$, $\mathbf{\boldsymbol{\mathcal{I}}}_{{\mathbf{s}}{\boldsymbol{\sigma}}_{\ }}^{(\rm j)}(\boldsymbol{\mathrm{\theta}})$, and $\mathbf{\boldsymbol{\mathcal{I}}}_{{\boldsymbol{\sigma}}{\boldsymbol{\sigma}}}^{(\rm j)}(\boldsymbol{\mathrm{\theta}})$ for $\rm j=1, ..., N_t$.  
					To compute $\mathbf{\boldsymbol{\mathcal{I}}}_{{\boldsymbol{\mathrm{\beta}}}{\boldsymbol{\mathrm{\beta}}}_{\mathrm{\ }}^{\mathrm{\ }}}^{\mathrm{(j)}}(\boldsymbol{\mathrm{\theta}})$,
					we have
					\begin{equation}\label{eqb14}
						\begin{split}
							\resizebox{.87\hsize}{!}{$	{\big[\mathbf{\boldsymbol{\mathcal{I}}}_{{\boldsymbol{\mathrm{\beta}}}{\boldsymbol{\mathrm{\beta}}}_{\mathrm{\ }}^{\mathrm{\ }}}^{\mathrm{(j)}}(\boldsymbol{\mathrm{\theta}})\big]}_{\mathrm{mn}}=\mathrm{E}\left\{ {\bigg[\frac{{\mathrm{\partial}}\mathcal{L}\left({\boldsymbol{\mathrm{\theta}}}_{\mathrm{r}},{\boldsymbol{\mathrm{\theta}}}_{\mathrm{s}}\mathrm{;}\boldsymbol{\mathrm{\ }}{\boldsymbol{\mathrm{z}}}_{\mathrm{j}}\right)}{\mathrm{\partial}{\boldsymbol{\mathrm{\beta}}}_{j}^{\mathrm{*}}}\bigg]}_{\mathrm{m}}{{\bigg[\frac{{\mathrm{\partial}}\mathcal{L}\left({\boldsymbol{\mathrm{\theta}}}_{\mathrm{r}},{\boldsymbol{\mathrm{\theta}}}_{\mathrm{s}}\mathrm{;}\boldsymbol{\mathrm{\ }}{\boldsymbol{\mathrm{z}}}_{\mathrm{j}}\right)}{\mathrm{\partial}{\boldsymbol{\mathrm{\beta}}}_{j}^{\mathrm{*}}}\bigg]}_{\mathrm{n}}^{\mathrm{H}}}_{\ }\right\}$} \end{split}
					\end{equation}
					where
					\begin{equation}\label{eqb15}
						\bigg[ \frac{\mathrm{\partial \mathcal{L}}\left( \mathrm{\mathbf{\theta 
							}}_{\mathrm{r}}\mathrm{\mathbf{,}}\mathrm{\mathbf{\theta 
							}}_{\mathrm{s}}\mathrm{\mathbf{;\, }}\mathrm{\mathbf{z}}_{\mathrm{j}} 
							\right)}{\mathrm{\partial }\mathrm{\mathbf{\beta
							}}_{\mathrm{j}}^{\mathrm{\ast }}} \bigg]_{\mathrm{m}}\mathrm{=\, 
						}\frac{\mathrm{1}}{\mathrm{\sigma }^{\mathrm{2}}_{\mathrm{jm}}}( 
						\mathrm{\mathbf{s}}_{\mathrm{j}}^{\mathrm{H}}\mathrm{\boldsymbol{y}}_{\mathrm{jm}}\mathrm{-}\mathrm{\beta 
						}_{\mathrm{jm}}\mathrm{\mathbf{s}}_{\mathrm{j}}^{\mathrm{H}}\mathrm{\mathbf{s}}_{\mathrm{j}} 
						)
					\end{equation}
					Using $\mathrm{\boldsymbol{y}}_{\mathrm{jk}}\sim\mathcal{CN}({\mathrm{\beta}}_{\mathrm{jk}}\mathbf{s}_{\mathrm{j}},{\sigma}_{\mathrm{jk}}^{2}\mathbf{I}_{\mathrm{L}})$, after some algebra, yields
					\begin{equation}\label{eqb16}
						\mathbf{\boldsymbol{\mathcal{I}}}_{{\boldsymbol{\mathrm{\beta}}}{\boldsymbol{\mathrm{\beta}}}_{\ }^{\mathrm{\ }}}^{(\rm j)}(\boldsymbol{\mathrm{\theta}})={\left\Vert {{\mathbf{s}}}_{{\mathrm{j}}}\right\Vert }^{\mathrm{2}}{\boldsymbol{\mathfrak{O}}}^2_{\rm j}
					\end{equation}
					The m-th row of matrix $\mathbf{\boldsymbol{\mathcal{I}}}_{{\boldsymbol{\mathrm{\beta}}}{\mathbf{s}}}^{(\rm j)}(\boldsymbol{\mathrm{\theta}})$ can be represented as
					\begin{equation}\label{eqb18}
						\resizebox{.87\hsize}{!}{$
							\begin{split}
								{\big[\mathbf{\boldsymbol{\mathcal{I}}}_{{\boldsymbol{\mathrm{\beta}}}{\mathbf{s}}}^{\mathrm{(j)}}(\boldsymbol{\mathrm{\theta}})\big]}_{\mathrm{m*}}=\mathrm{E}\Bigg\{ {\bigg[\frac{{\mathrm{\partial}}\mathcal{L}\left({\boldsymbol{\mathrm{\theta}}}_{\mathrm{r}},{\boldsymbol{\mathrm{\theta}}}_{\mathrm{s}}\mathrm{;}\boldsymbol{\mathrm{\ }}{\boldsymbol{\mathrm{z}}}_{\mathrm{j}}\right)}{\mathrm{\partial}{\boldsymbol{\mathrm{\beta}}}_{j}^{\mathrm{*}}}\bigg]}_{\mathrm{m}}{{\bigg[{\bigg[\frac{{\mathrm{\partial}}\mathcal{L}\left({\boldsymbol{\mathrm{\theta}}}_{\mathrm{r}},{\boldsymbol{\mathrm{\theta}}}_{\mathrm{s}}\mathrm{;}\boldsymbol{\mathrm{\ }}{\boldsymbol{\mathrm{z}}}_{\mathrm{j}}\right)}{\mathrm{\partial}{\mathbf{s}}_{\mathrm{j}}}\bigg]}^{\mathrm{T}}\bigg]}_{*}}_{\ }\Bigg\} 
							\end{split}$}
					\end{equation}
					where
					\begin{equation}\label{eqb19}
						\resizebox{.87\hsize}{!}{$	\bigg[ \frac{\mathrm{\partial }\mathcal{L}\left( 
								\mathrm{\mathbf{\theta 
								}}_{\mathrm{r}}\mathrm{\mathbf{,}}\mathrm{\mathbf{\theta 
								}}_{\mathrm{s}}\mathrm{\mathbf{;\, }}\mathrm{\mathbf{z}}_{\mathrm{j}} 
								\right)}{\mathrm{\partial }\mathrm{\mathbf{s}}_{\mathrm{j}}^{\mathrm{\ast 
							}}} 
							\bigg]^{\mathrm{H}}\mathrm{=}\sum\limits_{\mathrm{k=1}}^\mathrm{N_r} 
							\frac{\mathrm{\boldsymbol{\alpha} 
								}_{\mathrm{jk}}\mathrm{\boldsymbol{x}}_{\mathrm{jk}}^{\mathrm{H}}\mathrm{+}\mathrm{\beta 
								}_{\mathrm{jk}}\mathrm{\boldsymbol{y}}_{\mathrm{jk}}^{\mathrm{H}}\mathrm{-}\left| \mathrm{\boldsymbol{\alpha} 
								}_{\mathrm{jk}} 
								\right|^{\mathrm{2}}\mathrm{\mathbf{s}}_{\mathrm{j}}^{\mathrm{H}}\mathrm{-}\left| 
								\mathrm{\beta }_{\mathrm{jk}} 
								\right|^{\mathrm{2}}\mathrm{\mathbf{s}}_{\mathrm{j}}^{\mathrm{H}}}{\mathrm{\sigma 
								}^{\mathrm{2}}_{\mathrm jk}}$}
					\end{equation}
					Substituting (\ref{eqb15}) and (\ref{eqb19}) into (\ref{eqb18}), after some algebra, produces $	\mathbf{\boldsymbol{\mathcal{I}}}_{{\boldsymbol{\mathrm{\beta}}}{\mathbf{s}}}^{\mathrm{(j)}}(\boldsymbol{\mathrm{\theta}})={\boldsymbol{\mathfrak{O}}}^2_{\rm j}{{\boldsymbol{\mathrm{\beta}}}}_{\mathrm{j}}{{\mathbf{s}}}_{\boldsymbol{\mathrm{j}}}^{{\mathrm{H}}}$. 
					To calculate $\mathbf{\boldsymbol{\mathcal{I}}}_{{\boldsymbol{\mathrm{\beta}}}{\boldsymbol{\sigma}}}^{(\rm j)}(\boldsymbol{\mathrm{\theta}})$, we have
					\begin{equation}\label{eqb14x}
						\begin{split}
							\resizebox{.87\hsize}{!}{$	{\big[\mathbf{\boldsymbol{\mathcal{I}}}_{{\boldsymbol{\mathrm{\beta}}}{\boldsymbol{\sigma}}_{\mathrm{\ }}^{\mathrm{\ }}}^{\mathrm{(j)}}(\boldsymbol{\mathrm{\theta}})\big]}_{\mathrm{mn}}=\mathrm{E}\left\{ {\bigg[\frac{{\mathrm{\partial}}\mathcal{L}\left({\boldsymbol{\mathrm{\theta}}}_{\mathrm{r}},{\boldsymbol{\mathrm{\theta}}}_{\mathrm{s}}\mathrm{;}\boldsymbol{\mathrm{\ }}{\boldsymbol{\mathrm{z}}}_{\mathrm{j}}\right)}{\mathrm{\partial}{\boldsymbol{\mathrm{\beta}}}_{j}^{\mathrm{*}}}\bigg]}_{\mathrm{m}}{{\bigg[\frac{{\mathrm{\partial}}\mathcal{L}\left({\boldsymbol{\mathrm{\theta}}}_{\mathrm{r}},{\boldsymbol{\mathrm{\theta}}}_{\mathrm{s}}\mathrm{;}\boldsymbol{\mathrm{\ }}{\boldsymbol{\mathrm{z}}}_{\mathrm{j}}\right)}{\mathrm{\partial}{\boldsymbol{\sigma}_{\rm j}}}\bigg]}_{\mathrm{n}}^{\mathrm{H}}}_{\ }\right\}$} \end{split}
					\end{equation}
					where
					\begin{equation}\label{eqb14w}
						\resizebox{.87\hsize}{!}{${{\bigg[\frac{{\mathrm{\partial }}\mathcal{L}\left({\boldsymbol{\mathrm{\theta }}}_{\mathrm{r}},{\boldsymbol{\mathrm{\theta }}}_{\mathrm{s}}\mathrm{;}\boldsymbol{\mathrm{\ }}{\boldsymbol{\mathrm{z}}}_{\mathrm{j}}\right)}{\mathrm{\partial }{\boldsymbol{\sigma }}_{\mathrm{j}}}\bigg]}}_{\mathrm{n}}\mathrm{=}{\mathrm{-}}\frac{\mathrm{2L}}{{{\mathrm{\sigma }}^{\mathrm{2}}_{\mathrm{jn}}}^{\mathrm{\ }}}\mathrm{+}\frac{{\left\|{\boldsymbol{{x}}}_{\mathrm{jn}}\mathrm{-}{\mathrm{\alpha }}_{\mathrm{jn}}{\mathbf{s}}_{\mathrm{j}}\right\|}^{\mathrm{2}}}{{\mathrm{\sigma }}^{\mathrm{4}}_{\mathrm{jn}}}\mathrm{+}\frac{{\left\|{\boldsymbol{{y}}}_{\mathrm{jn}}\mathrm{-}{\mathrm{\beta }}_{\mathrm{jn}}{\mathbf{s}}_{\mathrm{j}}\right\|}^{\mathrm{2}}}{{\mathrm{\sigma }}^{\mathrm{4}}_{\mathrm{jn}}}$}
					\end{equation}
					Substituting (\ref{eqb15}) and (\ref{eqb14w}) into (\ref{eqb14x}), after some algebra,
					produces 
					\begin{equation}\label{eqb14f}
						\resizebox{.87\hsize}{!}{$	{\big[{\boldsymbol{\mathcal{I}}}^{\mathrm{(j)}}_{{\boldsymbol{\mathrm{\beta }}}\boldsymbol{\mathrm{\sigma }}}\left(\boldsymbol{\mathrm{\theta }}\right)\big]}_{\mathrm{mn}}\mathrm{=}\frac{\mathrm{{\mathbf{s}}^{\mathrm{H}}_{\mathrm{jm}}}}{{\mathrm{\sigma }}^{\mathrm{2}}_{\mathrm{jm}}}\mathrm{E}\left\{\big({\boldsymbol{{y}}}_{\mathrm{jm}}\mathrm{-}{\mathrm{\beta }}_{\mathrm{jm}}{\mathbf{s}}_{\mathrm{jm}}\big)\frac{{\left\|{\boldsymbol{{y}}}_{\mathrm{jn}}\mathrm{-}{\mathrm{\beta }}_{\mathrm{jn}}{\mathbf{s}}_{\mathrm{j}}\right\|}^{\mathrm{2}}}{{{\mathrm{\sigma }}^{\mathrm{4}}_{\mathrm{jn}}}}\right\}$}
					\end{equation}
					Since ${\boldsymbol{{y}}}_{\mathrm{jk}}\mathrm{-}{\mathrm{\beta }}_{\mathrm{jk}}\mathbf{s}_{\mathrm{j}}\mathrm{\sim}\mathcal{CN}(\boldsymbol{\mathrm{0}}\mathrm{,\ }{\mathrm{\sigma}}^{\mathrm{2}}_{\mathrm{jk}}\boldsymbol{\mathrm{I}}_{\rm L})$, one can obtain that ${\boldsymbol{\mathcal{I}}}^{\mathrm{(j)}}_{{\boldsymbol{\mathrm{\beta }}}_{}\boldsymbol{\mathrm{\sigma }}}\left(\boldsymbol{\mathrm{\theta }}\right)=\boldsymbol{0}_{\rm{N_r \times N_r}}$.
					The matrix $\mathbf{\boldsymbol{\mathcal{I}}}_{{\mathbf{s}}{\mathbf{s}}_{\mathrm{\ }}}^{\mathrm{(j)}}(\boldsymbol{\mathrm{\theta}})$
					can be computed as
					\begin{equation}\label{eqb22}
						\mathbf{\boldsymbol{\mathcal{I}}}_{{\mathbf{s}}{\mathbf{s}}_{\mathrm{\ }}}^{\mathrm{(j)}}(\boldsymbol{\mathrm{\theta}})=\mathrm{E}\Bigg\{ \frac{{\mathrm{\partial}}\mathcal{L}\left({\boldsymbol{\mathrm{\theta}}}_{\mathrm{r}},{\boldsymbol{\mathrm{\theta}}}_{\mathrm{s}}\mathrm{;}\boldsymbol{\mathrm{\ }}{\boldsymbol{\mathrm{z}}}_{\mathrm{j}}\right)}{\mathrm{\partial}{{\mathbf{s}}_{\mathrm{j}}}^{\mathrm{*}}}{\bigg[\frac{{\mathrm{\partial}}\mathcal{L}\left({\boldsymbol{\mathrm{\theta}}}_{\mathrm{r}},{\boldsymbol{\mathrm{\theta}}}_{\mathrm{s}}\mathrm{;}\boldsymbol{\mathrm{\ }}{\boldsymbol{\mathrm{z}}}_{\mathrm{j}}\right)}{\mathrm{\partial}{{\mathbf{s}}_{\mathrm{j}}}^{\mathrm{*}}}\bigg]}^{\mathrm{T}}\Bigg\} 
					\end{equation}
					Using  (\ref{eqb19}), followed by some algebra, yields 
					\begin{equation}\label{eqb23}
						\boldsymbol{\mathcal{I}}_{\mathrm{\mathbf{s}}_{\mathrm{\, 
							}}\mathrm{\mathbf{s}}_{\mathrm{\, }}}^{\mathrm{(j)}}\left( 
						\mathrm{\boldsymbol{\theta}} \right)\mathrm{=}\mathrm{\boldsymbol{\beta}}_{\mathrm{j}}^{\mathrm{H}}{\boldsymbol{\mathfrak{O}}}^2_{\rm j}\mathrm{\, 
						}\mathrm{\boldsymbol{\beta}}_{\mathrm{j}}^{\mathrm{\, 
						}}\mathrm{\mathbf{I}}_{\mathrm{{L}}}\mathrm{+}\mathrm{\mathbf{\boldsymbol{\alpha} }}_{\mathrm{j}}^{\mathrm{H}}{\boldsymbol{\mathfrak{O}}}^2_{\rm j}\mathrm{\, 
						}\mathrm{\mathbf{\boldsymbol{\alpha} }}_{\mathrm{j}}^{\mathrm{\, 
						}}\mathrm{\mathbf{I}}_{\mathrm{{L}}}
					\end{equation}
					In order to compute $\mathbf{\boldsymbol{\mathcal{I}}}_{{\mathbf{s}}{\boldsymbol{\sigma}}}(\boldsymbol{\mathrm{\theta}})$, we can write
					\begin{equation}\label{eqb12we}
						\begin{split}
							\resizebox{.87\hsize}{!}{$		{\big[\mathbf{\boldsymbol{\mathcal{I}}}_{{\mathbf{s}}{\boldsymbol{\sigma}}_{\mathrm{\ }}^{\mathrm{\ }}}^{\mathrm{(j)}}(\boldsymbol{\mathrm{\theta}})\big]}_{\mathrm{mn}}=\mathrm{E}\left\{ {\bigg[\frac{{\mathrm{\partial}}\mathcal{L}\left({\boldsymbol{\mathrm{\theta}}}_{\mathrm{r}},{\boldsymbol{\mathrm{\theta}}}_{\mathrm{s}}\mathrm{;}\boldsymbol{\mathrm{\ }}{\boldsymbol{\mathrm{z}}}_{\mathrm{j}}\right)}{\mathrm{\partial}{\mathbf{s}}_{j}^{\mathrm{*}}}\bigg]}_{\mathrm{m}}{{\bigg[\frac{{\mathrm{\partial}}\mathcal{L}\left({\boldsymbol{\mathrm{\theta}}}_{\mathrm{r}},{\boldsymbol{\mathrm{\theta}}}_{\mathrm{s}}\mathrm{;}\boldsymbol{\mathrm{\ }}{\boldsymbol{\mathrm{z}}}_{\mathrm{j}}\right)}{\mathrm{\partial}{\boldsymbol{\sigma}_{\rm j}}}\bigg]}_{\mathrm{n}}^{\mathrm{H}}}_{\ }\right\}$} \end{split}
					\end{equation}
					Substituting (\ref{eqb19}) and (\ref{eqb14w}) into (\ref{eqb12we}), after some algebra, yields $		{\boldsymbol{\mathcal{I}}}^{\mathrm{(j)}}_{{\boldsymbol{\mathrm{s }}}\boldsymbol{\mathrm{\sigma }}}\left(\boldsymbol{\mathrm{\theta }}\right)=\boldsymbol{0}_{\rm{L \times N_r}}$.
					Finally, $\mathbf{\boldsymbol{\mathcal{I}}}_{{\boldsymbol{\sigma}}{\boldsymbol{\sigma}}}(\boldsymbol{\mathrm{\theta}})$ can be represented as
					\begin{equation}\label{eqb23a}
						\begin{split}
							\resizebox{.87\hsize}{!}{${\big[\mathbf{\boldsymbol{\mathcal{I}}}_{{\boldsymbol{\sigma}}{\boldsymbol{\sigma}}_{\mathrm{\ }}^{\mathrm{\ }}}^{\mathrm{(j)}}(\boldsymbol{\mathrm{\theta}})\big]}_{\mathrm{mn}}=\mathrm{E}\left\{ {\bigg[\frac{{\mathrm{\partial}}\mathcal{L}\left({\boldsymbol{\mathrm{\theta}}}_{\mathrm{r}},{\boldsymbol{\mathrm{\theta}}}_{\mathrm{s}}\mathrm{;}\boldsymbol{\mathrm{\ }}{\boldsymbol{\mathrm{z}}}_{\mathrm{j}}\right)}{\mathrm{\partial}{\boldsymbol{\sigma}}_{j}}\bigg]}_{\mathrm{m}}{{\bigg[\frac{{\mathrm{\partial}}\mathcal{L}\left({\boldsymbol{\mathrm{\theta}}}_{\mathrm{r}},{\boldsymbol{\mathrm{\theta}}}_{\mathrm{s}}\mathrm{;}\boldsymbol{\mathrm{\ }}{\boldsymbol{\mathrm{z}}}_{\mathrm{j}}\right)}{\mathrm{\partial}{\boldsymbol{\sigma}_{\rm j}}}\bigg]}_{\mathrm{n}}^{\mathrm{H}}}_{\ }\right\}$} \end{split}
					\end{equation}
					Substituting (\ref{eqb14w}) into (\ref{eqb23a}), after some algebra,
					yields
					\begin{equation}\label{eqb23c}
						\begin{split}
							{\big[{\boldsymbol{\mathcal{I}}}^{\mathrm{(j)}}_{\boldsymbol{\mathrm{\sigma }}\boldsymbol{\mathrm{\sigma }}}(\boldsymbol{\mathrm{\theta}})\big]}_{\mathrm{nm}}
							\mathrm{=}&\boldsymbol{\mathrm{-}}\frac{\mathrm{4L}}{{{\mathrm{\sigma }}^{\mathrm{2}}_{\mathrm{jm}}}^{\mathrm{\ }}}\frac{\mathrm{L}}{{{\mathrm{\sigma }}^{\mathrm{2}}_{\mathrm{jn}}}^{\mathrm{\ }}}\\&\mathrm{+}\frac{\mathrm{1}}{{({\mathrm{\sigma }}^{\mathrm{2}}_{\mathrm{jn}})}^{\mathrm{2}}{({\mathrm{\sigma }}^{\mathrm{2}}_{\mathrm{jm}})}^{\mathrm{2}}}\mathrm{E}\left\{{\left\|{\boldsymbol{{n}}}_{\mathrm{jn}}\right\|}^{\mathrm{2}}{\left\|{\boldsymbol{{n}}}_{\mathrm{jm}}\right\|}^{\mathrm{2}}\right\}\\&\mathrm{+}\frac{\mathrm{1}}{{({\mathrm{\sigma }}^{\mathrm{2}}_{\mathrm{jn}})}^{\mathrm{2}}{({\mathrm{\sigma }}^{\mathrm{2}}_{\mathrm{jm}})}^{\mathrm{2}}}\mathrm{E}\left\{{\left\|{\boldsymbol{{n}}}_{\mathrm{jn}}\right\|}^{\mathrm{2}}{\left\|{\boldsymbol{{e}}}_{\mathrm{jm}}\right\|}^{\mathrm{2}}\right\}\\&\mathrm{+}\frac{\mathrm{1}}{{({\mathrm{\sigma }}^{\mathrm{2}}_{\mathrm{jn}})}^{\mathrm{2}}{({\mathrm{\sigma }}^{\mathrm{2}}_{\mathrm{jm}})}^{\mathrm{2}}}\mathrm{E}\left\{{\left\|{\boldsymbol{{e}}}_{\mathrm{jn}}\right\|}^{\mathrm{2}}{\left\|{\boldsymbol{{n}}}_{\mathrm{jm}}\right\|}^{\mathrm{2}}\right\}\\&\mathrm{+}\frac{\mathrm{1}}{{({\mathrm{\sigma }}^{\mathrm{2}}_{\mathrm{jn}})}^{\mathrm{2}}{({\mathrm{\sigma }}^{\mathrm{2}}_{\mathrm{jm}})}^{\mathrm{2}}}\mathrm{E}\left\{{\left\|{\boldsymbol{{e}}}_{\mathrm{jn}}\right\|}^{\mathrm{2}}{\left\|{\boldsymbol{{e}}}_{\mathrm{jm}}\right\|}^{\mathrm{2}}\right\}	
						\end{split}
					\end{equation}
					Using $\mathrm{E}\left\{{\left\|{\boldsymbol{{n}}}_{\mathrm{jk}}\right\|}^{\mathrm{4}}\right\}\mathrm{=}{\mathrm{\sigma }}^{\mathrm{4}}_{\mathrm{jk}}{\mathrm{L}}^{\mathrm{2}}\mathrm{+2}{\mathrm{\sigma }}^{\mathrm{4}}_{\mathrm{jk}}\mathrm{L}$ and $\mathrm{E}\left\{{\left\|{\boldsymbol{{n}}}_{\mathrm{jn}}\right\|}^{\mathrm{2}}{\left\|{\boldsymbol{{n}}}_{\mathrm{jm}}\right\|}^{\mathrm{2}}\right\}\mathrm{=}{\mathrm{L}}^{\mathrm{2}}{\mathrm{\sigma }}^{\mathrm{2}}_{\mathrm{j}\mathrm{n}}\mathrm{\ }{\mathrm{\sigma }}^{\mathrm{2}}_{\mathrm{j}\mathrm{m}}$ for $n\ne m$, we can come up with
					\begin{equation}\label{eqb23b}
						{\boldsymbol{\mathrm{I}}}^{\mathrm{(j)}}_{\boldsymbol{\mathrm{\sigma }}\boldsymbol{\mathrm{\sigma }}}\left(\boldsymbol{\mathrm{\theta}} \right)\mathrm{=diag}\bigg(\frac{\mathrm{8L}}{{{\mathrm{\sigma }}^4_{\mathrm{j1}}}_{\mathrm{\ }}}\mathrm{\ ,\dots }\frac{\mathrm{8L}}{{{\mathrm{\sigma }}^4_{{\mathrm{jN}}_{\mathrm{r}}}}_{\mathrm{\ }}}\bigg)	
					\end{equation}
					Another block of the matrix $\boldsymbol{\mathcal{J}}(\boldsymbol{\theta})$, denoted by $\boldsymbol{\mathcal{J}}_{\mathrm{rs}}(\boldsymbol{\theta})$ in (\ref{eqb5}), can be expressed more explicitly as
					\begin{equation}\label{eqb24}
						\boldsymbol{\mathcal{J}}_{\mathrm{rs}}(\boldsymbol{\mathrm{\theta}})=\big[\begin{array}{ccc}
							\mathbf{\boldsymbol{\mathcal{I}}}_{{\boldsymbol{\mathrm{\boldsymbol{\alpha}}}}{\boldsymbol{\mathrm{\beta}}}}(\boldsymbol{\mathrm{\theta}}) & \mathbf{\boldsymbol{\mathcal{I}}}_{{\boldsymbol{\mathrm{\boldsymbol{\alpha}}}}{\mathbf{s}}}(\boldsymbol{\mathrm{\theta}}) & \mathbf{\boldsymbol{\mathcal{I}}}_{{\boldsymbol{\mathrm{\boldsymbol{\alpha}}}}\sigma}(\boldsymbol{\mathrm{\theta}})\end{array}\big]
					\end{equation}
					Similarly $\mathbf{\boldsymbol{\mathcal{I}}}_{{\boldsymbol{\mathrm{\boldsymbol{\alpha}}}}{\boldsymbol{\mathrm{\beta}}}}(\boldsymbol{\mathrm{\theta}})$, $\mathbf{\boldsymbol{\mathcal{I}}}_{{\boldsymbol{\mathrm{\boldsymbol{\alpha}}}}{\mathbf{s}}}(\boldsymbol{\mathrm{\theta}})$ and $\mathbf{\boldsymbol{\mathcal{I}}}_{{\boldsymbol{\mathrm{\boldsymbol{\alpha}}}}\sigma}(\boldsymbol{\mathrm{\theta}})$ are all block-diagonal matrices with diagonal matrices $\mathbf{\boldsymbol{\mathcal{I}}}_{{\boldsymbol{\mathrm{\boldsymbol{\alpha}}}}{\boldsymbol{\mathrm{\beta}}}}^{(\rm j)}(\boldsymbol{\mathrm{\theta}})$, $\mathbf{\boldsymbol{\mathcal{I}}}_{{\boldsymbol{\mathrm{\boldsymbol{\alpha}}}}{\mathbf{s}}}^{(\rm j)}(\boldsymbol{\mathrm{\theta}})$, and $\mathbf{\boldsymbol{\mathcal{I}}}_{{\boldsymbol{\mathrm{\boldsymbol{\alpha}}}}{\boldsymbol{\sigma}}}^{(\rm j)}(\boldsymbol{\mathrm{\theta}})$ for $\rm j=1, ..., N_t$, respectively.
					Similar to our previous computations, it can be shown that $	\mathbf{\boldsymbol{\mathcal{I}}}_{{\boldsymbol{\mathrm{\boldsymbol{\alpha}}}}{\boldsymbol{\mathrm{\beta}}}}^{\left(\mathrm{j}\right)}\left({{\boldsymbol{\mathrm{\theta}}}}\right)=\mathbf{0}_{\rm{N_r \times N_r}}$, $	\mathbf{\boldsymbol{\mathcal{I}}}_{{\boldsymbol{\mathrm{\alpha}}}{\mathbf{s}}}^{\mathrm{(j)}}(\boldsymbol{\mathrm{\theta}})={\boldsymbol{\mathfrak{O}}}^2_{\rm j}{{\boldsymbol{\mathrm{\alpha}}}}_{\mathrm{j}}{{\mathbf{s}}}_{\boldsymbol{\mathrm{j}}}^{{\mathrm{H}}}$, and $	\mathbf{\boldsymbol{\mathcal{I}}}_{{\boldsymbol{\mathrm{\boldsymbol{\alpha}}}}{\boldsymbol{\sigma}}_{\ }}^{(\rm j)}(\boldsymbol{\mathrm{\theta}})=\mathbf{0}_{\rm{N_r \times N_r}}$. 
					\section{MLE Of Unknown Parameters}\label{C}
					In this Appendix, we will obtain the MLE of unknown parameters under both null and alternative hypotheses, where $\hat{\boldsymbol{\theta } }_{\mathbf{1}}$  and $ 
					\hat{\boldsymbol{\theta } }_{\mathbf{0}}$ are respectively the MLEs of the unknown parameters under $\mathcal{H}_{1}$ and $\mathcal{H}_{0}$. In our case,  $\hat{\boldsymbol{\mathrm{\theta}}}_\mathrm{i}=[\hat{\boldsymbol{\mathrm{\theta}}}_{\mathrm{ri}},\hat{\boldsymbol{\mathrm{\theta}}}_\mathrm{s,i}]^{\rm T}$ with $\hat{\boldsymbol{\mathrm{\theta}}}_{\rm r, 0}={\boldsymbol{\mathrm{\theta}}}_{\rm r, 0}=\boldsymbol{0}$,  $\hat{\boldsymbol{\mathrm{\theta}}}_{\rm r,1}=\hat{\boldsymbol{\boldsymbol{\alpha}}}$ and $\hat{\boldsymbol{\mathrm{\theta}}}_{\mathrm{s, i}}=[\hat{\boldsymbol{\beta}}^{\rm T},\hat{\mathbf{s}}_\mathrm{i}^{\rm T},\widehat{\boldsymbol{\sigma}}_\mathrm{i}^{\rm T}]^{\rm T}$ with  $\hat{{\boldsymbol{\beta}}}=[\hat{\boldsymbol{\beta}}_{11}^{\rm T},...,\hat{\boldsymbol{\beta}}_{\mathrm{{\mathrm{N_{t}}}1}}^{\rm T}]^{\rm T}$, $\hat{\mathrm{{\mathbf{s}}}}_{\rm i}=[\hat{\mathbf{s}}_{1,{\rm i}}^{\rm T},...,\hat{\mathbf{s}}_{\mathrm{\rm {\mathrm{N_{t}}},{\rm i}}}^{\rm T}]^{\rm T}$, and $\widehat{\mathrm{{\boldsymbol{\sigma}}}}_i=[\widehat{{\boldsymbol{\sigma}}}^{\rm T}_{1,i},...,\widehat{{\boldsymbol{\sigma}}}^{\rm T}_{\rm N_t,i}]^{\rm T}$ with $\widehat{\boldsymbol{\sigma}}_{\rm j,i}=[\widehat{{\sigma}^2_{\rm j1,i}},...,\widehat{{\sigma}^2_{\rm jN_r,i}}]^{\rm T}$ for $\rm i=0$ (i.e., $\mathcal{H}_{0}$), 1 (i.e., $\mathcal{H}_{1}$). According to  (\ref{eq3a5}), (\ref{eq3a6z}), (\ref{eq3a6h}), (\ref{eq3a6y}), (\ref{eq3a6vv}) and (\ref{eq3a6s}), we can obtain
					\begin{equation}\label{eqb3a}  
						\mathcal{H}_{0}:	\left\{ {\begin{array}{l}
								{\hat{\mathbf{s}}}_{\mathrm{j,0}}\mathrm{=}{\mathrm{e}}_{\mathrm{1}}({\boldsymbol{\mathcal{Y}}}_{\mathrm{j}}{\boldsymbol{\mathrm{\Xi }}}^{-1}_{\mathrm{j}}{{\boldsymbol{\mathcal{Y}}}_{\mathrm{j}}}^\mathrm{H})\\
								\widehat{{\mathrm{\sigma }}^{\mathrm{2}}_{\mathrm{j}\mathrm{k,0}}}=\frac{{\lVert  \boldsymbol{y}_{\mathrm{jk}}\rVert}^{\mathrm{2}}\mathrm{-}\frac{{\left|{{\hat{\mathbf{s}}}_{\mathrm{j,0}}}^{\mathrm{H}}\boldsymbol{y}_{\mathrm{jk}}\right|}^{\mathrm{2}}}{{\hat{\mathbf{s}}}^{\mathrm{H}}_{\mathrm{j,0}}{\hat{\mathbf{s}}}_{\mathrm{j,0}}}\mathrm{+}{\lVert  \boldsymbol{x}_{\mathrm{jk}}\rVert}^{\mathrm{2}}}{\mathrm{2L}}
						\end{array}} \right.
					\end{equation} 
					\begin{equation}\label{eqb3b}  
						\mathcal{H}_{1}:	\left\{ {\begin{array}{l}
								{\hat{\boldsymbol{\mathrm{\alpha }}}}^{\mathrm{T}}_{\mathrm{j1}}\mathrm{=}\frac{{\hat{\mathbf{s}}}^{\mathrm{H}}_{\mathrm{j,1}}}{{\hat{\mathbf{s}}}^{\mathrm{H}}_{\mathrm{j,1}}{\hat{\mathbf{s}}}_{\mathrm{j,1}}}{\boldsymbol{\mathcal{X}}}_{\mathrm{j}} \\ 
								{\hat{\boldsymbol{\mathrm{\beta }}}}^{\mathrm{T}}_{\mathrm{j1}}\mathrm{=}\frac{{\hat{\mathbf{s}}}^{\mathrm{H}}_{\mathrm{j,1}}}{{\hat{\mathbf{s}}}^{\mathrm{H}}_{\mathrm{j,1}}{\hat{\mathbf{s}}}_{\mathrm{j,1}}}{\boldsymbol{\mathcal{Y}}}_{\mathrm{j}} \\
								{\hat{\mathbf{s}}}_{\mathrm{j,1}}={\mathrm{e}}_{\mathrm{1}}({\boldsymbol{\mathcal{Y}}}_{\mathrm{j}}{\boldsymbol{\mathrm{\Xi }}}^{-1}_{\mathrm{j}}{{\boldsymbol{\mathcal{Y}}}_{\mathrm{j}}}^\mathrm{H}\mathrm{+}{\boldsymbol{\mathrm{\chi }}}_{\mathrm{j}}{\boldsymbol{\mathrm{\Xi }}}^{-1}_{\mathrm{j}}{{\boldsymbol{\mathrm{\chi }}}_{\mathrm{j}}}^\mathrm{H})\\
								\widehat{{\mathrm{\sigma }}^{\mathrm{2}}_{\mathrm{j}\mathrm{k,}\mathrm{1}}}\mathrm{=}\frac{{\lVert  \boldsymbol{y}_{\mathrm{jk}}\rVert}^{\mathrm{2}}\mathrm{-}\frac{{\left|{{\mathbf{s}}^{\mathrm{H}}_{\mathrm{j,1}}}\boldsymbol{y}_{\mathrm{jk}}\right|}^{\mathrm{2}}}{{\lVert {\mathbf{s}}_{\mathrm{j,1}}\rVert}^{\mathrm{2}}}\mathrm{+}{\lVert  \boldsymbol{x}_{\mathrm{jk}}\rVert}^{\mathrm{2}}\mathrm{-}\frac{{\left|{{\mathbf{s}}^{\mathrm{H}}_{\mathrm{j,1}}}\boldsymbol{x}_{\mathrm{jk}}\right|}^{\mathrm{2}}}{{\lVert {\mathbf{s}}_{\mathrm{j,1}}\rVert}^{\mathrm{2}}}}{\mathrm{2L}} 
						\end{array}} \right.
					\end{equation} 
					where $\mathrm{e}_{\mathrm{1}}(.)$ is the eigenvector associated with the largest eigenvalue of its argument.
					\section{Wald Detector Derivation}\label{D}
					This Appendix presents the derivation of the usual (standard) and alternative Wald detectors. In our case,  we have $\mathrm{\boldsymbol{\theta }}_{\mathrm{r0}}\mathrm{\mathbf{=0}}$ and $\hat{\boldsymbol{\theta} }_{\mathrm{r1}}\mathrm{\mathrm{=}}\hat{\boldsymbol{\alpha}}$. The forms of these test statistics are then given as
					\begin{equation}\label{eqb2}
						\mathrm{\Lambda }_{\mathrm{SW}}\mathrm{(\mathbf{z})}=2 \hat{\boldsymbol{\alpha}}^{\mathrm{H}}( [ \boldsymbol{\mathcal{J}}^{\mathrm{\mathbf{-1}}}( 
						\hat{\boldsymbol{\theta } }_{\rm{1}} ) ]_{\mathrm{rr}} 
						)^{\mathbf{-1}}\hat{\boldsymbol{\alpha}}
					\end{equation} 
					\begin{equation}\label{eqb2b}
						\mathrm{\Lambda }_{\mathrm{AW}}\mathrm{(\mathbf{z})}=2 \hat{\boldsymbol{\alpha}}^{\mathrm{H}}( [ \boldsymbol{\mathcal{J}}^{\mathrm{\mathbf{-1}}}( 
						\hat{\boldsymbol{\theta } }_{\rm{0}} ) ]_{\mathrm{rr}} 
						)^{\mathbf{-1}}\hat{\boldsymbol{\alpha}}
					\end{equation} 
					where $\hat{\boldsymbol{\alpha}}=[\hat{\boldsymbol{\alpha}}_{11}^{\rm T},...,\hat{\boldsymbol{\alpha}}_{\mathrm{{\mathrm{N_{t}}}1}}^{\rm T}]^{\rm T}$. In general,  ${[\boldsymbol{\mathcal{J}}^{\boldsymbol{\mathrm{-}}\boldsymbol{\mathrm{1}}}(\boldsymbol{\mathrm{\theta}})]}_{\mathrm{rr}}$ refers to the submatrix located at the top-left corner of the inverse of the FIM evaluated at $\boldsymbol{\mathrm{\theta}}$. This submatrix has a dimension of $\rm N_tN_r\times N_tN_r$, and it can be obtained from 
					\begin{equation}\label{eqb6}
						\begin{split}({[\boldsymbol{\mathcal{J}}^{\boldsymbol{\mathrm{-}}\boldsymbol{\mathrm{1}}}(\boldsymbol{\mathrm{\theta}})]}_{\mathrm{rr}})^{\mathrm{-}\mathrm{1}}={\boldsymbol{\mathcal{J}}_{\mathrm{rr}}(\boldsymbol{\mathrm{\theta}})\mathrm{-}\boldsymbol{\mathcal{J}}_{\mathrm{rs}}(\boldsymbol{\mathrm{\theta}})\boldsymbol{\mathcal{J}}_{\mathrm{ss}}^{\boldsymbol{\mathrm{-}}\boldsymbol{\mathrm{1}}}(\boldsymbol{\mathrm{\theta}})\boldsymbol{\mathcal{J}}_{\mathrm{rs}}^{\mathrm{H}}(\boldsymbol{\mathrm{\theta}})}\end{split}
					\end{equation}
					It can be shown that $({[\boldsymbol{\mathcal{J}}^{\boldsymbol{\mathrm{-}}\boldsymbol{\mathrm{1}}}(\hat{\boldsymbol{\theta } }_{\mathbf{1}})]}_{\mathrm{rr}})^{\mathrm{-}\mathrm{1}}$ is a block diagonal matrix, whose the j-th matrix block can be found as  
					\begin{equation}
						\begin{split}
							\label{eqb30}
							&[{\boldsymbol{\mathcal{J}}_{\mathrm{rr}}(\hat{\boldsymbol{\theta } }_{\mathbf{1}})\mathrm{-}\boldsymbol{\mathcal{J}}_{\mathrm{rs}}(\hat{\boldsymbol{\theta } }_{\mathbf{1}})\boldsymbol{\mathcal{J}}_{\mathrm{ss}}^{\boldsymbol{\mathrm{-}}\boldsymbol{\mathrm{1}}}(\hat{\boldsymbol{\theta } }_{\mathbf{1}})\boldsymbol{\mathcal{J}}_{\mathrm{rs}}^{\mathrm{H}}(\hat{\boldsymbol{\theta } }_{\mathbf{1}})}
							]^{\mathrm{(j)}}=\lVert  \mathrm{\hat{\mathbf{s}}}_{\mathrm{{j,1}}} 
							\rVert^{\mathrm{2}}\\& \hskip 2cm \times \widehat{\boldsymbol{\mathfrak{O}}}_{\rm j, 1}\Big( 
							\mathrm{\mathbf{I}}_{\mathrm{N}_{\mathrm{r}}}-\frac{{\widehat{\boldsymbol{\mathfrak{O}}}_{\rm j, 1}}\mathrm{\mathbf{\boldsymbol{\hat{\alpha}} 
								}}_{\mathrm{\rm j1}}\mathrm{\mathbf{\boldsymbol{\hat{\hat{\alpha}}} }}_{\mathrm{\rm j1}}^{\mathrm{H}}{\widehat{\boldsymbol{\mathfrak{O}}}_{\rm j, 1}}}{\lVert  
								{\widehat{\boldsymbol{\mathfrak{O}}}_{\rm j, 1}}\mathrm{\mathbf{\boldsymbol{\hat{\alpha}} }}_{\mathrm{\rm j1}}^{\mathrm{\, }} \rVert^{2}} \Big){\widehat{\boldsymbol{\mathfrak{O}}}_{\rm j, 1}}
						\end{split}
					\end{equation}
					where $	{\widehat{\boldsymbol{\mathfrak{O}}}_{\rm j, 1}}\rm{=}\mathrm{diag}\Big(\frac{\mathrm{1}}{\widehat{\mathrm{\sigma }_{\mathrm{j1,1}}}},...,\frac{\mathrm{1}}{\widehat{\mathrm{\sigma }_{\mathrm{jN_r,1}}}}\Big)$.
					Applying (\ref{eqb30}) into (\ref{eqb2}) yields 
					\begin{equation}\label{eqb2XD}
						\mathrm{\Lambda }_{\mathrm{SW}}\mathrm{(\mathbf{z})}=2\sum^{{\mathrm{N}}_{\mathrm{t}}}_{\mathrm{j=1}} \lVert  \mathrm{\hat{\mathbf{s}}}_{\mathrm{{j,1}}} 
						\rVert^{\mathrm{2}}({\widehat{\boldsymbol{\mathfrak{O}}}_{\rm j, 1}}\hat{\boldsymbol{\alpha}})^{\mathrm{H}}{\mathrm{\Pi }}^{\bot }_{{\widehat{\boldsymbol{\mathfrak{O}}}_{\rm j, 1}}\hat{\boldsymbol{\alpha}}}       ({\widehat{\boldsymbol{\mathfrak{O}}}_{\rm j, 1}}\hat{\boldsymbol{\alpha}})
					\end{equation} 
					where
					\begin{equation}\label{eqb2XD}
						{\mathrm{\Pi }}^{\bot }_{{\widehat{\boldsymbol{\mathfrak{O}}}_{\rm j, 1}}\hat{\boldsymbol{\alpha}}}      =
						\mathrm{\mathbf{I}}_{\mathrm{N}_{\mathrm{r}}}-\frac{{\widehat{\boldsymbol{\mathfrak{O}}}_{\rm j, 1}}\mathrm{\mathbf{\boldsymbol{\hat{\alpha}} 
							}}_{\mathrm{\rm j1}}\mathrm{\mathbf{\boldsymbol{\hat{\hat{\alpha}}} }}_{\mathrm{\rm j1}}^{\mathrm{H}}{\widehat{\boldsymbol{\mathfrak{O}}}_{\rm j, 1}}}{\lVert  
							{\widehat{\boldsymbol{\mathfrak{O}}}_{\rm j, 1}}\mathrm{\mathbf{\boldsymbol{\hat{\alpha}} }}_{\mathrm{\rm j1}}^{\mathrm{\, }} \rVert^{2}}
					\end{equation} 
					resulting in $\mathrm{\Lambda }_{\mathrm{SW}}\mathrm{(\mathbf{z})}=0$. However, the alternative Wald test statistic to be computed is now given by $\mathrm{\Lambda }_{\mathrm{AW}}\mathrm{(\mathbf{z})}=\hat{\boldsymbol{\boldsymbol{\alpha}}
					}^{\mathrm{H}}\mathbf{\boldsymbol{\mathcal{J}}}_{\mathrm{rr}}( 
					\hat{\boldsymbol{\theta } }_{{0}})\hat{\boldsymbol{\boldsymbol{\alpha}} }$.
					Applying (\ref{eqb10}) evaluating at $ 
					\boldsymbol{\hat{\boldsymbol{\theta } }}_{{0}}$ into $\mathrm{\Lambda }_{\mathrm{AW}}\mathrm{(\mathbf{z})}$ produces
					\begin{equation}\label{eq17}
						\mathrm{\Lambda }_{\mathrm{AW}}\left( \mathrm{\mathbf{z}} 
						\right)=\sum^{{\mathrm{N}}_{\mathrm{t}}}_{\mathrm{j=1}}{\sum^{{\mathrm{N}}_{\mathrm{r}}}_{\mathrm{k=1}}{\frac{\frac{{\hat{\mathbf{s}}}^{\mathrm{H}}_{\mathrm{j,1}}\boldsymbol{x}_{\mathrm{jk}}\boldsymbol{x}^{\mathrm{H}}_{\mathrm{jk}}{\hat{\mathbf{s}}}_{\mathrm{j,1}}}{{\lVert \boldsymbol{y}_{\mathrm{jk}}\rVert}^{\mathrm{2}}\mathrm{+}{\lVert \boldsymbol{x}_{\mathrm{jk}}\rVert}^{\mathrm{2}}\mathrm{-}\frac{{\hat{\mathbf{s}}}^{\mathrm{H}}_{\mathrm{j,0}}\boldsymbol{y}_{\mathrm{jk}}\boldsymbol{y}^{\mathrm{H}}_{\mathrm{jk}}{\hat{\mathbf{s}}}_{\mathrm{j,0}}}{{\hat{\mathbf{s}}}^{\mathrm{H}}_{\mathrm{j,0}}{\hat{\mathbf{s}}}_{\mathrm{j,0}}}}}{{\hat{\mathbf{s}}}^{\mathrm{H}}_{\mathrm{j,1}}{\hat{\mathbf{s}}}_{\mathrm{j,1}}}\frac{{\lVert {\hat{\mathbf{s}}}_{\mathrm{j,0}}\rVert}^{\mathrm{2}}}{{\lVert {\hat{\mathbf{s}}}_{\mathrm{j,1}}\rVert}^{\mathrm{2}}}}}
					\end{equation}
					In our case, as shown in (\ref{eq31bxt}), the resemblance between the received signals in the reference and surveillance channels implies that the surveillance channel may not offer additional information useful in distinguishing between the null and alternative hypotheses when exploiting estimations of the alternative hypothesis in the standard Wald test, resulting in a zero test statistic for the standard Wald test.  
					\section{Rao Detector Derivation}\label{E}
					According the Rao test of (\ref{eq17X}), the Rao test statistic can be rewritten as
					\begin{align}\label{eq10h}
						\resizebox{.87\hsize}{!}{$\Lambda_{\mathrm{R}}(\mathbf{z})=2\frac{\partial \mathcal{L}(\boldsymbol{\mathrm{\theta}} ; {{\mathbf{z}}})}{\partial \boldsymbol{\mathrm{\alpha}}^*}^{\mathrm{H}}\bigg|_{{\boldsymbol{\theta}=\hat{\boldsymbol{\theta}}_0}}
							[\boldsymbol{\mathcal{J}}^{-\mathbf{1}}(\hat{\boldsymbol{\mathrm{\theta}}}_0)]_{\mathrm{rr}}\frac{\partial \mathcal{L}(\boldsymbol{\mathrm{\theta}}; {{\mathbf{z}}})}{\partial \boldsymbol{\mathrm{\alpha}}^*}\bigg|_{{\boldsymbol{\theta}=\hat{\boldsymbol{\theta}}_0}}$}
					\end{align}
					where $\boldsymbol{\mathcal{J}}^{-1}(\hat{\boldsymbol{\mathrm{\theta}}}_0)=\mathrm{Diag}(\boldsymbol{\mathcal{J}}^{-1}_{\mathrm{rr}}(\hat{\boldsymbol{\mathrm{\theta}}}_0), \boldsymbol{\mathcal{J}}^{-1}_{\mathrm{ss}}(\hat{\boldsymbol{\mathrm{\theta}}}_0))$.
					This implies that
					\begin{equation}\label{eqc5x}
						\resizebox{.98\hsize}{!}{$	[\mathbf{\boldsymbol{\mathcal{J}}}^{-\mathbf{1}}(\hat{\boldsymbol{\mathrm{\theta}}}_0)]_{\mathrm{rr}}=\boldsymbol{\mathcal{J}}_{\mathrm{rr}}^{-1}(\hat{\boldsymbol{\mathrm{\theta}}}_0)=\mathrm{Diag}^{-1}({\boldsymbol{\mathcal{J}}_{\mathrm{rr}}^{(1)}}(\hat{\boldsymbol{\mathrm{\theta}}}_0),\dots,{\boldsymbol{\mathcal{J}}_{\mathrm{rr}}^{(\rm {\mathrm{N_{t}}})}}(\hat{\boldsymbol{\mathrm{\theta}}}_0))$}
					\end{equation}
					where $	\boldsymbol{\mathcal{J}}_{{\mathrm{rr}}}^{\mathrm{(j)}}(\hat{\boldsymbol{\mathrm{\theta}}}_0)$ can be obtained from (\ref{eqb10}), given by
					\begin{equation}\label{eqb10x}
						\boldsymbol{\mathcal{J}}_{{\mathrm{rr}}}^{\mathrm{(j)}}(\hat{\boldsymbol{\mathrm{\theta}}}_0)={\left\Vert {\hat{\mathbf{s}}}_{{\mathrm{j, 0}}}\right\Vert }^{\mathrm{2}}\widehat{\boldsymbol{\mathfrak{O}}}^{2}_{\rm j, 0}
					\end{equation}
					where
					\begin{equation}\label{eqb10xy}
						\widehat{\boldsymbol{\mathfrak{O}}}^{2}_{\rm j,0}\boldsymbol{=}\mathrm{diag}\Big(\frac{\mathrm{1}}{\widehat{{\mathrm{\sigma }}^{\mathrm{2}}_{\mathrm{j}\mathrm{1,0}}}}\mathrm{,\dots ,\ }\frac{\mathrm{1}}{\widehat{{\mathrm{\sigma }}^{\mathrm{2}}_{\mathrm{j}\mathrm{N_r,0}}}}\Big)
					\end{equation}
					In (\ref{eq10h}),   $\frac{\partial \mathcal{L}\left(\boldsymbol{\mathrm{\theta}}_{\mathrm{r}}, \boldsymbol{\mathrm{\theta}}_{\mathrm{s}}; {{\mathbf{z}}}\right)}{\partial {\boldsymbol{\mathrm{\alpha}}}^*}$ represents the gradient of the LLF with respect to $\boldsymbol{\mathrm{\alpha}}^*$. This gradient operator can be partitioned as $\frac{\partial \mathcal{L}(\boldsymbol{\mathrm{\theta}}; {{\mathbf{z}}})}{\partial \boldsymbol{\mathrm{\alpha}}^*}\bigg|_{{\boldsymbol{\theta}=\hat{\boldsymbol{\theta}}_0}}=\big[	\mathrm{\mathbf{u}}_{\mathrm{1,0}}^{\mathrm{T}}, ...,			\mathrm{\mathbf{u}}_{\mathrm{N}_{\mathrm{t}},0}^{\mathrm{T}} \big]^{\mathrm{T}}$
					where $	\mathrm{\mathbf{u}}_{\mathrm{j,0}}=\frac{\mathrm{\partial \mathcal L}( 
						\hat{\boldsymbol{\theta } }_{\mathrm{{r,0}}}, \hat{\boldsymbol{\theta } 
						}_{\mathrm{{s,0}}}\mathrm{\mathbf{;\, 
						}}\mathrm{\mathbf{z}}_{\mathrm{j}} )}{\mathrm{\partial 
						}\mathrm{\boldsymbol{\alpha }}_{\mathrm{j}}^{\mathrm{\ast 
					}}}\mathrm{=}\widehat{\boldsymbol{\mathfrak{O}}}^{2}_{\rm j, 0}\big[ 
					\hat{\mathbf{s}}_{\mathrm{j,0}}^{\mathrm{H}}\mathrm{\boldsymbol{x}}_{\mathrm{j,1}}, ..., 
					\hat{\mathbf{s}}_{\mathrm{j,0}}^{\mathrm{H}}\mathrm{\boldsymbol{x}}_{{\mathrm{jN}}_{\mathrm{r}}} 
					\big]^{\mathrm{T}}$.
					Inserting the MLE of the unknown parameters under the ${\mathcal{H}}_0$ hypothesis, given in (\ref{eqb3a}), into (\ref{eqb10xy}), and plugging  (\ref{eqc5x}) into (\ref{eq10h}), after some algebraic manipulations, we arrive at
					\begin{equation}\label{eqc7}
						\mathrm{\Lambda }_{\mathrm{R}}\left( \mathrm{\mathbf{z}} 
						\right)=\sum^{{\mathrm{N}}_{\mathrm{t}}}_{\mathrm{j=1}}{\sum^{{\mathrm{N}}_{\mathrm{r}}}_{\mathrm{k=1}}{\frac{\frac{{\hat{\mathbf{s}}}^{\mathrm{H}}_{\mathrm{j},0}\frac{\boldsymbol{x}_{\mathrm{jk}}\boldsymbol{x}^{\mathrm{H}}_{\mathrm{jk}}}{{\lVert \boldsymbol{x}_{\mathrm{jk}}\rVert}^{\mathrm{2}}+{\lVert \boldsymbol{y}_{\mathrm{jk}}\rVert}^{\mathrm{2}}}{\hat{\mathbf{s}}}_{\mathrm{j},0}}{{\lVert {\hat{\mathbf{s}}}_{\mathrm{j},0}\rVert}^{\mathrm{2}}}}{\mathrm{1-}\frac{{\hat{\mathbf{s}}}^{\mathrm{H}}_{\mathrm{j},0}\frac{\boldsymbol{y}_{\mathrm{jk}}\boldsymbol{y}^{\mathrm{H}}_{\mathrm{jk}}}{{\lVert \boldsymbol{x}_{\mathrm{jk}}\rVert}^{\mathrm{2}}+{\lVert \boldsymbol{y}_{\mathrm{jk}}\rVert}^{\mathrm{2}}}{\hat{\mathbf{s}}}_{\mathrm{j},0}}{{\lVert {\hat{\mathbf{s}}}_{\mathrm{j},0}\rVert}^{\mathrm{2}}}}}}
					\end{equation}
					
					\section{Gradient Detector Derivation}\label{F}
					The Gradient test statistic of (\ref{eq16}) can be rewritten as
					\begin{equation}\label{eq11x}
						\Lambda_{\mathrm{G}}(\mathbf{z})=2\mathcal{Q}\Big({\frac{\partial{\mathcal {L }}\left(\boldsymbol{\mathrm{\theta}}; {{\mathbf{z}}}\right)}{\partial \boldsymbol{\mathrm{\alpha}}^*}}\bigg|_{{\boldsymbol{\theta}=\hat{\boldsymbol{\theta}}_0}}\hat{\boldsymbol{\alpha}}\Big)
					\end{equation}
					In (\ref{eq11x}), $\hat{\boldsymbol{\alpha}}$ can be represented as $	\hat{\boldsymbol{\alpha}}=\big[ 
					\hat{\boldsymbol{\alpha} }_{\mathrm{11}}^{\mathrm{T}},..., 
					\hat{\boldsymbol{\alpha} }_{\mathrm{N}_{\mathrm{t}}1}^{\mathrm{T}}  
					\big]^{\mathrm{T}}$
					in which $	\hat{\boldsymbol{\alpha} }_{\mathrm{j1}}^{\mathrm{T}}\mathrm{=}\Big[ 
					\frac{\hat{\mathbf{s}}_{\mathrm{j,1}}^{\mathrm{H}}\mathrm{\boldsymbol{x}}_{\mathrm{j1}}}{\hat{\mathbf{s}}_{\mathrm{j,1}}^{\mathrm{H}}\hat{\mathbf{s}}_{\mathrm{j,1}}}, ...,			
					\frac{\hat{\mathbf{s}}_{\mathrm{j,1}}^{\mathrm{H}}\mathrm{\boldsymbol{x}}_{\mathrm{j1}}}{\hat{\mathbf{s}}_{\mathrm{j,1}}^{\mathrm{H}}\hat{\mathbf{s}}_{\mathrm{j,1}}} 
					\Big]$
					Using these, after some algebraic manipulations, leads to
					\begin{equation}\label{eqd6}
						\mathrm{\Lambda }_{\mathrm{G}}( \mathrm{\mathbf{z}} 
						)\mathrm{=}2\mathcal{Q}{\Bigg(\sum^{{\mathrm{N}}_{\mathrm{t}}}_{\mathrm{j=1}}{\sum^{{\mathrm{N}}_{\mathrm{r}}}_{\mathrm{k=1}}{\frac{\frac{{\hat{\mathbf{s}}}^{\mathrm{H}}_{\mathrm{j},0}\frac{\boldsymbol{x}_{\mathrm{jk}}\boldsymbol{x}^{\mathrm{H}}_{\mathrm{jk}}}{{\lVert \boldsymbol{x}_{\mathrm{jk}}\rVert}^{\mathrm{2}}+{\lVert \boldsymbol{y}_{\mathrm{jk}}\rVert}^{\mathrm{2}}}{\hat{\mathbf{s}}}_{\mathrm{j},0}}{{\lVert {\hat{\mathbf{s}}}_{\mathrm{j},0}\rVert}^{\mathrm{2}}}}{\mathrm{1-}\frac{{\hat{\mathbf{s}}}^{\mathrm{H}}_{\mathrm{j},0}\frac{\boldsymbol{y}_{\mathrm{jk}}\boldsymbol{y}^{\mathrm{H}}_{\mathrm{jk}}}{{\lVert \boldsymbol{x}_{\mathrm{jk}}\rVert}^{\mathrm{2}}+{\lVert \boldsymbol{y}_{\mathrm{jk}}\rVert}^{\mathrm{2}}}{\hat{\mathbf{s}}}_{\mathrm{j},0}}{{\lVert {\hat{\mathbf{s}}}_{\mathrm{j},0}\rVert}^{\mathrm{2}}}}}}\Bigg)}
					\end{equation}
					
					\section{Durbin Detector Derivation} \label{G}
					In this Appendix, we would like to make an interesting statement, namely that the Rao test is truely equivalent to the Durbin test. To do so, let    
					$\hat{\boldsymbol{\mathrm{\theta}}}_{\mathrm{r}, 10}$ be the MLE of $\boldsymbol{\mathrm{\theta}}_{\rm r}$ under $\mathcal{H}_{1}$ with $\boldsymbol{\mathrm{\theta}}_{\rm s}=\hat{\boldsymbol{\mathrm{\theta}}}_{\rm s,0}$, given by
					\begin{equation} \label{eqe2}
						\hat{\boldsymbol\theta }_{\mathrm{r,10}}\mathrm{=}\sup_{\mathrm{\boldsymbol{\theta }}_{\mathrm{r}}}\big\{ \mathcal{L}( 
						\mathrm{\boldsymbol{\theta }}_{\mathrm{r}}\mathrm{\mathbf{,}}\hat{\boldsymbol\theta 
						}_{\mathrm{s0}}\mathrm{\mathbf{;}}\mathbf{\, }\mathrm{\mathbf{z}} ) 
						\big\}
					\end{equation} 
					In our case, one can show that $	[\mathbf{\boldsymbol{\mathcal{J}}}(\hat{\boldsymbol{\mathrm{\theta}}}_0)]_{\mathrm{rr}}[\mathbf{\boldsymbol{\mathcal{J}}}^{-1}(\hat{\boldsymbol{\mathrm{\theta}}}_0)]_{\mathrm{rr}}[\mathbf{\boldsymbol{\mathcal{J}}}(\hat{\boldsymbol{\mathrm{\theta}}}_0)]_{\mathrm{rr}}\mathrm{=}\boldsymbol{\mathcal{J}}_{\mathrm{rr}}(\hat{\boldsymbol{\mathrm{\theta}}}_0)$,
					resulting in $\mathrm{\Lambda }_{\mathrm{D}}\left( \mathrm{\mathbf{z}} 
					\right)\mathrm{=}\hat{\boldsymbol{\theta }
					}_{\mathrm{r,10}}^{\mathrm{H}}\boldsymbol{\mathcal{J}}_{\mathrm{rr}}(\hat{\boldsymbol{\mathrm{\theta}}}_0)\hat{\boldsymbol{\theta } }_{\mathrm{r,10}}$
					where $\hat{\boldsymbol{\theta } }_{\mathrm{r,10}}\mathrm{=}\big[ 
					\hat{\boldsymbol{\alpha} }_{\mathrm{1,10}}^{\mathrm{T}}  , ..., 
					\hat{\boldsymbol{\alpha} }_{\mathrm{N}_{\mathrm{t}},10}^{\mathrm{T}} 
					\big]^{\mathrm{T}}$  
					with $	\hat{\boldsymbol{\alpha} }_{\mathrm{j,10}}^{\mathrm{T}}  \mathrm{=}\Big[ 			
					\frac{\hat{\mathbf{s}}_{\mathrm{j,0}}^{\mathrm{H}}\mathrm{\boldsymbol{x}}_{\mathrm{j1}}}{\hat{\mathbf{s}}_{\mathrm{j,0}}^{\mathrm{H}}\hat{\mathbf{s}}_{\mathrm{j,0}}}, ...,			
					\frac{\hat{\mathbf{s}}_{\mathrm{j,0}}^{\mathrm{H}}\mathrm{\boldsymbol{x}}_{{\mathrm{jN}}_{\mathrm{r}}}}{\hat{\mathbf{s}}_{\mathrm{j,0}}^{\mathrm{H}}\hat{\mathbf{s}}_{\mathrm{j,0}}} 
					\Big]$. 
					Applying these leads to the same test statistic as the Rao one. Thus, in our case, we show that the Durbin test coincides
					with the Rao test. 

				\end{document}